\documentclass[aps,prd,longbibliography,nofootinbib]{revtex4-2}
\usepackage{graphicx}
\usepackage{dcolumn}
\usepackage{amsmath,amssymb,epsfig}
\usepackage{paralist}
\usepackage{comment}
\usepackage{multirow}
\usepackage{color,soul}
\usepackage{suffix}
\usepackage{mathtools}

\usepackage[breaklinks=true,colorlinks=true,linkcolor=magenta,citecolor=blue]{hyperref}

\usepackage{wasysym}

\renewcommand{\vec}[1]{\boldsymbol{\mathrm{#1}}}

\usepackage{booktabs}
\usepackage{siunitx} 
\usepackage[framemethod=tikz]{mdframed} 

\newcommand{\TT}{\ensuremath{\mathrm{TT}}}
\newcommand{\TCG}{\ensuremath{\mathrm{TCG}}}
\newcommand{\TDB}{\ensuremath{\mathrm{TDB}}}
\newcommand{\TCB}{\ensuremath{\mathrm{TCB}}}
\newcommand{\TCL}{\ensuremath{\mathrm{TCL}}}
\newcommand{\TL}{\ensuremath{\mathrm{TL}}}

\newcommand{\LG}{6.969\,290\,134\times10^{-10}}
\newcommand{\LC}{1.480\,826\,867\,41\times10^{-8}}
\newcommand{\LB}{1.550\,519\,767\,72\times10^{-8}}

\def\s{{\rm s}}

\def\ve#1{{\bf #1}}

\def\vx{\ve{x}}

\def\vX{\ve{X}}

\def\cO{{\cal O}}

\def\beqn{\begin{eqnarray*}}
\def\eeqn{\end{eqnarray*}}

\newcommand{\be}{\begin{equation}}
\newcommand{\ee}{\end{equation}}
\newcommand{\ba}{\begin{eqnarray}}
\newcommand{\ea}{\end{eqnarray}}

\begin{document}

\title{Relativistic framework for high-precision GNSS processing in GCRS/BCRS\\ with extension to cislunar space}


\author{Slava G. Turyshev, Yoaz E. Bar-Sever  and William I. Bertiger
}   

\affiliation{ 
Jet Propulsion Laboratory, California Institute of Technology,\\
4800 Oak Grove Drive, Pasadena, CA 91109-0899, USA
}%

\date{\today}

\begin{abstract}

We present an implementation-oriented relativistic modeling framework for high-precision {\tt GNSS} processing consistent with the IAU-adopted Barycentric and Geocentric Celestial Reference Systems ({\tt BCRS}/{\tt GCRS}) and their associated time scales ({\tt TCB/TDB} and {\tt TCG/TT}). We derive explicit ${\cal O}(c^{-2})$ transformations for position, velocity, and acceleration between {\tt TT}-compatible {\tt GCRS} quantities and {\tt TDB}-compatible {\tt BCRS} quantities, and provide screened operational forms with conservative remainder bounds that quantify state-map truncation errors for cm-class orbit modeling. For $10^{-16}$-class fractional-frequency transfer, the ${\cal O}(c^{-4})$ clock-rate terms identified below must be retained or explicitly included in the observable error budget. We implement a {\tt BCRS}-native processing option in JPL's GipsyX and verify it internally via a 24~h round-trip {\tt GCRS}$\rightarrow${\tt BCRS}$\rightarrow${\tt GCRS} propagation-and-transform closure test at the few-mm level, demonstrating consistency of the implemented dynamical model and state transformations under matched force-model assumptions. To support emerging Earth--Moon applications, we define a Lunicentric Celestial Reference System ({\tt LCRS}), its coordinate time ({\tt TCL}), and a scaled lunar-surface time ({\tt TL}), and specify a minimal near-rectilinear halo orbit ({\tt NRHO})-like regression test that exercises the {\tt BCRS}$\leftrightarrow${\tt LCRS} transformation chain together with the 1PN barycentric light-time model. End-to-end cislunar navigation performance additionally depends on signal availability and estimation strategy; the present work provides the relativistic reference-frame and time-transfer infrastructure needed to model observables at the centimeter and tens-of-picoseconds level.
 
\end{abstract}

\maketitle


\section{Introduction}

High‐precision spacecraft navigation requires a fully relativistic treatment of both time and space observables \cite{Moyer:2003,Montenbruck-Gill:2005,Soffel-Han:2019}.  Central to this treatment is the choice of an appropriate coordinate reference system.  The Barycentric Celestial Reference System ({\tt BCRS}), with origin at the solar‐system barycenter, underlies deep‐space applications such as lunar laser ranging ({\tt LLR}), very long baseline interferometry ({\tt VLBI}), and the construction of planetary and lunar ephemerides \cite{Park-etal:2021}.  By contrast, the Geocentric Celestial Reference System ({\tt GCRS}), centered on Earth’s mass center, is the natural frame for precise orbit determination of {\tt GNSS} and other Earth‐orbiting satellites \cite{Montenbruck-Gill:2005}.  Transforming observables between these frames requires inclusion of all \(O(c^{-2})\) corrections—gravitational time dilation, velocity‐dependent terms, multipolar potential perturbations, and frame‐dragging effects—to avoid systematic errors in both timing and positioning.

For Earth-orbiting spacecraft, {\tt GCRS}-based models incorporate relativistic corrections to the Newtonian equations of motion, including adjustments to central gravitational acceleration and terms arising from Lense-Thirring precession (caused by Earth's angular momentum), geodesic precession (due to spacetime curvature), and Coriolis effects \cite{Petit-Luzum:2010, Kopeikin-book-2011}. These relativistic corrections, while relatively small compared to non-relativistic forces such as atmospheric drag and solar radiation pressure, are critical to achieving the cm-level positional accuracy required for modern navigation systems. Over time, accumulated relativistic errors could cause significant positioning drift in long-duration satellite missions, making it essential to integrate these terms into orbit determination models to prevent systematic errors.

For solar-system navigation beyond Earth orbit it is common to express spacecraft and planetary dynamics in a global barycentric reference system; the IAU-adopted BCRS with coordinate time TCB (or its scaled form TDB) provides the standard realization for modern ephemerides and interplanetary tracking \cite{Moyer:2003,Soffel-etal:2003,Turyshev-etal:2025}.
The mean rate offset between {\tt TT} and {\tt TCB} is $L_{\tt B}=1.55051976772\times10^{-8}$, corresponding to $\simeq1.34$ ms/day; inconsistent handling of this rate biases modeled one-way and two-way coordinate light-time at the millisecond-per-day level near Earth \cite{Kaplan:2005,Turyshev-etal:2025}.

As {\tt GNSS} technology advances with the goal of achieving cm-level positional accuracy, the inclusion of general relativistic corrections in orbit determination models becomes indispensable \cite{Turyshev-etal:2013, Turyshev:2014dea, Sosnica:2021,Turyshev-Toth:2023-grav-phase,Sosnica:2025}.  Modern cm-class GNSS solutions rely on precise orbit and clock products and precise point positioning (PPP) strategies;
see, e.g., \cite{ZumbergeEtAl1997PPP,KoubaHeroux2001PPP}.
These corrections affect multiple aspects of orbit determination, including light-time equations, clock synchronization, station coordinates, and signal propagation delays  \cite{Montenbruck-Gill:2005,Teunissen-Montenbruck:2017,Tapley-Schultz-Born:2004}. For {\tt GNSS} links, the dominant Shapiro contribution from the Sun is at the few-nanosecond level (meters in range) with mm-level annual modulation, while Earth's contribution is at the \(\lesssim 50\)~ps level (Sec.~\ref{sec:lighttime}). 
Inconsistent handling of the {\tt TT}--{\tt TCB} (or {\tt TT}--{\tt TDB}) rate introduces a secular \(\sim1.34~{\rm ms/day}\) bias in coordinate light-time near Earth \cite{Kaplan:2005,Turyshev-etal:2025}. Without these corrections, accumulated errors could lead to significant positional offsets, potentially corresponding to meter‐level range errors if uncorrected \cite{Turyshev-etal:2025}.  

For interplanetary and lunar applications, navigation observables and dynamical models are often formulated in barycentric coordinates using a first post-Newtonian ($1$PN) $N$-body description consistent with the IAU {\tt BCRS} metric \cite{MTW,Moyer:2003}. The secular rate difference between {\tt TT} and {\tt TCB} is $L_{\tt B}=1.5505\times10^{-8}$, producing a drift of $\simeq 1.34$~ms/day, while the periodic part of ${\tt TDB}-{\tt TT}$ reaches $\simeq 1.6$~ms \cite{Kaplan:2005}. These time-scale effects correspond to $\mathcal{O}(10^{2}\!-\!10^{5})\,\mathrm{m}$ one-way range-equivalent biases if left untreated, and thus must be handled explicitly (or absorbed consistently) in any mixed {\tt GCRS}/{\tt BCRS} estimation problem. At the link level, the $1$PN Shapiro delay contributes up to $\sim 2.7$~ns from the Sun and $\sim 40$~ps from the Earth for a {\tt GNSS} MEO signal path \cite{Petit-Luzum:2010}; near solar conjunction on interplanetary links it can reach $\mathcal{O}(10\!-\!10^{2})\,\mu\mathrm{s}$ \cite{Moyer:2003}. Accordingly, the key requirement for high-precision processing is an internally consistent set of (i) reference-frame and time-scale transformations, (ii) dynamical models, and (iii) signal-propagation models; the particular choice of integration frame ({\tt GCRS} vs {\tt BCRS}) is a design choice rather than a physical necessity.

A consistent treatment across {\tt GCRS} and {\tt BCRS} is therefore required for cm/ps-class processing when combining geocentric {\tt GNSS} models with barycentric ephemerides and relativistic light-time. While standard practice propagates {\tt GNSS} spacecraft in the {\tt GCRS} and transforms states/observables as needed, a {\tt BCRS}-native processing option becomes convenient when a single
estimation pipeline must also accommodate cislunar or interplanetary geometries.

In this work, we report on the implementation of  the {\tt BCRS} framework in JPL's GipsyX \cite{Bertiger-etal:2020}, demonstrating mm-level closure results. Specifically, we demonstrate 24 h round-trip {\tt GCRS}$\rightarrow${\tt BCRS} propagation-and-transform closure at the few-mm level (Fig.~\ref{fig:closure}), i.e., internal consistency of the dynamical model and state transformations when implemented in GipsyX. 
The relativistic reference systems used here follow the IAU/IERS conventions and are not new; the contribution of this work is to provide an implementation-oriented, error-budgeted, and software-verified formulation at the levels relevant to modern {\tt GNSS} processing and its extension to Earth--Moon geometries. 
In particular, we:
\begin{enumerate}
\item Give explicit closed-form ${\cal O}(c^{-2})$ transformations for position, velocity, and acceleration between {\tt TT}-compatible {\tt GCRS} quantities and {\tt TDB}-compatible {\tt BCRS} quantities, with conservative remainder bounds and screening thresholds;
\item Provide an internal {\tt GCRS}$\leftrightarrow${\tt BCRS} frame-closure experiment in JPL's GipsyX demonstrating few-mm agreement over 24~h for a GPS-like orbit when using dynamically consistent force models; and
\item Extend the same construction to a Lunicentric Celestial Reference System ({\tt LCRS}), including its coordinate time ({\tt TCL}) and a scaled lunar-surface time ({\tt TL}), together with a minimal NRHO-like regression test that exercises the {\tt BCRS}$\leftrightarrow${\tt LCRS} transformation chain and the $1$PN light-time model.
\end{enumerate}

While the underlying {\tt BCRS}/{\tt GCRS} framework is standard (IAU/IERS), this work contributes:
(i) closed-form ${\cal O}(c^{-2})$ state maps written in an implementation-oriented form with
\emph{screened operational subsets} and explicit conservative remainder bounds,
(ii) a software-verified 24~h {\tt GCRS}$\rightarrow${\tt BCRS}$\rightarrow${\tt GCRS} frame-closure test at the few-mm level in GipsyX,
and (iii) a minimal {\tt BCRS}$\leftrightarrow${\tt LCRS} cislunar regression configuration that exercises the transformation chain together with
the 1PN barycentric light-time model.

In this paper, ``screening'' means evaluating each displayed post-Newtonian state-transformation term over a prescribed state envelope and omitting only those terms whose combined worst-case contribution remains below the adopted transformation-remainder budget. The omitted terms are not assumed to vanish; they define the conservative remainders reported below. For the GPS-like MEO envelope used in Sec.~\ref{sec:GCRS-pvac}, the resulting remainder budgets are \(7.01\times10^{-5}\)~m for position, \(1.29\times10^{-7}\)~m\,s\(^{-1}\) for velocity, and \(6.68\times10^{-14}\)~m\,s\(^{-2}\) for acceleration. Other orbit classes require re-screening with the appropriate \((|\vec X|,|\dot{\vec X}|,|\ddot{\vec X}|)\) envelope.

This paper is structured as follows: In Section~\ref{sec:GNSS}, we present the relativistic framework necessary for processing {\tt GNSS} data from the {\tt BCRS}, accounting for current measurement precision. This section addresses the relativistic time scales used in astronomical practice (such as {\tt TT}, {\tt TCG}, {\tt TCB} and {\tt TDB}) and provides expressions to transform positions, velocities, and accelerations between various  practically relevant reference frames. 
In Section~\ref{sec:implement}, we demonstrate the practical implementation of processing {\tt GNSS} data within the {\tt BCRS}. This section focuses on applying relativistic corrections, including the transformation of observables between reference frames, and addresses the challenges associated with data analysis when accounting for both the {\tt BCRS} and {\tt GCRS} systems. Finally, in Section~\ref{sec:conclude}, we summarize the results and discuss their implications for future spacecraft navigation, including applications for extending {\tt GNSS} capabilities beyond Earth's vicinity and into the Earth-Moon system. In Appendix~\ref{sec:time-M}, we introduce and discuss the Lunicentric Celestial Reference System ({\tt LCRS}) with the associated times {\tt TCL} and {\tt TL}, which is essential for accurately modeling {\tt GNSS} receivers operating in cislunar space, where gravitational and relativistic effects must be well-modeled.  

\section{Relativistic framework for processing {\tt GNSS} data}
\label{sec:GNSS}

\subsection{Coordinate transformations between {\tt GCRS} and {\tt BCRS}}

\subsubsection{Coordinate transformation from {\tt BCRS} to {\tt GCRS}}
\label{sec:BCRS-to-GCRS}

According to the IAU conventions, the {\tt GCRS} is realized by the geocentric metric tensor \(G_{mn}\) in coordinates \((T,\mathbf{X})\), where  $T \equiv {\tt TCG}$ 
denotes Geocentric Coordinate Time \cite{Soffel-etal:2003}.  Following the approach of \cite{Turyshev-etal:2025,Turyshev-scales:2025}, we express \(G_{mn}\) to the order required for modern {\tt GNSS} timing applications.  

Since current {\tt GNSS} clocks reach fractional frequency stabilities of order \(3\times10^{-15}\), the standard transformation\footnote{The notational conventions employed in this paper are those used in \cite{Moyer:2003}. Letters from the second half of the Latin alphabet, $m, n,...=0...3$ denote spacetime indices. Greek letters $\alpha, \beta,...=1...3$ denote spatial indices. The metric $\eta_{mn}={\rm diag}(+1,-1,-1,-1)$ is the Minkowski spacetime in the Cartesian representation. We employ the Einstein summation convention with indices being lowered or raised using $\eta_{mn}$.  The scalar \(\gamma\) is reserved for the PPN curvature parameter. We adopt the IAU/IERS positive-potential convention \(U,W,w>0\): for a monopole \(U=GM/R\), \(\vec\nabla U=-GM\,\vec R/R^{3}\), so the Newtonian acceleration is written \(\vec a=\vec\nabla U\). Readers using the field-theory convention \(\Phi=-U\) should translate \(U\mapsto-\Phi\). Powers of \(G\) and negative powers of \(c\) are used as order-counting devices.} from a {\tt BCRS} event, represented by \(\bigl(t={\tt TCB},\,\vec r_{\tt E}=\vec x-\vec x_{\tt E}(t)\bigr)\), to its {\tt GCRS} coordinates \((T={\tt TCG},\,\vec X)\) is required at ${\cal O}(c^{-2})$. With the notation of Refs.~\cite{Turyshev-etal:2025,Turyshev-scales:2025}, the transformation takes the form:
{}
\begin{eqnarray}
T &=& t-c^{-2}\Big\{\int^{t}_{t_{0}}\Big({\textstyle\frac{1}{2}}v_{\tt E}^{2} +
\sum_{{\tt B\not= E}} {G M_{\tt B} \over r_{\tt BE}} \Big)dt + (\vec v_{\tt E}  \cdot \vec r_{\tt E} ) \Big\}+ {\cal O}\Big(1.10\times 10^{-16}(t-t_0)\Big),
 \label{eq:coord-tr-T1-RR}\\
\vec X &=&  \vec r_{\tt E}  + c^{-2} \Big\{ {\textstyle\frac{1}{2}}( \vec v_{\tt E}  \cdot\vec r_{\tt E} )\vec v_{\tt E}  +\sum_{{\tt B\not= E}}{G M_{\tt B} \over r_{\tt BE}} \vec r_{\tt E}  +(\vec a_{\tt E} \cdot \vec r_{\tt E} )\vec r_{\tt E} - {\textstyle\frac{1}{2}}r^2_{\tt E} \vec a_{\tt E} \Big\}  + {\cal O}\Big(1.28\times 10^{-12}~{\rm m}\Big),
 \label{eq:coord-tr-XrecRR}
\end{eqnarray}
where $\vec x_{\tt E}$  and $\vec v_{\tt E}=d\vec x_{\tt E}/dt$ being the Earth's position and velocity vectors  in the {\tt BCRS}, with their magnitudes known to be $x_{\tt E}=|\vec x_{\tt E}|\equiv {\rm AU}\simeq 1.49\,597\,870\,700\times 10^{11}$\,m, and $v_{\tt E}=|\vec v_{\tt E}|=29\, 784.7\,$\,m/s, see Table~\ref{tab:nom-values}. Here $t_0$ is an arbitrary reference epoch (e.g., J2000) that fixes the integration constant in the time transformation, and $T_0\equiv T(t_0)$ denotes the corresponding {\tt TCG} epoch. All modeled observables depend only on time differences, so the choice of $t_0$ cancels provided it is used consistently throughout the transformation chain. Note that in their most accurate form (see \cite{Soffel-etal:2003}), the time transformations  (\ref{eq:coord-tr-T1-RR}) also include terms $\propto{\cal O}(c^{-4})$. Therefore, the uncertainty in (\ref{eq:coord-tr-T1-RR}) comes from the omitted ${\cal O}(c^{-4})$ rate terms. For the dominant solar contribution, with \(r_{\tt SE}=|\vec x_{\tt S}-\vec x_{\tt E}|\), these terms scale  as $\sim c^{-4}(-\frac{1}{8}v_{\tt E}^4-\frac{3}{2}v^2_{\tt E}G M_{\tt S}/r_{\tt SE}+{\textstyle\frac{1}{2}}(G M_{\tt S}/r_{\tt SE})^2)\simeq -1.10\times 10^{-16}$, corresponding to an accumulated time magnitude of approximately \(9.50\)~ps/d. 

Consider a GPS spacecraft at an altitude of $h_{\tt GPS}=20\,200$\,km, thus its distance from the origin of the {\tt GCRS} is $r_{\tt GPS}=R_{\tt E}+h_{\tt GPS}=26\,578$\,km and the geocentric velocity is $v_{\tt GPS}\simeq \sqrt{GM_{\tt E}/r_{\tt GPS}}\simeq 3\,872.64$\, m/s. Then, the $c^{-2}$ terms under the integral  in (\ref{eq:coord-tr-T1-RR}) were evaluated to contribute up to  $c^{-2}({\textstyle\frac{1}{2}}v_{\tt E}^{2} +\sum_{{\tt B\not= E}} {G M_{\tt B} / r_{\tt BE}})\simeq c^{-2}(\tfrac32GM_{\tt S}/{\rm AU}+GM_{\tt M}/r_{\tt EM})\simeq 1.48061\times 10^{-8}$, while position-dependent term has the magnitude of  $c^{-2}(\vec v_{\tt E}  \cdot \vec r_{\tt E} ) \lesssim c^{-2}v_{\tt E} r_{\tt GPS}\approx 8.81\times 10^{-6}\,{\rm s}=8.81 \mu$s.  In (\ref{eq:coord-tr-XrecRR}), the velocity-dependent term  $c^{-2} {\textstyle\frac{1}{2}} (\vec v_{\tt E}  \cdot \vec r_{\tt E} )\vec v_{\tt E} \simeq c^{-2} {\textstyle\frac{1}{2}} v^2_{\tt E}  r_{\tt GPS} \simeq 13.12$\,cm reflects the Lorentz contraction, while the term $c^{-2}\sum_{{\tt B\not= E}} ({G M_{\tt B}/ r_{\tt BE}}) \vec r_{\tt E}\simeq c^{-2}(G M_{\tt S}/ {\rm AU}+GM_{\tt M}/r_{\tt EM}) r_{\tt GPS}\simeq 26.24$\, cm corresponds to an isotropic rescaling of length due to the external potential. Both terms are significant and must be included in the model. 

The acceleration-dependent terms in (\ref{eq:coord-tr-XrecRR}) may contribute up to $2.68 \times 10^{-6}$~m at a ground station and $4.66 \times 10^{-5}$~m at $r_{\tt GPS}$. Although these offsets are small, they become significant when comparing spacecraft accelerations in the {\tt BCRS} and the {\tt GCRS}. The dominant uncertainty in (\ref{eq:coord-tr-XrecRR}) stems from the neglected solar quadrupole moment, \(J_2 = 2.25\times10^{-7}\) \cite{Park:2017,MecheriMeftah:2021}, which induces corrections of order
$c^{-2}w_{2,{\tt S}}(t,\vx) \vec r_{\tt E}  \simeq c^{-2} (GM_{\tt S} J_2R_{\tt S}^2/{\rm AU}^3)R_{\tt GPS}\sim  1.28 \times 10^{-12}$\,m, a value that is entirely negligible and therefore serves as a conservative error bound.

\subsubsection{Inverse transformation from {\tt GCRS} to {\tt BCRS}}
\label{sec:GCRS-to-BCRS}

To compare the relativistic description of spacecraft acceleration across different coordinate reference systems, we also require the inverse map, namely the transformation from {\tt GCRS} coordinates to the simultaneous geocentric {\tt BCRS} difference vector. This inverse to Eqs.~(\ref{eq:coord-tr-T1-RR})--(\ref{eq:coord-tr-XrecRR}) is obtained by noting that the relativistic time transformation at the {\tt GCRS} origin (\(\mathbf{X}=0\)) is
{}
\begin{equation}
T = t^* - {1 \over c^2}  \int_{t_0}^{t^*} \Big({\textstyle{\textstyle{\textstyle{1 \over 2}}}}v_{\tt E} ^2 + \sum_{{\tt B\not= E}} {G M_{\tt B} \over r_{\tt BE}}\Big)dt+{\cal O}\Big(1.10\times 10^{-16}(t^*-t_0)\Big).
 \label{eq:tT+}
\end{equation}
and the {\tt TCG} representation of the motion of the Earth with respect to the {\tt BCRS} given as below
\begin{equation}
\bar{\vec x}_{\tt E} (T) = \vec x_{\tt E} (t^*),
\label{eq:EeqM}
\end{equation}
Eqs.~(\ref{eq:tT+})--(\ref{eq:EeqM}) may be evaluated either by numerical quadrature and inversion or, when appropriate, by an analytic approximation. For the forward \({\tt BCRS}\rightarrow{\tt GCRS}\) transformation, the integral in Eq.~(\ref{eq:coord-tr-T1-RR}) is evaluated along the Earth's barycentric ephemeris. For the inverse \({\tt GCRS}\rightarrow{\tt BCRS}\) transformation, the barycentric epoch \(t^\ast=t^\ast(T)\) is obtained by inverting Eq.~(\ref{eq:tT+}). Eq.~(\ref{eq:tT+}) is simply Eq.~(\ref{eq:coord-tr-T1-RR}) evaluated at the geocenter, \(\vec X=0\).

We can connect the Earth's center world line, $\vec x_{\tt E} (t)$, to an observer's world line, $\vec x(t)$, using a space-time curve that is orthogonal to both world lines.  Note that in data-processing programs the relevant coordinate differences are not simply $\Delta \vec r$, but rather the simultaneous barycentric differences $\Delta \vec r = \vec x(t^*) - \vec x_{\tt E}(t^*)$ evaluated at a common barycentric coordinate time $t^\ast$.\footnote{
Throughout, $t^\ast$ denotes the {\tt BCRS} coordinate time at which the Earth center and the event on the
observer worldline are taken to be simultaneous in the adopted {\tt BCRS} slicing (i.e., the subtraction
$\mathbf{x}(t^\ast)-\mathbf{x}_{\tt E}(t^\ast)$ is formed at equal barycentric coordinate time).
This is the quantity that is directly compatible with ephemeris states tabulated versus {\tt TCB}/{\tt TDB}.}
 
Accordingly, in the inverse transformations of a geocentric position vector $\vec X$ from the {\tt GCRS} (compatible with $\tt TCG$, $\tt TAI$, or $\tt TT$ for ground stations) to the solar system barycentric frame (compatible with $\tt TDB$), one must account for the Lorentz contraction of $\vec X$ and make relativistic adjustments to the scale of $\vec X$, as illustrated below:
{}
\begin{eqnarray}
t &=& T + {1 \over c^2} \Big\{ \int_{T_0}^T \Big({\textstyle{\textstyle{\textstyle{1 \over 2}}}} v_{\tt E} ^2 + \sum_{{\tt B\not= E}} {G M_{\tt B} \over r_{\tt BE}}\Big)dT + ({\vec  v}_{\tt E}  \cdot {\vec X})\Big\} + {\cal O}\Big(1.10\times 10^{-16}\Big)(T-T_0), 
\label{eq:coord-tr-T1-rec2++}\\
\vec r_{\tt E}  &=&\vec X-c^{-2}\Big\{ {\textstyle\frac{1}{2}} ( \vec v_{\tt E}  \cdot\vec X)\vec v_{\tt E} +\sum_{{\tt B\not= E}} {G M_{\tt B} \over r_{\tt BE}} \vec X+ (\vec a_{\tt E} \cdot \vec X)\vec X- {\textstyle\frac{1}{2}}X^2\vec a_{\tt E} \Big\}  + 
{\cal O}\Big(1.28\times 10^{-12}~{\rm m}\Big).
 \label{eq:coord-tr-Xrec2++}
\end{eqnarray} 
\noindent Here $T_0$ is the conventional reference epoch used in the IAU scaling definitions (see Sec.~\ref{sec:time-TT-TDB} and  Table~\ref{tab:timescales} and notations sued throughout shown in Table~\ref{tab:frames-times}); the corresponding {\tt TCB} epoch at the geocenter is $t_0=t^\ast(T_0)$.

As a result, Eqs.~(\ref{eq:coord-tr-T1-RR})--(\ref{eq:coord-tr-XrecRR}) give the {\tt BCRS}$\rightarrow${\tt GCRS} event transformation, while Eqs.~(\ref{eq:coord-tr-T1-rec2++})--(\ref{eq:coord-tr-Xrec2++}) give the inverse {\tt GCRS}$\rightarrow${\tt BCRS} transformation to the same post-Newtonian order. Note that a similar set of transformations, albeit without acceleration-dependent terms, has been implemented for processing LLR data, as discussed in \cite{Williams-Boggs:2015}. Special attention must be given when dealing with either ground stations or Earth-orbiting spacecraft, as the coordinates \(\vec X\) will differ \cite{Turyshev-scales:2025}. Specifically, terms proportional to \(1/c^4\) in the time transformations (\ref{eq:coord-tr-T1-RR}) and (\ref{eq:coord-tr-T1-rec2++}) are necessary if a frequency transfer accuracy of approximately \(1.0\times10^{-16}\) is required, and acceleration-dependent terms in position transformations (\ref{eq:coord-tr-XrecRR}) and (\ref{eq:coord-tr-Xrec2++}) are needed for positional accuracy within a few millimeters.

\subsection{Times $\tt TT$ and $\tt TDB$ and related scaling constants}
\label{sec:time-TT-TDB}

\subsubsection{Relativistic time scales}
\label{sec:time-scale}

To compare clocks following different world-lines in the solar system one needs to account for the differences in  relevant dynamical environment. We first consider  the relationship between $\tt TT$  and $\tt TCG$. $\tt TT$ was defined by IAU Resolution A4 (1991) \cite{Guinot:1992} as: `a time scale differing from $\tt TCG$ by a constant rate, with the unit of measurement of $\tt TT$ chosen so that it matches the SI second on the geoid.' According to the transformation between proper and coordinate time, this constant rate is expressed as 
{}
\begin{equation}
\frac{d{\tt TT}}{ d {\tt TCG}} = 1 - \frac{1}{c^2}W_{\tt g} 
= 1 - L_{\tt G}, 
\label{eq:LG}
\end{equation}
where $W_{\tt g}$ is the combined gravitational and rotational potential on the geoid, determined as $W_{\tt g}=(62636856.0\pm0.5)~{\rm m}^2{\rm s}^{-2}$ \cite{Groten:2004}. The IAU value for $L_{\tt G}$ is $6.969\,290\,134 \times 10^{-10}\approx 60.2$ microseconds/day $(\mu{\rm s/d})$,  a defining constant as set by IAU 2000 Resolution B1.9, Table 1.1 in \cite{Petit-Luzum:2010}. 

According to the IAU Resolutions \cite{Brumberg-Groten:2001}, practical geocentric modeling uses scaled \({\tt TT}\)-compatible quantities rather than unscaled \({\tt TCG}\)-compatible coordinate time, spatial coordinates, and gravitational parameters. Denoting \(\mu\equiv GM\), the scaling is \cite{Turyshev-etal:2025,Turyshev-scales:2025}
{}
\begin{equation}
{\tt TT} ={\tt TCG}- L_{\tt G}({\tt TCG}-{\tt T}_0), \qquad
\vec X_{\tt TT} = (1-L_{\tt G})\vec X_{\tt TCG}, \qquad
(GM)_{\tt TT} = (1-L_{\tt G})(GM)_{\tt TCG},
\label{eq:TCGT}
\end{equation}
where $T_0$ is the conventional reference epoch used in the IAU scaling definitions; in this work we adopt
${\tt T}_0 = 2443144.5003725$ JD, as in Eq.~(\ref{eq:TDBC}).

Another constant $L_{\tt C}$ removes the constant rate between  ${\tt TCG}$ and ${\tt TCB}$. It is determined as the long time average of the rate computed from the full post-Newtonian version of (\ref{eq:coord-tr-T1-rec2++}) accurate to ${\cal O}(c^{-4})$ (see  \cite{Turyshev-etal:2025}):
{}
\begin{equation}
\Big<\frac{d{\tt TCG}}{d{\tt TCB}}\Big>=1-L_{\tt C}.
\label{eq:constTCGTCB}
\end{equation}
According to the IAU 2000 Resolution B1.5,  \( L_{\tt C} = 1.480\,826\,867\,41 \times 10^{-8} \pm 2 \times 10^{-17} \approx 1.279\,434\,4~\text{ms/d} \pm 1.7~\text{ps/d} \), introducing it as a fundamental constant  (refer to Table 1.1 in \cite{Petit-Luzum:2010}).  

The IAU 2000 Resolution B1.5, recommended the following relationship between the constants $L_{\tt B}$, $L_{\tt G}$, and $L_{\tt C}$:
{}
\begin{equation}
\Big<\frac{d{\tt TT}}{d{\tt TCB}}\Big>=\Big(\frac{d{\tt TT}}{d{\tt TCG}}\Big)\Big<\frac{d{\tt TCG}}{d{\tt TCB}}\Big> \qquad
\Rightarrow \qquad 1-L_{\tt B}=(1-L_{\tt G})(1-L_{\tt C}),
\label{eq:constLBLCLG}
\end{equation}
where $L_{\tt B}$ is determined as $L_{\tt B} = L_{\tt G}+L_{\tt C}-L_{\tt G}L_{\tt C}=1.550\,519\,767\,72 \times 10^{ - 8}\approx 1.34$~ms/d.

Finally, {\tt TDB} is a timescale rescaled from {\tt TCB}, as defined by IAU 2006 Resolution B3 and  IAU 2009 Resolution 3 \cite{IAU2009ResB3,Petit-Luzum:2010}, given by the following set of expressions:
{}
\begin{equation}
{\tt TDB} = {\tt TCB} - L_{\tt B}({\tt TCB}-{\tt T}_0)+{\tt TDB}_0, \qquad
\vec x_{\tt TDB} = (1-L_{\tt B})\vec x_{\tt TCB}, \qquad
(GM)_{\tt TDB} = (1-L_{\tt B})(GM)_{\tt TCB},
\label{eq:TDBC}
\end{equation}
where \(L_{\tt B}=1.550519768\times10^{-8}\), \({\tt T}_0=2443144.5003725\,\mathrm{JD}\), and \({\tt TDB}_0=-65.5\,\mu\mathrm{s}\) are the conventional constants entering the IAU 2006/2009 definition of \({\tt TDB}\). This linear scaling makes \({\tt TDB}\) advance, on average, at the same rate as \({\tt TT}\) for an observer near the geocenter. The same constants are used in the JPL ephemeris time argument convention historically associated with DE405/Teph \cite{Standish:DE405-1998}, providing operational continuity with existing planetary ephemerides. The offset \({\tt TDB}_0\) aligns the conventional \(({\tt TDB}-{\tt TT})\) relation \cite{Fairhead-Bretagnon:1990}; consequently, \({\tt TDB}\) is not synchronized with \({\tt TT}\), \({\tt TCG}\), or \({\tt TCB}\) at 1977-01-01 00:00:32.184 TAI at the geocenter (see discussion in \cite{Kaplan:2005,Turyshev-etal:2025,Turyshev-scales:2025}).

\begin{table*}[t]
\centering
\caption{Reference time-scale constants and implied mean rates. The first block lists official IAU/IERS constants used in this work; the second block lists lunar coefficients derived or adopted for the manuscript-level \({\tt TCL}/{\tt TL}\) realization.}
\label{tab:timescales}
\begin{tabular}{@{}cllll@{}}
\hline
Symbol & Definition & Value (s/s) & Mean rate per day & Note \\
\hline\hline
\multicolumn{5}{@{}l}{\textit{Official IAU/IERS time-scale constants}}\\
$L_{\tt G}$ & $d{\tt TT}/d{\tt TCG}=1-L_{\tt G}$ & $\LG$ & 60.2146\,$\mu{\rm s\, d}^{-1}$ & IAU 2000 B1.9 \\
$L_{\tt C}$ & $\langle d{\tt TCG}/d{\tt TCB}\rangle=1-L_{\tt C}$ & $\LC$ & \SI{1.279434}{\milli\second\per\day} & IAU 2000 B1.5 \\
$L_{\tt B}$ & $1-L_{\tt B}=(1-L_{\tt G})(1-L_{\tt C})$ & $\LB$ & \SI{1.339649}{\milli\second\per\day} & IAU 2006/2009 \\
\hline
\multicolumn{5}{@{}l}{\textit{Lunar coefficients used in this manuscript}}\\
$L_{\tt L}$ & $d{\tt TL}/d{\tt TCL}=1-L_{\tt L}$ & $3.1390541\times10^{-11}$ & 2.712143\,$\mu{\rm s\, d}^{-1}$ & this work, Eq.~(\ref{eq:(29)}) \\
$L_{\tt H}$ & $\langle d{\tt TCL}/d{\tt TCB}\rangle=1-L_{\tt H}$ & $1.4825362\times10^{-8}$ & \SI{1.280911}{\milli\second\per\day} & this work, Eq.~(\ref{eq:constTCGTCBH}) \\
$L_{\tt M}$ & $\langle d{\tt TL}/d{\tt TCB}\rangle=1-L_{\tt M}$ & $1.485675294\times10^{-8}$ & \SI{1.283624}{\milli\second\per\day} & this work, Eq.~(\ref{eq:consTM}) \\
\hline
\end{tabular}
\end{table*}
\begin{table}[t]
\centering
\caption{Reference systems and associated coordinate times used in this work.}
\label{tab:frames-times}
\begin{tabular}{@{}llll@{}}
\toprule
System & Origin & Coordinates & Coordinate time (scaled) \\
\midrule
{\tt BCRS} & Solar-system barycenter & $(t,\mathbf{x})$ & $\TCB$ ($\TDB$) \\
{\tt GCRS} & Geocenter & $(T,\mathbf{X})$ & $\TCG$ ($\TT$) \\
{\tt LCRS} & Selenocenter & $(\mathcal{T},\boldsymbol{\mathcal{X}})$ & $\TCL$ ($\TL$) \\
\bottomrule
\end{tabular}
\end{table}

\subsubsection{Transformations from {\tt TDB}- to {\tt TT}-compatible quantities}
\label{sec:TDBtoTT}

To process the {\tt GNSS} data, one needs transformations between {\tt GCRS} quantities scaled compatibly with \({\tt TT}\), \(({\tt TT},\vec X_{\tt TT})\), and {\tt BCRS} quantities scaled compatibly with \({\tt TDB}\), \(({\tt TDB},(\vec r_{\tt E})_{\tt TDB})\), where \((\vec r_{\tt E})_{\tt TDB}=(\vec x-\vec x_{\tt E}(t))_{\tt TDB}\). Applying the scaling relationships in Eqs.~(\ref{eq:TCGT}) and (\ref{eq:TDBC}), and introducing
\((\vec v_{\tt E}\cdot\vec r_{\tt E})_{\tt TDB}=\vec v_{\tt E}\cdot(\vec r_{\tt E})_{\tt TDB}\), gives the following {\tt TDB}$\rightarrow${\tt TT} map (see details in \cite{Turyshev-scales:2025}):
{}
\begin{eqnarray}
{\tt TT} &=& {\tt TDB}-{\tt TDB}_0+L_{\tt C}\Big({\tt TDB}-{\tt T}_0-{\tt TDB}_0\Big)-\nonumber\\
&-&
c^{-2}\Big\{
\int^{\tt TDB}_{{\tt T}_0+{\tt TDB}_0}\Big({\textstyle\frac{1}{2}}v_{\tt E}^{2} +\sum_{{\tt B\not= E}} {G M_{\tt B} \over r_{\tt BE}} \Big)dt+ \big(\vec v_{\tt E}  \cdot \vec r_{\tt E} \big)\Big\}_{\tt TDB} +
{\cal O}\big(2.19\times 10^{-16}\big)\Big({\tt TDB}-{\tt T}_0-{\tt TDB}_0\Big),~~
 \label{eq:coord-tr-RR1a}\\
\vec X_{\tt TT} &=& \big(1+ L_{\tt C}\big)\vec r_{\tt E} {}_{\tt TDB}+c^{-2} \Big\{ {\textstyle\frac{1}{2}}( \vec v_{\tt E}  \cdot\vec r_{\tt E} )\vec v_{\tt E}  +\sum_{{\tt B\not= E}}{G M_{\tt B} \over r_{\tt BE}} \vec r_{\tt E}  +(\vec a_{\tt E} \cdot \vec r_{\tt E} )\vec r_{\tt E} - {\textstyle\frac{1}{2}}r^2_{\tt E} \vec a_{\tt E} \Big\}_{\tt TDB} +
 {\cal O}\Big(5.83\times 10^{-9}~{\rm m}\Big).~~~~
 \label{eq:coord-tr-RR2b}
\end{eqnarray}
Note that the errors in (\ref{eq:coord-tr-RR1a}) are determined by the omitted products of the $L_{\tt C}^2$ term. If included, the error would be  $\sim {\cal O}(1.46\times 10^{-16})$, resulting from the product of $L_{\tt C}$ and the external gravitational potential. Similarly, the error in (\ref{eq:coord-tr-RR2b}) is set by the omitted term $L_{\tt C}^2(\vec r_{\tt E} )_{\tt TDB}\simeq {\cal O}(5.83\times 10^{-9}~{\rm m})$.

\subsubsection{Transformations from {\tt TT}- to {\tt TDB}-compatible quantities}
\label{sec:TTtoTDB}

Similarly, the inverse map gives {\tt TDB}-compatible {\tt BCRS} quantities as functions of {\tt TT}-compatible {\tt GCRS} quantities. For this, we use the inverse {\tt GCRS}$\rightarrow${\tt BCRS} transformation given by Eqs.~(\ref{eq:coord-tr-T1-rec2++})--(\ref{eq:coord-tr-Xrec2++}). Then, applying the scaling relationships (\ref{eq:TCGT}) and (\ref{eq:TDBC}), we have:
{}
\begin{eqnarray}
{\tt TDB}&=&(1-L_{\tt C}){\tt TT} +L_{\tt C}{\tt T}_0 + {\tt TDB}_0 +{1 \over c^2} \Big\{ \int_{{\tt T}_0}^{\tt TT} \Big({\textstyle{\textstyle{\textstyle{1 \over 2}}}} v_{\tt E} ^2 + \sum_{{\tt B\not= E}} {G M_{\tt B} \over r_{\tt BE}}\Big)dt+ ({\vec  v}_{\tt E}  \cdot {\vec X})\Big\}_{\tt TT}  + {\cal O}\Big(1.46\times 10^{-16}\Big)\big({\tt TT}-{\tt T}_0\big),~~~~~
\label{eq:coord-TTa1a}\\
(\vec r_{\tt E})_{\tt TDB} &=&\big(1 -L_{\tt C}\big)\vec X_{\tt TT}-c^{-2}\Big\{ {\textstyle\frac{1}{2}} ( \vec v_{\tt E}  \cdot\vec X)\vec v_{\tt E} +
\sum_{{\tt B\not= E}} {G M_{\tt B} \over r_{\tt BE}} \vec X
 + (\vec a_{\tt E} \cdot \vec X)\vec X-{\textstyle\frac{1}{2}}X^2\vec a_{\tt E} \Big\}_{\tt TT}   + {\cal O}\Big(5.83\times 10^{-9}~{\rm m}\Big).~~~~~
 \label{eq:coord-TTa2b}
\end{eqnarray} 
Note that the error in (\ref{eq:coord-TTa1a}) is set by the product of $L_{\tt C}$ and the external gravitational potential. Similar to (\ref{eq:coord-tr-RR2b}), the errors in (\ref{eq:coord-TTa2b}) are due to omitted $L_{\tt C}^2\vec X_{\tt TT}\simeq 5.83\times 10^{-9}~{\rm m}$ term.
 
Result (\ref{eq:coord-TTa1a}) yields the following useful derivative:
{}
\begin{eqnarray}
\frac{d{\tt TDB}}{d{\tt TT}} &=& 1-L_{\tt C}+c^{-2}\Big\{
{\textstyle\frac{1}{2}}v_{\tt E}^{2}+\sum_{{\tt B\not= E}} {G M_{\tt B} \over r_{\tt BE}}  + \big( \vec a_{\tt E}  \cdot \vec X\big)+\big(\vec v_{\tt E}  \cdot \dot {\vec X}\big)\Big\}_{\tt TT}+ {\cal O}\Big(1.46\times 10^{-16}\Big).
 \label{eq:coord-TTa2c}
\end{eqnarray}

Now we have all the expressions needed to establish  the transformation rules for velocities and accelerations.  

\subsection{Transforming positions, velocity and acceleration}
\label{sec:pos-vel-acc}

\subsubsection{Preliminary derivations}

We now present the transformation required for the $\tt TT$-compatible {\tt GCRS} position vector $\vec X_{\tt TT}$, velocity $\vec V_{\tt TT} = \dot{\vec X}_{\tt TT} = d\vec X/d{\tt TT}$, and acceleration $\vec A_{\tt TT} = \ddot{\vec X}_{\tt TT} = d\dot{\vec X}_{\tt TT}/d{\tt TT} = d^2\vec X/d{\tt TT}^2$. These transformations will convert these measurements to the {\tt BCRS} with ${\tt TDB}$-compatible position $(\vec r_{\tt E})_{\tt TDB} = (\vec x - \vec x_{\tt E} )_{\tt TDB}$, velocity $(\dot{\vec r}_{\tt E})_{\tt TDB} = d(\vec r_{\tt E})_{\tt TDB}/d{\tt TDB} = (\vec v - \vec v_{\tt E} )_{\tt TDB}$, and acceleration $(\ddot{\vec r}_{\tt E})_{\tt TDB} = d(\dot{\vec r}_{\tt E})_{\tt TDB}/d{\tt TDB} = d^2(\vec r_{\tt E})_{\tt TDB}/d{\tt TDB}^2 = (\vec a - \vec a_{\tt E} )_{\tt TDB}$.

We begin with the position transformation that is given by  (\ref{eq:coord-TTa2b}), expressing it, for convenience,  as below
{}
\begin{eqnarray}
(\vec r_{\tt E})_{\tt TDB} &=&(\vec x-\vec x_{\tt E} )_{\tt TDB}=\vec X_{\tt TT}-\Delta \vec X_{\tt TT}, 
\label{eq:coord-TTa2b1}
\end{eqnarray} 
where the corrective term $\Delta \vec X_{\tt TT}$ is of the order of ${\cal O}(c^{-2})$ and is given as 
{}
\begin{eqnarray}
\Delta \vec X_{\tt TT} &=& L_{\tt C}\vec X_{\tt TT}+c^{-2}\Big\{ {\textstyle\frac{1}{2}} ( \vec v_{\tt E}  \cdot\vec X)\vec v_{\tt E} +
\gamma U_{\rm ext}\, \vec X
 + (\vec a_{\tt E} \cdot \vec X)\vec X-{\textstyle\frac{1}{2}}X^2\vec a_{\tt E} \Big\}_{\tt TT}   + {\cal O}\Big(5.83\times 10^{-9}~{\rm m}\Big),
  \label{eq:coord-TTa2b2}
\end{eqnarray} 
where, for convenience,  we use 
{}
\begin{equation}
U_{\rm ext}\equiv 
\sum_{{\tt B\not= E}} {G M_{\tt B} \over r_{\tt BE}} +{\cal O}(c^{-2}).
\label{eq:Uext0}
\end{equation} 
The scalar \(\gamma\) in (\ref{eq:coord-TTa2b2}) is the PPN curvature parameter. We retain it only in the terms whose coefficient changes directly under the standard PPN extension of the IAU local/global reference-frame transformation \cite{Kopeikin-Vlasov:2004}; all numerical results in this paper set \(\gamma=1\). Thus, the displayed state maps are not intended as a complete multi-parameter alternative-gravity model. Rather, they are the general-relativistic IAU/IERS transformations with the leading \(\gamma\)-dependence left explicit for sensitivity tracking. We evaluate derivatives along \({\tt TT}\) unless noted:
\[
\dot U_{\rm ext}
= -\sum_{{\tt B}\ne{\tt E}}GM_{\tt B}\,
\frac{(\vec v_{\tt B}-\vec v_{\tt E})\cdot(\vec x_{\tt B}-\vec x_{\tt E})}{r_{\tt BE}^{3}},
\]
and \(\ddot U_{\rm ext}\) analogously by differentiating once more; see Eqs.~(\ref{eq:coord-tr-TT-TDB2})--(\ref{eq:coord-tr-TT-TDB3}). We explicitly retain the PPN parameter $\gamma$ (with $\gamma=1$ in general relativity) to make clear which terms in the {\tt GCRS}$\leftrightarrow${\tt BCRS} state transformations would change in a PPN analysis; all numerical evaluations in this work set $\gamma=1$.

Similarly, recognizing from (\ref{eq:coord-TTa2b}) that $\vec r_{\tt E} {}_{\tt TDB} =\vec X_{\tt TT}+{\cal O}(c^{-2})$ and $\dot{\vec r}_{\tt E} {}_{\tt TDB} =\dot{\vec X}_{\tt TT}+{\cal O}(c^{-2})$ we present 
{}
\begin{eqnarray}
\frac{d{\tt TT}}{d{\tt TDB}} &=& 1+L_{\tt C}-c^{-2}\Big\{
{\textstyle\frac{1}{2}}v_{\tt E}^{2} +U_{\rm ext}  + \big( \vec a_{\tt E}  \cdot \vec X\big)+\big(\vec v_{\tt E}  \cdot \dot {\vec X}\big)\Big\}_{\tt TT}+ {\cal O}\Big(2.19\times 10^{-16}\Big).
 \label{eq:coord-tr-TT-TDB}
\end{eqnarray}

Numerically at the geocenter, $GM_\odot/{\rm AU}\simeq 8.87\times 10^{8}$ m$^2$/s$^2$ while
$GM_{\tt M}/r_{\tt EM}\simeq 1.28\times 10^{4}$ m$^2$/s$^2$; thus the solar term dominates $U_{\rm ext}$ by $\sim 7\times 10^{4}$,
with other planetary contributions smaller still.

We can now develop the time derivatives of $(\vec r_{\tt E} )_{\tt TDB}$:
 {} 
 \begin{eqnarray}
 (\dot {\vec r}_{\tt E})_{\tt TDB}&=& \frac{d(\vec r_{\tt E})_{\tt TDB}}{d{\tt TDB}}\equiv (\dot{\vec x}-\vec v_{\tt E} )_{\tt TDB}, 
 \qquad
 ( \ddot {\vec r}_{\tt E})_{\tt TDB}= \frac{d^2 (\vec r_{\tt E})_{\tt TDB}}{d{\tt TDB}^2}\equiv (\ddot{\vec x}-\vec a_{\tt E} )_{\tt TDB}.
 \label{eq:trans-vel-acc}
 \end{eqnarray}
For that, similarly to \cite{Huang-etal:1990} and using expressions (\ref{eq:coord-TTa2b1})--(\ref{eq:coord-TTa2b2}), we have
 {} 
 \begin{eqnarray}
\frac{d(\vec r_{\tt E})_{\tt TDB}}{d{\tt TDB}}&=&\frac{d\vec X_{\tt TT}}{d{\tt TT}}\frac{d{\tt TT}}{d{\tt TDB}}-\frac{d\vec \Delta \vec X_{\tt TT}}{d{\tt TT}}\frac{d{\tt TT}}{d{\tt TDB}},
 \label{eq:trans-vel}\\
\frac{d^2(\vec r_{\tt E})_{\tt TDB}}{d{\tt TDB}^2} &=&\frac{d^2\vec X_{\tt TT}}{d{\tt TT}^2}\Big(\frac{d{\tt TT}}{d{\tt TDB}}\Big)^2+\frac{d\vec X_{\tt TT}}{d{\tt TT}}\frac{d^2{\tt TT}}{d{\tt TDB}^2}-\frac{d^2\vec \Delta \vec X_{\tt TT}}{d{\tt TT}^2}\Big(\frac{d{\tt TT}}{d{\tt TDB}}\Big)^2-\frac{d\vec \Delta \vec X_{\tt TT}}{d{\tt TT}}\frac{d^2{\tt TT}}{d{\tt TDB}^2}.
 \label{eq:trans-acc}
 \end{eqnarray}
 
From (\ref{eq:coord-tr-TT-TDB}), we see that ${d{\tt TT}}/{d{\tt TDB}} =1+ {\cal O}(c^{-2})$. In addition, $\Delta \vec X_{\tt TT}$ is of the order of ${\cal O}(c^{-2})$. Given this, we may simplify  (\ref{eq:trans-vel}) and (\ref{eq:trans-acc}), as below
{} 
 \begin{eqnarray}
\frac{d(\vec r_{\tt E})_{\tt TDB}}{d{\tt TDB}}&=&\frac{d\vec X_{\tt TT}}{d{\tt TT}}\frac{d{\tt TT}}{d{\tt TDB}}-\frac{d\vec \Delta \vec X_{\tt TT}}{d{\tt TT}}+{\cal O}(c^{-4}),
 \label{eq:trans-vel-appr}\\
\frac{d^2(\vec r_{\tt E})_{\tt TDB}}{d{\tt TDB}^2} &=&\frac{d^2\vec X_{\tt TT}}{d{\tt TT}^2}\Big(\frac{d{\tt TT}}{d{\tt TDB}}\Big)^2+\frac{d\vec X_{\tt TT}}{d{\tt TT}}\frac{d^2{\tt TT}}{d{\tt TDB}^2}-\frac{d^2\vec \Delta \vec X_{\tt TT}}{d{\tt TT}^2}+{\cal O}(c^{-4}).
 \label{eq:trans-acc-appr}
 \end{eqnarray}
 
 Using (\ref{eq:coord-tr-TT-TDB}), we have
 {}
\begin{eqnarray}
\frac{d{\tt TT}}{d{\tt TDB}} &=& 1+L_{\tt C}-c^{-2}\Big\{
{\textstyle\frac{1}{2}}v_{\tt E}^{2} +U_{\rm ext}  + \big( \vec a_{\tt E}  \cdot \vec X\big)+\big(\vec v_{\tt E}  \cdot \dot {\vec X}\big)\Big\}_{\tt TT}+ {\cal O}\Big(2.19\times 10^{-16}\Big),
 \label{eq:coord-tr-TT-TDB1}\\
\Big( \frac{d{\tt TT}}{d{\tt TDB}}\Big)^2 &=& 1+2L_{\tt C}-c^{-2}\Big\{
{v_{\tt E}^{2}} +2U_{\rm ext}  + 2\big( \vec a_{\tt E}  \cdot \vec X\big)+2\big(\vec v_{\tt E}  \cdot \dot {\vec X}\big)\Big\}_{\tt TT}+ {\cal O}\Big(4.38\times 10^{-16}\Big),
 \label{eq:coord-tr-TT-TDB2}\\
 \frac{d^2{\tt TT}}{d{\tt TDB}^2} &=&-c^{-2}\Big\{
(\vec v_{\tt E}\cdot\vec a_{\tt E} ) +\dot U_{\rm ext}  + \big( \dot{\vec a}_{\tt E}  \cdot \vec X\big)+2\big( \vec a_{\tt E}  \cdot \dot{\vec X}\big)+\big(\vec v_{\tt E}  \cdot \ddot {\vec X}\big)\Big\}_{\tt TT}+ {\cal O}\Big(2.91\times 10^{-23}\,{\s}^{-1}\Big),~~~
 \label{eq:coord-tr-TT-TDB3}
\end{eqnarray}
 where the error in $ {d^2{\tt TT}}/{d{\tt TDB}^2}$ is set by the magnitude of the omitted $c^{-4}$ terms (\ref{eq:coord-tr-RR1a}), see \cite{Turyshev-scales:2025}.

In addition, from (\ref{eq:coord-TTa2b2}), we compute 
 \begin{eqnarray}
\frac{d\vec \Delta \vec X_{\tt TT}}{d{\tt TT}}&=& L_{\tt C}\dot{\vec X}_{\tt TT}+c^{-2}\Big\{ {\textstyle\frac{1}{2}} \Big(( \vec a_{\tt E}  \cdot\vec X)+( \vec v_{\tt E}  \cdot \dot{\vec X})\Big)\vec v_{\tt E} + \Big({\textstyle\frac{1}{2}}(\vec v_{\tt E}  \cdot\vec X)-({\vec X}\cdot\dot{\vec X})\Big)\vec a_{\tt E} +\Big((\vec a_{\tt E} \cdot \vec X)+\gamma U_{\rm ext}\Big) \dot{\vec X}+ \nonumber\\
&&\hskip 0pt + \, 
\Big( (\dot {\vec a}_{\tt E} \cdot \vec X)+(\vec a_{\tt E} \cdot \dot{\vec X})+\gamma \dot U_{\rm ext}\Big)\vec X-{\textstyle\frac{1}{2}}X^2\dot{\vec a}_{\tt E} \Big\}_{\tt TT}   + {\cal O}\Big(8.49\times 10^{-13}~{\rm m/s}\Big),~~~
  \label{eq:coord-Dvel}
\end{eqnarray}
where the errors in (\ref{eq:coord-Dvel})  are set by the term  $L_{\tt C}^2(\vec r_{\tt E})_{\tt TDB}=L_{\tt C}^2\,\vec X_{\tt TT}+{\cal O}(c^{-2})\simeq 5.83\times 10^{-9}~{\rm m}$  omitted in (\ref{eq:coord-Dvel}), that yields $L_{\tt C}^2\,\dot {\vec X}_{\tt TT}\simeq 8.49\times 10^{-13}~{\rm m/s}$ for a MEO orbiter.  Also, $\dot U_{\rm  ext}={dU_{\rm  ext}}/{d{\tt TT}}+ {\cal O}(c^{-2})$ has the form
{}
\begin{eqnarray}
\dot U_{\rm  ext}\equiv\frac{dU_{\rm  ext}}{d{\tt TT}}=-\sum_{{\tt B\not= E}}\frac{GM_{\tt B}\big((\vec v_{\tt B}-\vec v_{\tt E} )\cdot(\vec x_{\tt B}-\vec x_{\tt E} )\big)}{r^3_{\tt BE}}+ {\cal O}(c^{-2}),
 \label{eq:Uext}
 \end{eqnarray}
 with $GM_{\tt B}, \vec r_{\tt B}$ and $\vec v_{\tt B}$ being the mass parameter and barycentric position and velocity vectors of body $\tt B$,  $\tt E$ is the index for Earth and $r_{\tt BE}$ is the geocentric range of body $\tt B$ and $r_{\tt BE}=|\vec x_{\tt B}-\vec x_{\tt E} |$.

Finally, we use (\ref{eq:coord-Dvel}) to compute
{}
 \begin{eqnarray}
\frac{d^2\vec \Delta \vec X_{\tt TT}}{d{\tt TT}^2}&=& L_{\tt C}\ddot{\vec X}_{\tt TT}+ c^{-2}\Big\{ 
{\textstyle\frac{1}{2}} \Big(( \dot{\vec a}_{\tt E}  \cdot\vec X)+2( \vec a_{\tt E}  \cdot \dot{ \vec X})+( \vec v_{\tt E}  \cdot \ddot{\vec X})\Big)\vec v_{\tt E} + 
\Big(( \vec a_{\tt E}  \cdot\vec X)+ ( \vec v_{\tt E}  \cdot \dot{\vec X})-(\dot{\vec X}\cdot\dot{\vec X})-({\vec X}\cdot\ddot{\vec X})\Big)\vec a_{\tt E} +
\nonumber\\
&+&
\Big( {\textstyle\frac{1}{2}} ( \vec v_{\tt E}  \cdot\vec X)-2({\vec X}\cdot\dot{\vec X})\Big)\dot{\vec a}_{\tt E} +\Big(\gamma\ddot U_{\rm ext}+ 
 (\ddot {\vec a}_{\tt E} \cdot \vec X)+ 2(\dot {\vec a}_{\tt E} \cdot \dot{\vec X})+(\vec a_{\tt E} \cdot \ddot{\vec X})\Big)\vec X+
\nonumber\\
&+&
2\Big(\gamma\dot U_{\rm ext}+(\dot {\vec a}_{\tt E} \cdot \vec X)+ (\vec a_{\tt E} \cdot \dot{\vec X})\Big)\dot{\vec X}+
\Big(\gamma U_{\rm ext}+ (\vec a_{\tt E} \cdot \vec X)\Big)\ddot{\vec X}-
{\textstyle\frac{1}{2}}X^2\ddot{\vec a}_{\tt E} \Big\}_{\tt TT}   + {\cal O}\Big(1.24\times 10^{-16}~{\rm m/s}^2\Big),~~~~~~
  \label{eq:coord-Dacc}
\end{eqnarray}
where the errors in (\ref{eq:coord-Dacc})  are set by the omitted term  $L_{\tt C}^2(\vec r_{\tt E})_{\tt TDB}=L_{\tt C}^2\,\vec X_{\tt TT}+{\cal O}(c^{-2})\simeq 5.83\times 10^{-9}~{\rm m}$  omitted in (\ref{eq:coord-Dvel}), that yields $L_{\tt C}^2\,\ddot {\vec X}_{\tt TT}\simeq 1.24\times 10^{-16}~{\rm m/s}^2$ for a MEO orbiter. In addition, $\ddot U_{\rm  ext}={d^2U_{\rm  ext}}/{d{\tt TT}^2}+ {\cal O}(c^{-2})$ has the form
{}
\begin{eqnarray}
\ddot U_{\rm  ext}\equiv\frac{d^2U_{\rm  ext}}{d{\tt TT}^2}&=&-\sum_{{\tt B\not= E}}\frac{GM_{\tt B}}{r^3_{\tt BE}} \Big(
(\vec a_{\tt B}-\vec a_{\tt E} )\cdot(\vec x_{\tt B}-\vec x_{\tt E} )+\nonumber\\
&&+~(\vec v_{\tt B}-\vec v_{\tt E} )\cdot(\vec v_{\tt B}-\vec v_{\tt E} )-
\frac{3}{r^2_{\tt BE}}\big((\vec v_{\tt B}-\vec v_{\tt E} )\cdot(\vec x_{\tt B}-\vec x_{\tt E} )\big)^2\Big)+ {\cal O}(c^{-2}),
 \label{eq:Uext-dot}
 \end{eqnarray}
 with $GM_{\tt B}, \vec r_{\tt B}$ and $\vec v_{\tt B}$ being the mass parameter and barycentric position and velocity vectors of body $\tt B$,  $\tt E$ is the index for Earth and $r_{\tt BE}$ is the geocentric range of body $\tt B$ and $r_{\tt BE}=|\vec x_{\tt E}-\vec x_{\tt B}|$.

\subsubsection{Transforming velocity and acceleration}

Now, we can present the expressions that describe the velocity and acceleration transformations.  
To do this, starting with the velocity, we use (\ref{eq:trans-vel}), (\ref{eq:trans-vel-appr}), (\ref{eq:coord-tr-TT-TDB1}), and (\ref{eq:coord-Dvel}), which yield 
{} 
 \begin{eqnarray}
(\dot{\vec x}-\dot{\vec x}_{\tt E} )_{\tt TDB}&=&\dot{\vec X}_{\tt TT}-c^{-2}\Big\{\Big(
{\textstyle\frac{1}{2}} v_{\tt E}^{2} +(1+\gamma)U_{\rm ext}  + 2\big( \vec a_{\tt E}  \cdot \vec X\big)+\big(\vec v_{\tt E}  \cdot \dot {\vec X}\big)\Big)\dot{\vec X}+ {\textstyle\frac{1}{2}} \Big(( \vec a_{\tt E}  \cdot\vec X)+( \vec v_{\tt E}  \cdot \dot{\vec X})\Big)\vec v_{\tt E} +\nonumber\\
 &&\hskip -50pt +~
  \Big({\textstyle\frac{1}{2}}(\vec v_{\tt E}  \cdot\vec X)-({\vec X}\cdot\dot{\vec X})\Big)\vec a_{\tt E} +\Big( (\dot {\vec a}_{\tt E} \cdot \vec X)+(\vec a_{\tt E} \cdot \dot{\vec X})+\gamma \dot  U_{\rm ext}\Big)\vec X-{\textstyle\frac{1}{2}}X^2\dot{\vec a}_{\tt E} \Big\}_{\tt TT}  +
 {\cal O}\Big(8.49\times 10^{-13}~{\rm m/s}\Big).~~~~~
 \label{eq:trans-vel-appr2*}
 \end{eqnarray}
Note that all the quantities on the right-hand side of (\ref{eq:trans-vel-appr2*})  are {\tt TT}-compatible. 
  
Similarly, we have  the expression for transforming the acceleration  between the two reference systems. For this, we will use (\ref{eq:trans-vel}), (\ref{eq:trans-acc-appr}), (\ref{eq:coord-tr-TT-TDB2})--(\ref{eq:coord-tr-TT-TDB3}), and (\ref{eq:coord-Dacc}), which yield the following:
{} 
 \begin{eqnarray}
(\ddot{\vec x}-\ddot{\vec x}_{\tt E} )_{\tt TDB}&=&\ddot{\vec X}_{\tt TT} +L_{\tt C}\ddot{\vec X}_{\tt TT} -c^{-2}\Big\{\Big(
{v_{\tt E}^{2}} +(2+\gamma)U_{\rm ext}  + 3\big( \vec a_{\tt E}  \cdot \vec X\big)+2\big(\vec v_{\tt E}  \cdot \dot {\vec X}\big)\Big)\ddot{\vec X}+
\nonumber\\
 &+&
\Big(
(\vec v_{\tt E}\cdot\vec a_{\tt E} ) +(1+2\gamma)\dot U_{\rm ext}  + 3\big( \dot{\vec a}_{\tt E}  \cdot \vec X\big)+4\big( \vec a_{\tt E}  \cdot \dot{\vec X}\big)+\big(\vec v_{\tt E}  \cdot \ddot {\vec X}\big)\Big)\dot{\vec X}+
\nonumber\\
 &+&
{\textstyle\frac{1}{2}} \Big(( \dot{\vec a}_{\tt E}  \cdot\vec X)+2( \vec a_{\tt E}  \cdot \dot{ \vec X})+( \vec v_{\tt E}  \cdot \ddot{\vec X})\Big)\vec v_{\tt E} + 
\Big(( \vec a_{\tt E}  \cdot\vec X)+ ( \vec v_{\tt E}  \cdot \dot{\vec X})-(\dot{\vec X}\cdot\dot{\vec X})-({\vec X}\cdot\ddot{\vec X})\Big)\vec a_{\tt E} +
\nonumber\\
&+&
\Big( {\textstyle\frac{1}{2}} ( \vec v_{\tt E}  \cdot\vec X)-2({\vec X}\cdot\dot{\vec X})\Big)\dot{\vec a}_{\tt E} +\Big(\gamma\ddot U_{\rm ext}+ 
 (\ddot {\vec a}_{\tt E} \cdot \vec X)+ 2(\dot {\vec a}_{\tt E} \cdot \dot{\vec X})+(\vec a_{\tt E} \cdot \ddot{\vec X})\Big)\vec X-
{\textstyle\frac{1}{2}}X^2\ddot{\vec a}_{\tt E} \Big\}_{\tt TT}   +\nonumber\\[2pt]
&+&
 {\cal O}\Big(1.24\times 10^{-16}~{\rm m/s}^2\Big),
 \label{eq:trans-acc-apprTR}
 \end{eqnarray}
 where the error terms in these expressions are set by the omitted $L^2_{\tt C}$ terms. 

The absence of a standalone \(L_{\tt C}\dot{\vec X}_{\tt TT}\) term in Eq.~(\ref{eq:trans-vel-appr2*}) is a useful implementation check: the \(L_{\tt C}\dot{\vec X}_{\tt TT}\) contribution from \(d{\tt TT}/d{\tt TDB}\) cancels the derivative of the scale term \(L_{\tt C}\vec X_{\tt TT}\) in \(\Delta\vec X_{\tt TT}\). In contrast, an \(L_{\tt C}\ddot{\vec X}_{\tt TT}\) term remains in the acceleration map, Eq.~(\ref{eq:trans-acc-apprTR}).

\textit{Compatibility convention:} Unless explicitly indicated otherwise, all quantities on the right-hand side
of Eqs.~(\ref{eq:trans-vel-appr2*}) and (\ref{eq:trans-acc-apprTR}) are evaluated as {\tt TT}-compatible {\tt GCRS}
quantities at the same epoch.

The screening and remainder bounds in this section are computed using the parameter envelopes in Table~\ref{tab:nom-values}, representative of (i) a terrestrial station and (ii) a {\tt GNSS} MEO spacecraft. For other regimes (e.g., GEO, HEO, lunar transfer, {\tt NRHO}), the retained terms should be re-screened using the same procedure with updated envelopes
for $(\mathbf{X},\dot{\mathbf{X}},\ddot{\mathbf{X}})$ and the relevant external potentials and accelerations.

\subsection{Practical form for the relevant expressions}
\label{sec:GCRS-pvac}

Results  (\ref{eq:coord-TTa2b1})--(\ref{eq:coord-TTa2b2}), (\ref{eq:trans-vel-appr2*}), and (\ref{eq:trans-acc-apprTR}) have many terms that may be too small.  We estimate the magnitudes of the terms in these expressions using Table~\ref{tab:nom-values}, which provides the nominal values for the quantities involved. This will help us understand the changes in station or satellite coordinates relative to the Earth's center of mass in the barycentric frame. We also recognize that the solar gravitational potential provides the largest contribution to $U_{\rm ext}$. 

Numerically at the geocenter, $GM_{\tt S}/{\rm AU}\simeq 8.87\times 10^{8}\,{\rm m^2/s^2}$ while $GM_{\tt M}/r_{\tt EM}\simeq 1.28\times 10^{4}\,{\rm m^2/s^2}$, so the solar term dominates $U_{\rm ext}$ by $\sim 7\times 10^{4}$ (the remaining planetary terms are smaller still for Earth-orbiting applications).
 
\begin{table}[t]
\caption{Approximate values for various parameters involved.}
\label{tab:nom-values}
\begin{tabular}{rlc}
\hline
Notation & Meaning & Value \\
\hline\hline
$\gamma$ & Parametrized post-Newtonian (PPN) parameter &1 \\
$GM_{\tt S}$ & Solar gravitational parameter &  $1.32712442099 \times 10^{20}$~${\rm m}^{3}{\rm s}^{-2}$ \\
$|\vec x_{\tt E}|$ & Earth's position relative to barycenter  &  $1\,{\rm AU} = 149\,597\,870\,700.0~\mathrm{m}$\\
${\vec v}_{\tt E} $ & Earth's velocity relative to barycenter & $29.78\times10^3$~m/s \\
\hline
&{\it Ground-based station: }&\\ 
$\vec X_{\tt S}$ & position relative to the center of the Earth   & $6378.1\times10^3$~m\\
$\dot{\vec X}_{\tt S}$ & velocity relative to the center of the Earth   & $({2\pi}/{\rm day})R_{\tt E}=463.83$~m/s\\
$\ddot{\vec X}_{\tt S}$ & acceleration  relative to the center of the Earth   & ~~$({2\pi}/{\rm day})^2R_{\tt E}=3.37\times 10^{-2}$~m/s${}^2$\\
\hline
&{\it {\rm MEO} spacecraft at altitude of $h=20200$\,km}&\\ 
$\vec X_{\tt MEO}$ & position relative to the center of the Earth   & $26578.1\times10^3$~m\\
$\dot{\vec X}_{\tt MEO}$ & velocity relative to the center of the Earth   & $3.87\times10^3$~m/s\\
$\ddot{\vec X}_{\tt MEO}$ & acceleration  relative to the center of the Earth   & $0.56$~m/s${}^2$\\
\hline
\end{tabular}
\end{table}

As the characteristic magnitudes of $(|\vec X|,|{\dot {\vec X}}|,|{\ddot {\vec X}}|)$ for a MEO (GPS-like) spacecraft exceed those of a ground-based station (Table~\ref{tab:nom-values}), we screen the transformation terms using the MEO envelope to obtain conservative bounds. We consider the  transformations for a MEO orbiter (i.e., a GPS spacecraft), with the position, velocity and the acceleration present in  (\ref{eq:coord-TTa2b1})--(\ref{eq:coord-TTa2b2}), (\ref{eq:trans-vel-appr2*}), and (\ref{eq:trans-acc-apprTR}) take the relevant values of MEO orbiter (i.e., a GPS spacecraft), namely 
$({\vec X}, \dot{\vec X}, \ddot {\vec X})\rightarrow ({\vec X}_{\tt MEO}, \dot{\vec X}_{\tt MEO}, \ddot {\vec X}_{\tt MEO})$, that are given in Table~\ref{tab:nom-values}. With these values, we use (\ref{eq:coord-TTa2b2}) to estimate contribution of the various terms in the position transformation:
{}
\begin{equation}
 \Big(L_{\tt C}+c^{-2}\gamma U_{\rm ext}\Big){\vec X}   \approx  
0.66 \, {\rm m},
 \qquad c^{-2}\Big\{{\textstyle\frac{1}{2}}
( \vec v_{\tt E}  \cdot {\vec X}) \vec v_{\tt E} \Big\}  \approx 
0.13 \, {\rm m}, 
 \qquad
 c^{-2}\Big\{(\vec a_{\tt E} \cdot \vec X)\vec X-{\textstyle\frac{1}{2}}X^2\vec a_{\tt E} \Big\}\lesssim  7.00\times 10^{-5}~{\rm m}.
 \label{eq:transfrom-posL2} 
\end{equation}
Note that the acceleration-dependent terms in (\ref{eq:transfrom-posL2}) are very small and are currently considered practically irrelevant. Consequently, a reasonable approximation of  (\ref{eq:coord-TTa2b1})--(\ref{eq:coord-TTa2b2}) for a MEO orbiter would be, as outlined in \cite{Moyer:2003},
{}
\begin{eqnarray}
(\vec x-\vec x_{\tt E} )_{\tt TDB}  &=&\vec X_{\tt TT}- L_{\tt C}\vec X_{\tt TT}-c^{-2}\Big\{ {\textstyle\frac{1}{2}} ( \vec v_{\tt E}  \cdot\vec X)\vec v_{\tt E} +
\gamma U_{\rm ext}\, \vec X\Big\}_{\tt TT}   + {\cal O}\Big(7.01\times 10^{-5}~{\rm m}\Big),
\label{eq:pos-approx-LEO}
 \end{eqnarray}
 where the error bound is provided by the omitted acceleration-dependent terms shown in (\ref{eq:transfrom-posL2}). Note that the same expression is valid for a ground-based station with the error term being of the order of $ {\cal O}\big(4.03\times 10^{-6}~{\rm m}\big)$.
 
Eq.~(\ref{eq:pos-approx-LEO}) is closely related to the station-coordinate transformations used in DSN radiometric modeling (e.g., range and Doppler) \cite{Moyer:2003}. Relative to commonly used operational forms, we keep the explicit scale term \(L_{\tt C}\) and track the small acceleration-dependent contributions and their remainder bound, which become relevant for millimeter-level state-map closure. These screened state transformations do not by themselves complete a \(10^{-16}\)-class frequency-transfer model; at that level the \({\cal O}(c^{-4})\) clock-rate terms discussed in Sec.~\ref{sec:properT-TDB} must be retained or budgeted explicitly.  For {\tt GNSS} cm-class processing the additional screened terms provide a conservative, implementation-ready state map between {\tt TT}-compatible {\tt GCRS} and {\tt TDB}-compatible {\tt BCRS}.

Next, we evaluate the terms present in the velocity transformation in (\ref{eq:trans-vel-appr2*}) as below
{}
\begin{eqnarray}
c^{-2}\Big\{\Big(
{\textstyle\frac{1}{2}} v_{\tt E}^{2} +(1+\gamma)U_{\rm ext}\Big)\dot{\vec X}\Big\}
  \approx 9.56 \times 10^{-5} \, {\rm m/s},  
  \qquad  ~~~
c^{-2}\Big\{\big(\vec v_{\tt E}  \cdot \dot {\vec X}\big)\dot{\vec X}\Big\}
   &\approx& 4.96 \times 10^{-6} \, {\rm m/s},
 \label{eq:transfrom-velB2L} 
 \label{eq:transfrom-velB1L} \\
 c^{-2}\Big\{2\big( \vec a_{\tt E}  \cdot \vec X\big)\dot{\vec X}\Big\}
   \approx 1.36 \times 10^{-8} \, {\rm m/s},
   \qquad
c^{-2} \Big\{{\textstyle\frac{1}{2}} ( \vec v_{\tt E}  \cdot \dot{\vec X})\vec v_{\tt E} \Big\}  &\approx& 3.82 \times 10^{-5} \, {\rm m/s}~~~
\\
c^{-2} \Big\{{\textstyle\frac{1}{2}} ( \vec a_{\tt E}  \cdot\vec X)\vec v_{\tt E} \Big\}  \approx 2.61 \times 10^{-8} \, {\rm m/s},
\qquad~~~~
c^{-2}   \Big\{({\vec X}\cdot\dot{\vec X})\vec a_{\tt E} \Big\} &\lesssim& 3.40 \times 10^{-9} \, {\rm m/s}, 
\\
c^{-2}   \Big\{ {\textstyle\frac{1}{2}}(\vec v_{\tt E}  \cdot\vec X)\vec a_{\tt E} \Big\} \approx 2.61 \times 10^{-8} \, {\rm m/s}, 
\qquad \hskip 25pt 
c^{-2}    \Big\{\gamma \dot U_{\rm ext}\vec X\Big\} &\approx& 5.22 \times 10^{-8} \, {\rm m/s},
\\
c^{-2}    \Big\{\Big( (\dot {\vec a}_{\tt E} \cdot \vec X)+(\vec a_{\tt E} \cdot \dot{\vec X})\Big)\vec X\Big\} \approx 6.81 \times 10^{-9} \, {\rm m/s}, 
 \qquad~~~~~~~\,
c^{-2} \Big\{{\textstyle\frac{1}{2}}X^2\dot{\vec a}_{\tt E} \Big\}  &\approx& 9.28 \times 10^{-12} \, {\rm m/s} . 
 \label{eq:transfrom-velBd2*L} 
\end{eqnarray}

As a result, limiting the terms at $4.96 \times 10^{-6}$~m/s, from (\ref{eq:trans-vel-appr2*}), we have the following  for the velocity transformation: 
   {}
\begin{eqnarray}
(\dot{\vec x}-\dot{\vec x}_{\tt E} )_{\tt TDB}&=&\dot{\vec X}_{\tt TT}-c^{-2}\Big\{\Big(
{\textstyle\frac{1}{2}} v_{\tt E}^{2} +(1+\gamma)U_{\rm ext}  
+\big(\vec v_{\tt E}  \cdot \dot {\vec X}\big)\Big)\dot{\vec X}+ 
{\textstyle\frac{1}{2}}( \vec v_{\tt E}  \cdot \dot{\vec X})\vec v_{\tt E} 
\Big\}_{\tt TT} +{\cal O}\Big(1.29\times 10^{-7}~{\rm m/s}\Big),
\label{eq:vel-tran2}
\end{eqnarray}
where the error bound results from the omitted acceleration-dependent and $\dot U_{\rm ext}$ terms. It is worth noting that result (\ref{eq:vel-tran2}) includes two additional terms compared to \cite{Moyer:2003}. The same expression is also applicable to ground-based stations, with the corresponding error term being on the order of ${\cal O}(9.73 \times 10^{-8}~{\rm m/s})$.
 
Finally, we evaluate the terms involved in the acceleration transformations  (\ref{eq:trans-acc-apprTR}) as below
 {}
 \begin{eqnarray}
L_{\tt C}\ddot{\vec X} \approx 8.36 \times 10^{-9} \, {\rm m/s}^2,
\qquad ~~\,
c^{-2}\Big\{{\textstyle\frac{1}{2}} \Big(2( \vec a_{\tt E}  \cdot \dot{ \vec X})+( \vec v_{\tt E}  \cdot \ddot{\vec X})\Big)\vec v_{\tt E} \Big\} &\approx&  2.97 \times 10^{-9} \, {\rm m/s}^2,    \\
c^{-2}\Big\{\Big(
{v_{\tt E}^{2}} +(2+\gamma)U_{\rm ext}  +3\big( \vec a_{\tt E}  \cdot \vec X\big)+2\big(\vec v_{\tt E}  \cdot \dot {\vec X}\big)\Big)\ddot{\vec X}\Big\}&\approx& 2.37 \times 10^{-8} \, {\rm m/s}^2,
 \label{eq:transfrom-accBd1L} \\
c^{-2}\Big\{
\Big(
(\vec v_{\tt E}\cdot\vec a_{\tt E} ) + (1+2\gamma)\dot U_{\rm ext}  +4\big( \vec a_{\tt E}  \cdot \dot{\vec X}\big)+\big(\vec v_{\tt E}  \cdot \ddot {\vec X}\big)\Big)\dot{\vec X}\Big\} &\approx&
7.58 \times 10^{-10} \, {\rm m/s}^2,\\
c^{-2}\Big\{3\big( \dot{\vec a}_{\tt E}  \cdot \vec X\big)\dot{\vec X}\Big\} \approx 8 .11 \times 10^{-15} \, {\rm m/s}^2,
\qquad ~~~
c^{-2}\Big\{{\textstyle\frac{1}{2}} ( \dot{\vec a}_{\tt E}  \cdot\vec X)\vec v_{\tt E} \Big\} &\approx&  1.40 \times 10^{-14} \, {\rm m/s}^2,
\\
c^{-2}\Big\{\Big(( \vec a_{\tt E}  \cdot\vec X)+(\vec v_{\tt E}  \cdot \dot{\vec X})-(\dot{\vec X}\cdot\dot{\vec X})-({\vec X}\cdot\ddot{\vec X})\Big)\vec a_{\tt E} \Big\} &\approx& 9.60 \times 10^{-12} \, {\rm m/s}^2,
\\
c^{-2}\Big\{ \Big( {\textstyle\frac{1}{2}} ( \vec v_{\tt E}  \cdot\vec X)-2({\vec X}\cdot\dot{\vec X})\Big)\dot{\vec a}_{\tt E} \Big\} &\approx&1.58 \times 10^{-14} \, {\rm m/s}^2,
\\
c^{-2}\Big\{\Big(\gamma\ddot U_{\rm ext}+ 
 (\ddot {\vec a}_{\tt E} \cdot \vec X)+ 2(\dot {\vec a}_{\tt E} \cdot \dot{\vec X})\Big)\vec X\Big\}&\approx&
3.70 \times 10^{-14} \, {\rm m/s}^2,\\
c^{-2}\Big\{(\vec a_{\tt E} \cdot \ddot{\vec X})\vec X\Big\}\approx
9.90 \times 10^{-13} \, {\rm m/s}^2,
\qquad ~~~~~~~~~
c^{-2}\Big\{{\textstyle\frac{1}{2}}X^2\ddot{\vec a}_{\tt E} \Big\}  &\approx& 7.35 \times 10^{-18} \, {\rm m/s}^2.
 \label{eq:transfrom-accBd2L} 
 \end{eqnarray}
 
As a result, limiting the terms to ${\cal O}(1\times 10^{-13}\, {\rm m/s}^2)$, we have the following expression to transform acceleration:
{}
\begin{eqnarray}
  (\ddot{\vec x}-\ddot{\vec x}_{\tt E} )_{\tt TDB}&=&\ddot{\vec X}_{\tt TT} +L_{\tt C}\ddot{\vec X}_{\tt TT} -c^{-2}\Big\{\Big(
{v_{\tt E}^{2}} +(2+\gamma)U_{\rm ext}  + 3\big( \vec a_{\tt E}  \cdot \vec X\big)+2\big(\vec v_{\tt E}  \cdot \dot {\vec X}\big)\Big)\ddot{\vec X}+
\nonumber\\
 &+&
\Big(
(\vec v_{\tt E}\cdot\vec a_{\tt E} ) +(1+2\gamma)\dot U_{\rm ext}  
+4\big( \vec a_{\tt E}  \cdot \dot{\vec X}\big)+\big(\vec v_{\tt E}  \cdot \ddot {\vec X}\big)\Big)\dot{\vec X}+
\Big(( \vec a_{\tt E}  \cdot \dot{ \vec X})+{\textstyle\frac{1}{2}} ( \vec v_{\tt E}  \cdot \ddot{\vec X})\Big)\vec v_{\tt E} +
\nonumber\\
 &+& 
\Big(( \vec a_{\tt E}  \cdot\vec X)+ ( \vec v_{\tt E}  \cdot \dot{\vec X})-(\dot{\vec X}\cdot\dot{\vec X})-({\vec X}\cdot\ddot{\vec X})\Big)\vec a_{\tt E} +
(\vec a_{\tt E} \cdot \ddot{\vec X}) \vec X \Big\}_{\tt TT}   +
 {\cal O}\Big(6.68\times 10^{-14}~{\rm m/s}^2\Big),
  \label{eq:coord-acc_sum-L}
 \end{eqnarray}
 where the error bound is from the group of the omitted terms involving $\dot {\vec a}_{\tt E} $ and $\ddot U_{\rm ext}$ derivatives. It is worth noting that result (\ref{eq:coord-acc_sum-L}) includes several additional terms compared to \cite{Moyer:2003}. Also, the same expression is also applicable to ground-based stations, with the corresponding error term being on the order of ${\cal O}\big(5.27\times 10^{-15}~{\rm m/s}^2\big)$.
 
Therefore, considering realistic measurement accuracies, we refer to the magnitudes of various terms conducted for a MEO orbiter. Consequently, the expressions (\ref{eq:pos-approx-LEO}), (\ref{eq:vel-tran2}), and (\ref{eq:coord-acc_sum-L}) are recommended for position, velocity, and acceleration transformations when {\tt GNSS} data processing is concerned. These results represent the set of the expressions recommended to transform position, velocity and acceleration vectors as needed to process {\tt GNSS} data from either {\tt GCRS} or {\tt BCRS}.  These formulae are accurate to order $c^{-2}$ and the corrective terms can be evaluated using for $\vec X_{\tt TT}, \vec V_{\tt TT}$ and $\vec A_{\tt TT}$ in TT-compatible  {\tt GCRS} or {\tt TDB}-compatible {\tt BCRS} quantities translated to the geocenter. These formulae are accurate to ${\cal O}(c^{-2})$. We adopt GR (i.e., $\gamma=1$) for all numerical results.

We note that expressions in \cite{Moyer:2003} are derived in the context of DSN radiometric observables (two-way range, Doppler, and $\Delta$DOR) and is typically used in models where station acceleration terms are negligible at the required accuracy. In contrast, {\tt GNSS} processing uses one-way code and carrier-phase observables, and our goal here is dynamical consistency of the full state mapping (position--velocity--acceleration) at the millimeter level; therefore we retain the acceleration-dependent terms that are dropped in the traditional DSN form.

\textit{Recommended  transformations:}
For cm/ps-class GNSS, use Eq.~(\ref{eq:pos-approx-LEO}) for position with the two leading $c^{-2}$ terms (scale and Lorentz), Eq.~(\ref{eq:vel-tran2}) for velocity with the three leading $c^{-2}$ terms, and Eq.~(\ref{eq:coord-acc_sum-L}) for acceleration retaining terms $\gtrsim 10^{-13}\,\mathrm{m\,s^{-2}}$. These screened sets were used in the 24\,h internal frame-closure test of Fig.~\ref{fig:closure}; the bounds below quantify the truncation error introduced solely by screening of the closed-form transformations.

\begin{table*}[t]
\caption{Screened operational transformations and conservative remainder bounds using the envelopes in Table~\ref{tab:nom-values}.}
\label{tab:screenedBounds}
\begin{tabular}{|l|c|c|c|}
\hline
Transformation & Position & Velocity & Acceleration \\
\hline\hline
Screened form (TT$\rightarrow$TDB) & Eq.~(\ref{eq:pos-approx-LEO}) & Eq.~(\ref{eq:vel-tran2}) & Eq.~(\ref{eq:coord-acc_sum-L}) \\
Remainder bound (ground) & $4.03\times10^{-6}$~m & $9.73\times10^{-8}$~m/s & $5.27\times10^{-15}$~m/s$^{2}$ \\
Remainder bound (MEO)    & $7.01\times10^{-5}$~m & $1.29\times10^{-7}$~m/s & $6.68\times10^{-14}$~m/s$^{2}$ \\
\hline
\end{tabular}
\end{table*}

To the retained order, the inverse transformations to Eqs.~(\ref{eq:pos-approx-LEO}), (\ref{eq:vel-tran2}), and (\ref{eq:coord-acc_sum-L}) are obtained by interchanging
\((\vec r_{\tt E})_{\tt TDB}=(\vec x-\vec x_{\tt E})_{\tt TDB}\) with \(\vec X_{\tt TT}\),
\((\dot{\vec r}_{\tt E})_{\tt TDB}=(\dot{\vec x}-\dot{\vec x}_{\tt E})_{\tt TDB}\) with \(\vec V_{\tt TT}\), and
\((\ddot{\vec r}_{\tt E})_{\tt TDB}=(\ddot{\vec x}-\ddot{\vec x}_{\tt E})_{\tt TDB}\) with \(\vec A_{\tt TT}\), and reversing the sign of the displayed \({\cal O}(c^{-2})\) corrective terms. Products of corrective terms are \({\cal O}(c^{-4})\); therefore this sign reversal is an inverse only to the stated post-Newtonian order. Applications requiring a strict \(10^{-16}\) rate model should use the unscreened transformations and include the corresponding \({\cal O}(c^{-4})\) time terms.

For 24\,h spans with cm/ps-class observables, enforce
\[
\sup_t \|\delta\mathbf{r}\| \leq 7.01 \times10^{-5}\,\mathrm{m},\qquad
\sup_t \|\delta\mathbf{v}\| \leq 1.29\times10^{-7}\,\mathrm{m\,s^{-1}},\qquad
\sup_t \|\delta\mathbf{a}\| \leq 6.68\times10^{-14}\,\mathrm{m\,s^{-2}},
\]
matching the remainders from Eqs.~(\ref{eq:pos-approx-LEO}), (\ref{eq:vel-tran2}), and (\ref{eq:coord-acc_sum-L}) used in our closures. Here \(\|\cdot\|\) denotes the Euclidean norm, and \(\delta\vec r(t),\delta\vec v(t),\delta\vec a(t)\) are the sums of the omitted terms in the screened position, velocity, and acceleration transformations, respectively. These conservative limits are intended as regression thresholds for the transformations themselves; the end-to-end closure residuals of Sec.~\ref{sec:implement} additionally include numerical-integration and force-model consistency errors.
For $\Delta T=24$\,h, these bounds correspond to worst-case state impacts of
$\delta r \sim 7\times10^{-5}$\,m directly from the position remainder,
$\delta r \sim (\delta v)\Delta T \approx 1.1$\,cm from the velocity remainder,
and $\delta r \sim \tfrac12(\delta a)\Delta T^2 \approx 2.5\times10^{-4}$\,m ($0.25$\,mm) from the acceleration remainder. Thus the screened acceleration set is consistent with the millimeter-level closure target, while the screened velocity set remains within a centimeter-class envelope over one day.

Consequently, the screened velocity transformation in Eq.~(\ref{eq:vel-tran2}) is appropriate for cm-class one-day arcs. For sub-centimeter closure over a one-day arc, one should either retain the full velocity transformation Eq.~(\ref{eq:trans-vel-appr2*}) or include the omitted terms in the velocity-remainder budget rather than relying solely on Eq.~(\ref{eq:vel-tran2}).

Finally, mass parameters of celestial bodies, $GM$,  must also be transformed between their value in the {\tt TT}-compatible {\tt GCRS}, $(GM)_{\tt TT}$, and the {\tt TDB}-compatible value in {\tt BCRS}, $(GM)_{\tt TDB}$, as: 
{}
\begin{eqnarray}
(GM)_{\tt TDB} = (1-L_{\tt C})(GM)_{\tt TT},
\label{eq:GM_TDB}
\end{eqnarray}
which may be obtained from (\ref{eq:TCGT}) and (\ref{eq:TDBC}), by equating $(GM)_{\tt TCB}=(GM)_{\tt TCG}$ and using (\ref{eq:constLBLCLG}).

We stress that dynamical parameters tied to a TDB‑compatible ephemeris (e.g., DE440) \emph{must not} be re‑scaled when used with Eqs.~(\ref{eq:coord-acc_sum-L})–(\ref{eq:GM_TDB}); only the \emph{labels} of the time arguments change. The IAU constants used are $L_{\tt G}=6.969290134\times 10^{-10}$ and $L_{\tt B}=1.550519768\times 10^{-8}$, with the conventional ${\tt TDB}_0\simeq -65.5~\mu\mathrm{s}$.

Implementation guardrails (ephemeris argument and $GM$ scaling):
\begin{enumerate}[(i)]
\item \textit{Ephemeris argument:} query DE440 with {\tt TDB} as the independent variable. If your integrator uses {\tt TT}, convert the epoch to {\tt TDB} using Eq.~(\ref{eq:coord-TTa1a}) (or the equivalent TT$\leftrightarrow$TDB map in Secs.~\ref{sec:TDBtoTT}--\ref{sec:TTtoTDB}) before ephemeris calls.
\item \textit{$GM$ scaling:} use $GM$ values as delivered with the {\tt TDB}-based kernel (e.g., DE440) \emph{without} applying an additional $(1-L_{\tt C})$ factor.
Apply Eq.~(\ref{eq:GM_TDB}) only when converting a constant that is explicitly defined in a {\tt TT}-scaled context, i.e., \((GM)_{\tt TDB}=(1-L_{\tt C})(GM)_{\tt TT}\) and inversely
\((GM)_{\tt TT}=(1-L_{\tt C})^{-1}(GM)_{\tt TDB}\).
\item \textit{Avoid double scaling:} do not simultaneously (a) rescale $GM$ and (b) rescale the state/time arguments; choose one consistent convention.
\end{enumerate}

\subsection{Minimal cislunar regression test: NRHO-like geometry}
\label{sec:cislunar-demo}

The next two subsections form the cislunar extension track of the paper. They are not used in the Earth-orbit frame-closure test of Fig.~\ref{fig:closure}; rather, they specify the additional reference-frame, light-time, and clock-rate ingredients needed when the same \({\tt BCRS}\)-native infrastructure is extended to Earth--Moon geometries.

We give a compact, implementation-ready regression test that exercises the Earth ({\tt GCRS/BCRS}) and Moon ({\tt LCRS}) transformation chain and the 1PN barycentric light-time model. Here ``{\tt NRHO}'' denotes a near-rectilinear halo orbit about the Earth--Moon $L_2$ region. The test geometry consists of (i) an Earth-based tracking station with {\tt TT}-compatible coordinates ({\tt ITRF}$\rightarrow${\tt GCRS}), and (ii) a cislunar user spacecraft in an {\tt NRHO}-like orbit about the Moon, whose state is propagated in the {\tt TDB}-compatible {\tt BCRS} and mapped to the {\tt LCRS} using Appendix~\ref{sec:time-M} [Eqs.~(\ref{eq:pos-approx-LEO-rec})--(\ref{eq:vel-tran2-rec-ddot})]. This is a regression/consistency test of the relativistic reference-system infrastructure; end-to-end cislunar {\tt GNSS} user positioning performance additionally depends on signal availability and estimation strategy \cite{Delepaut-etal:2020,UNOOSA:2018}. Specifically, 

\textit{Order-of-magnitude check:} In Eqs.~(\ref{eq:pos-approx-LEO-rec})--(\ref{eq:vel-tran2-rec-ddot}), \(\vec v_{\tt M}=d\vec x_{\tt M}/dt\) is the Moon's barycentric velocity from the same {\tt TDB}-compatible ephemeris used for the {\tt BCRS} propagation. It should not be replaced by the Earth--Moon relative velocity \(\vec v_{\tt EM}\sim1\)~km/s. For an {\tt NRHO}-like envelope \(|\vec{\cal X}|\sim10^{8}\)~m and \(|\vec v_{\tt M}|\sim3\times10^{4}\)~m/s, the time-coupling term \(c^{-2}(\vec v_{\tt M}\!\cdot\!\vec{\cal X})\) is at the \(\sim3\times10^{-5}\)~s level, while the Lorentz-type spatial correction \(\frac12c^{-2}(\vec v_{\tt M}\!\cdot\!\vec{\cal X})\vec v_{\tt M}\) is at the \(\sim0.5\)~m level. The scale term \((L_{\tt B}-L_{\tt L}+U_{\tt M,\rm ext}/c^2)\vec{\cal X}\) is meter-class for this envelope. These terms must therefore be retained in Earth--Moon geometries; only the smaller acceleration-dependent and derivative terms should be screened.

\textit{State propagation and transformations:}
For this regression, propagate the spacecraft state in the {\tt BCRS} using the barycentric point-mass relativistic dynamics and the mission-specified lunar/terrestrial gravity, SRP, and orientation models summarized in Table~\ref{tab:cislunar-config}. This configuration is intended to exercise the reference-frame transformation chain, not to define a complete {\tt NRHO} POD force model. For the Earth leg, use the screened operational transformations in Eqs.~(\ref{eq:pos-approx-LEO}), (\ref{eq:vel-tran2}), and (\ref{eq:coord-acc_sum-L}). Map spacecraft and ground states to the {\tt LCRS} using Eqs.~(\ref{eq:pos-approx-LEO-rec}) and (\ref{eq:vel-tran2-rec}) for position and velocity; accelerations follow the same screening logic as in the Earth case.

\paragraph*{Observables and light-time:}
Compute one-way light-time between a ground station and the cislunar spacecraft with (\ref{eq:light-time}). Retain Sun and Earth Shapiro terms only; other bodies are negligible at the accuracy targeted here. (Note that most of the solar term behaves as a range scale and cancels in differenced observables, but keep it for absolute light-time consistency at the centimeter level over $\Delta T$.) Use the same {\tt TT/TDB} handling and GM scaling guidance as in Sec.~\ref{sec:GCRS-pvac} to avoid double-scaling when using {\tt TDB}-consistent ephemerides. The lunar Shapiro delay is negligible at the accuracy targeted here: a rough upper bound is $\Delta t^{\rm Moon}_{\rm Shapiro}\sim 2GM_{\tt M}c^{-3}\ln(\cdots)\lesssim \mathcal{O}(1~{\rm ps})$, corresponding to $\lesssim 0.3$ mm in range, i.e., well below the cm-class absolute light-time envelope adopted in this test.

\paragraph*{Closure metric \rm ({\tt BCRS$\rightarrow$LCRS$\rightarrow$BCRS}):}
Define the cislunar frame-closure for position and velocity over an analysis span $\Delta T$ as
\begin{equation}
\Delta\vec r(t) \equiv \vec r_{\tt BCRS\to LCRS\to BCRS}(t) - \vec r_{\tt BCRS}(t), \qquad
\Delta\vec v(t) \equiv \vec v_{\tt BCRS\to LCRS\to BCRS}(t) - \vec v_{\tt BCRS}(t),
\label{eq:cislunar-closure}
\end{equation}
and require that the sup norms over \(t\in[t_0,t_0+\Delta T]\) do not exceed remainders computed for the adopted cislunar envelope. The Earth-MEO bounds in Table~\ref{tab:screenedBounds} must not be reused unchanged for {\tt NRHO} or lunar-transfer states. For example, with \(|\vec{\cal X}|\leq 7\times10^{7}\)~m and \(|\vec a_{\tt M}|\lesssim6\times10^{-3}\,{\rm m\,s^{-2}}\), the omitted acceleration-dependent position term in the screened {\tt LCRS} map is bounded conservatively by
\[
\delta r_{\rm LCRS,pos}\lesssim \frac{3}{2}\frac{|\vec a_{\tt M}|\,|\vec{\cal X}|^2}{c^2}\lesssim5\times10^{-4}\ {\rm m}.
\]
Thus, for a millimeter-class cislunar closure test one should either use the unscreened {\tt LCRS} transformations or publish a separate {\tt NRHO}-specific screening table for \((|\vec{\cal X}|,|\dot{\vec{\cal X}}|,|\ddot{\vec{\cal X}}|)\).

As an order-of-magnitude check for Earth--Moon geometries, consider a representative {\tt NRHO} with \(|\vec{\cal X}|\sim7\times10^{7}\)~m in the {\tt LCRS} and the Moon's barycentric speed \(|\vec v_{\tt M}|\sim3\times10^{4}\)~m/s. The leading \({\cal O}(c^{-2})\) time term in the {\tt BCRS}$\leftrightarrow${\tt LCRS} map scales as \((\vec v_{\tt M}\cdot\vec{\cal X})/c^{2}\sim 2\times10^{-5}\)~s, while the associated Lorentz-type spatial correction scales as \(\frac12(v_{\tt M}^{2}/c^{2})|\vec{\cal X}|\sim0.3\)~m. For an Earth--Moon light path, the Earth Shapiro delay is typically $\mathcal{O}(10\!-\!10^{2})\,\mathrm{ps}$ (cm range-equivalent), whereas the Moon Shapiro delay is bounded by $(1+\gamma)GM_{\tt M}c^{-3}\ln(\cdots)\lesssim \mathcal{O}(1)$~ps ($<1$~mm) and is neglected here consistently with the cm-class absolute light-time envelope adopted in this regression test.

\begin{table}[t]
\caption{Configuration for the cislunar check in Sec.~\ref{sec:cislunar-demo} (NRHO-like geometry, Earth relay).}
\label{tab:cislunar-config}
\begin{tabular}{ll}
\hline
Ephemerides and time scales & DE440 ({\tt TDB}); {\tt TT/TDB} transformations as in Eqs.~(\ref{eq:coord-tr-RR1a})--(\ref{eq:coord-TTa2b}) \\
State propagation frame & {\tt BCRS} \\
Dynamics (spacecraft) & BCRS point-mass relativistic dynamics plus mission-specified lunar/terrestrial  \\
& gravity, SRP, and orientation models; this table specifies a transformation-chain \\
& regression, not a complete NRHO POD force model \\
Operational transformations (Earth) & Position/velocity/accel: Eqs.~(\ref{eq:pos-approx-LEO}), (\ref{eq:vel-tran2}), (\ref{eq:coord-acc_sum-L}) \\
{\tt LCRS} transformations (Moon) & Position/velocity (\ref{eq:pos-approx-LEO-rec}), (\ref{eq:vel-tran2-rec}); acceleration (\ref{eq:vel-tran2-rec-ddot}). \\
Light-time and delays & Eq.~(\ref{eq:light-time}); Shapiro bodies: Sun, Earth \\
Closure metric & Eq.~(\ref{eq:cislunar-closure}); thresholds from an {\tt NRHO}-specific screening of Eqs.~(\ref{eq:pos-approx-LEO-rec})--(\ref{eq:vel-tran2-rec-ddot}) \\
Span $\Delta T$ & 24 h (nominal); longer spans permissible for operational testing \\
\hline
\end{tabular}
\end{table}

\paragraph*{Practical notes:}
(i) Use the same station coordinates and Earth-orientation handling as in the MEO test so that any differences trace solely to the frame transformations.
(ii) When constructing differenced observables, the solar Shapiro contribution behaves dominantly as a scale; retain it for absolute range, document its cancellation otherwise.
(iii) Publish the reference states and resulting $\Delta\vec r,\Delta\vec v$ statistics for this case alongside those of the MEO closure to enable external regression tests.

\subsection{Transformation between the proper time and TDB  in {\tt BCRS}}
\label{sec:properT-TDB} 

This subsection supplies the clock-rate component of the cislunar extension. The formulas below are included to define the \({\tt TDB}\)-referenced proper-time model needed for range-rate and frequency-transfer observables; they are not independently validated by the MEO state-closure test in Sec.~\ref{sec:implemnt}.

\subsubsection{Direct transformations: {\tt TDB} $\rightarrow$ proper time}

Considering {\tt GNSS} applications in cislunar environment we may use a truncated version of the metric tensor of the {\tt BCRS}  (see details in \cite{Turyshev-scales:2025}). Then, to sufficient accuracy, the  transformation between the proper time of an observer $\tau$ and the coordinate time in the {\tt BCRS}, $\tt TCB$,  is given as below:
{}
\begin{equation}
\frac{d \tau}{d{\tt TCB}} = 1-{1\over c^2}\Big\{
{\textstyle\frac{1}{2}}v_0^2
+\sum_{\tt B} {G M_{\tt B} \over r_{\tt B0}}\Big(1- J_{\tt B2} \Big(\frac{R_{\tt B}}{r_{\tt B0}}\Big)^2 P_{20}(\cos\theta_{\tt B}) \Big) \Big\}_{\tt TCB}+{\cal O}\Big(8.77 \times 10^{-17}\Big),
\label{eq:tauTCB1+=}
\end{equation}
where ${\vec x}_0$, $\vec v_0=\dot {\vec x}_0$, and $v_0=|\vec v_0|$ represent the {\tt BCRS} position and velocity vectors of the spacecraft; $\vec r_{\tt B0}=\vec x_B-\vec x_0$ is the separation vector between a body $B$ and the spacecraft, $r_{\tt B0}=|\vec r_{\tt B0}|$; also we define $\cos\theta_{\tt B}= (\vec n_{\tt B0}\cdot\vec s_{\tt B})$, where $\vec s_{\tt B}$ is the unit vector along the axis of rotation for the body {\tt B}.  

Note that we retained only the quadrupole ($J_{2,{\tt B}}$) contribution in the multipolar correction.
Higher harmonics scale as $J_{\ell,{\tt B}}(R_{\tt B}/r_{{\tt B}0})^{\ell}$ and are therefore suppressed by at least $(R_{\tt B}/r_{{\tt B}0})^{2}$ relative to the $J_{2,{\tt B}}$ term. For the Earth at a GPS altitude, $(R_{\tt E}/r)^2\simeq 0.058$ and $|J_{4,{\tt E}}/J_{2,{\tt E}}|\sim10^{-3}$, implying a fractional size $\lesssim10^{-4}$ of the $J_2$ correction, i.e., $\lesssim10^{-18}$ in $d\tau/dt$, below the
$5\times10^{-18}$ metric cutoff used in Sec.~\ref{sec:rec-GCRS}. See, e.g., Refs.~\cite{Petit-Luzum:2010,Soffel-etal:2003} for conventions and typical harmonic magnitudes. Ultimately, the error bound in (\ref{eq:tauTCB1+=})  came from the omitted contribution of the  $1/c^4$--terms (see \cite{Turyshev-scales:2025}), that for a GPS spacecraft evaluated to have magnitude of $c^{-4}\big(-{\textstyle\frac{1}{8}}v_0^4
-{\textstyle\frac{3}{2}}v_0^2 w_0 + 4(\vec v_0\cdot \vec w)
+{\textstyle\frac{1}{2}}w_0^2 + \Delta\big)\simeq 8.77 \times 10^{-17}.$ 
For cm-class orbit modeling and tens-of-picoseconds light-time budgets over daily arcs, these terms are below the accuracy targeted here. They are, however, comparable to a \(10^{-16}\)-class fractional-frequency objective and therefore must be retained, or carried as an explicit systematic term, in absolute frequency-transfer analyses at that level.

Taking into account (\ref{eq:TDBC}), the relationships between the $\tt TDB$-- and $\tt TCB$--compatible quantities yield a relation for the rates of an observer's proper time and $\tt TDB$. Specifically, by applying (\ref{eq:TDBC}) and using the chain rule for derivatives, $d \tau/d{\tt TDB} =(d \tau/d{\tt TCB})(d {\tt TCB}/d{\tt TDB})=(d \tau/d{\tt TCB})/(1-L_{\tt B})$, we obtain the following result:
{}
\begin{equation}
\frac{d \tau}{d{\tt TDB}} =\frac{1}{1-L_{\tt B}}\Big(
1-{1\over c^2}\Big\{{\textstyle\frac{1}{2}}v_0^2+\sum_{\tt B} {G M_{\tt B} \over r_{\tt B0}}\Big(1- \frac{J_{\tt B2} R_{\tt B}^2}{2r_{\tt B0}^2} \big(3(\vec n_{\tt B0}\cdot\vec s_{\tt B})^2-1\big)\Big) \Big\}_{\tt TDB}\Big)+{\cal O}\Big(8.77 \times 10^{-17}\Big).
\label{eq:tauTCB1+}
\end{equation}

Integrating (\ref{eq:tauTCB1+}) provides an expression that relates  the interval of the elapsed proper time $ \tau$ during a particular interval of  $\tt TDB\in [\tt TDB,\tt TDB_\tau]$:
{}
\begin{eqnarray}
 \tau &=& \frac{1}{1-L_{\tt B}}\Big({\tt TDB}-{\tt TDB}_\tau-
\frac{1}{c^2}
\int^{\tt TDB}_{{\tt TDB_\tau}}\Big\{{\textstyle\frac{1}{2}}v_0^2+\sum_{\tt B} {G M_{\tt B} \over r_{\tt B0}}\Big(1- \frac{J_{\tt B2} R_{\tt B}^2}{2r_{\tt B0}^2} \big(3(\vec n_{\tt B0}\cdot\vec s_{\tt B})^2-1\big)\Big) \Big\}_{\tt TDB} dt\Big)
+\nonumber\\
&+&
{\cal O}\Big(8.77\times 10^{-17}\Big)\big({\tt TDB}-{\tt TDB}_\tau\big),
 \label{eq:prop-TDB}
\end{eqnarray}
where $\vec r_{0}$ and  $\vec v_0=d\vec r_{0}/dt$ are the barycentric position and velocity of the spacecraft,
$\vec r_{\tt B0}=\vec r_{\tt B}-\vec r_{0}$ is the separation vector from the spacecraft to body $B$, with
$\vec n_{\tt B0}=\vec r_{\tt B0}/r_{\tt B0}$ being the corresponding unit vector, and $\vec s_{\tt B}$ is the unit vector along the axis of the rotation, $\tt TDB_\tau$ is the beginning of the integration interval.

\subsubsection{Inverse transformations: proper time $\rightarrow$ {\tt TDB}}

Similarly to (\ref{eq:tauTCB1+=}), we determine the inverse transformation from the proper time to the {\tt TCB}:
{}
\begin{equation}
\frac{d{\tt TCB}}{d \tau} = 1+{1\over c^2}\Big\{
{\textstyle\frac{1}{2}} v^2_0 +\sum_{\tt B} {G M_{\tt B} \over r_{\tt B0}}\Big(1- 
J_{\tt B2} \Big(\frac{R_{\tt B}}{r_{\tt B0}}\Big)^2 P_{20}(\cos\theta_{\tt B})\Big) \Big\}_{\tt \tau}+{\cal O}\Big(1.37 \times 10^{-16}\Big),
\label{eq:tauTCB1+inv}
\end{equation}
where then error bound came from the omitted contribution of the $c^{-4}$-terms in the inverse transformation that  were estimated to contribute   effect on the order of   
$ c^{-4}({\textstyle\frac{3}{8}}v^4 + {\textstyle\frac{5}{2}}v^2 w_0 - 4(\vec v\cdot \vec w) +{\textstyle{\textstyle{\textstyle{1 \over 2}}}}w_0^2 - \Delta )\simeq 1.37 \times 10^{-16}.$  Thus, these terms are negligible for the cm/ps-class state and light-time closure tests considered here, but they are not negligible for a strict \(10^{-16}\)-class absolute frequency-transfer budget unless explicitly retained or bounded in the observable model.

Taking into account the relationship between {\tt TCB} and {\tt TDB}, given by (\ref{eq:TDBC}), we develop an expression relating the rates of $\tt TDB$ and observer's proper time, which from $d{\tt TDB}/{d \tau} =(d{\tt TDB}/d{\tt TCB})\, d{\tt TCB}/d\tau=\big(1-L_{\tt B}\big)\, d{\tt TCB}/d\tau$, results in:
{}
\begin{equation}
\frac{d{\tt TDB}}{d \tau} =(1-L_{\tt B})\Big(1
+{1\over c^2}\Big\{{\textstyle\frac{1}{2}} v^2_0+\sum_{\tt B} {G M_{\tt B} \over r_{\tt B0}}\Big(1- J_{\tt B2} \Big(\frac{R_{\tt B}}{r_{\tt B0}}\Big)^2 P_{20}(\cos\theta_{\tt B})\Big) \Big\}_{\tt \tau}\Big)+{\cal O}\Big(1.37 \times 10^{-16}\Big).
\label{eq:tauTCB1+*}
\end{equation}

Integrating (\ref{eq:tauTCB1+*}) over $\tau$ from $\tau_0$ to $\tau$ provides a relation between the proper time and $\tt TDB$:
{}
\begin{equation}
{\tt TDB} = \big(1-L_{\tt B}\big)\Big(\Delta \tau+
\frac{1}{c^2}
\int^{\tau}_{\tau_0}\Big\{{\textstyle\frac{1}{2}} v^2_0+\sum_{\tt B} {G M_{\tt B} \over r_{\tt B0}}\Big(1- J_{\tt B2} \Big(\frac{R_{\tt B}}{r_{\tt B0}}\Big)^2 P_{20}(\cos\theta_{\tt B})\Big) \Big\}_\tau dt\Big)
+{\cal O}\Big(1.37\times 10^{-16}\Big)\Delta\tau,
 \label{eq:prop-TDB+}
\end{equation}
where the relativistic time equation at the clock is given  as
{}
\begin{equation}
\tau = t^* - {1 \over c^2}  \int_{t_0}^{t^*} \Big\{{\textstyle\frac{1}{2}} v^2_0+ \sum_{\tt B} {G M_{\tt B} \over r_{\tt B0}}\Big(1- J_{\tt B2} \Big(\frac{R_{\tt B}}{r_{\tt B0}}\Big)^2 P_{20}(\cos\theta_{\tt B})\Big)\Big\}dt+{\cal O}\Big(1.10\times 10^{-16}\Big)(t^*-t_0),
 \label{eq:tT+=}
\end{equation}
with the representation of the motion of the clock with respect to the {\tt BCRS} given as below
\begin{equation}
\bar{\vec x}_0(\tau) = \vec x_0(t^*),
\label{eq:EeqM=}
\end{equation}
that may be solved either numerically or analytically, in the analogy to the approach used in (\ref{eq:tT+})--(\ref{eq:EeqM}). For direct transformations (\ref{eq:prop-TDB}), one should integrate the integral either numerically or analytically. For the inverse transformation (\ref{eq:prop-TDB+}), to express the right-hand members in terms of $\tau$, one must obtain the function $t^* = t^*(\tau)$ by inverting the equation (\ref{eq:tT+=}), also either numerically or analytically.  

\section{BCRS implementation and frame-closure validation for {\tt GNSS} dynamics}
\label{sec:implement}

In this Section, we verify the {\tt BCRS}-native implementation and the {\tt GCRS}$\leftrightarrow${\tt BCRS} state transformations developed in Sec.~\ref{sec:GCRS-pvac} via an internal \emph{frame-closure} experiment. Specifically, we compare a GPS-like trajectory obtained by integrating the {\tt TT}-compatible {\tt GCRS} equations of motion to the trajectory obtained by
(i) transforming the same initial state into the {\tt TDB}-compatible {\tt BCRS},
(ii) integrating the corresponding barycentric equations of motion, and
(iii) transforming the resulting state back into the {\tt GCRS} at each epoch.
The resulting closure residuals quantify numerical and modeling consistency of the transformation chain and force models; they are not intended as an assessment of end-to-end user positioning accuracy from observations.

For completeness, we summarize the equations of motion used in each frame and highlight the treatment of the dominant Newtonian, post-Newtonian, and geopotential terms needed to reproduce the closure experiment.

\subsection{Equations of motion in the scaled {\tt TDB}-compatible {\tt BCRS}}

\subsubsection{{\tt BCRS} metric in the mass monopole approximation}
\label{sec:mass-monopoles}

For many applications, it suffices to consider only the mass monopoles of all solar system bodies, retaining solely the masses which represents the mass monopole approximation of the {\tt BCRS} metric,  for a system of $N$ point-like gravitational sources in four dimensions is outlined in \cite{Moyer:2003}:
{}
\begin{eqnarray}
\label{eq:metric}
g_{00}&=&1-\frac{2}{c^2}\sum_{j\not=i}\frac{\mu_j}{r_{ij}}+
\frac{2\beta}{c^4}\Big[\sum_{j\not=i}\frac{\mu_j}{r_{ij}}\Big]^2-
\frac{1+2\gamma}{c^4}\sum_{j\not=i}\frac{\mu_j{\dot r}^2_j}{r_{ij}}+
\frac{2(2\beta-1)}{c^4}\sum_{j\not=i}\frac{\mu_j}{r_{ij}}
\sum_{k\not=j}\frac{\mu_k}{r_{jk}}-
\frac{1}{c^4}\sum_{j\not=i}\mu_j\frac{\partial^2 r_{ij}}{\partial t^2}+
{\cal O}(c^{-5}),\nonumber\\
g_{0\alpha}&=& \frac{2(\gamma+1)}{c^3}\sum_{j\not=i}\frac{\mu_j{\dot {\bf r}}^\alpha_j}{r_{ij}}+
{\cal O}(c^{-5}),
\qquad 
g_{\alpha\beta}=\eta_{\alpha\beta}\Big(1+\frac{2\gamma}{c^2}\sum_{j\not=i}\frac{\mu_j}{r_{ij}}\Big)+{\cal O}(c^{-4}),   
\end{eqnarray}
where the indices $j$ and $k$ refer to the $N$ bodies, with $k$ including body $i$, whose motion is being investigated. The gravitational constant for body $j$, denoted as $\mu_j$, is defined as $\mu_j = Gm_j$, where $G$ is the universal Newtonian gravitational constant, and $m_j$ is the isolated rest mass of body $j$. Additionally, the vector $\vec r_i$ represents the barycentric radius-vector of this body. The vector $\vec r_{ij}=\vec r_{j}-\vec r_i$ points from body $i$ to body $j$, $r_{ij}=|\vec r_{j}-\vec r_i|$, and the vector $\vec n_{ij}=\vec r_{ij}/r_{ij}$ is the unit vector along this direction.

Furthermore, in (\ref{eq:metric}), we have reinstated the parameterized post-Newtonian (PPN) parameters \cite{Will_book93}, which have been adapted to relativistic reference frames in \cite{Kopeikin-Vlasov:2004}. The two PPN parameters, $\gamma$ and $\beta$, have specific physical interpretations: $\gamma$ quantifies the curvature of space-time produced by a unit rest mass, while $\beta$ quantifies the non-linearity in the superposition of gravitational fields in gravity theory. In general relativity, $\gamma=\beta=1$, and the other eight parameters are absent, situating the theory within a two-dimensional theoretical framework.

\subsubsection{Equations of motion in {\tt BCRS}: point-mass approximation}

The barycentric metric given by (\ref{eq:metric}) may be used to derive the post-Newtonian EIH equations of motion for a system of mass monopoles. These equations form the basis of modern solar system ephemerides. The relevant point-mass Newtonian and relativistic perturbative accelerations in the {\tt TDB}-compatible {\tt BCRS} are used to describe the motion of spacecraft in the solar system, including Earth orbiters and the Moon. The corresponding EIH equations are given as follows \cite{Moyer:2003}:
{}
\begin{eqnarray}
\ddot{\bf r}_i&=&\sum_{j\not=i}\frac{\mu_j({\bf r}_j-{\bf r}_i)}{r_{ij}^3}\bigg\{1-\frac{2(\beta+\gamma)}{c^2}\sum_{l\not=i}\frac{\mu_l}{r_{il}}-
\frac{2\beta-1}{c^2}\sum_{k\not=j}\frac{\mu_k}{r_{jk}}+
\gamma\Big(\frac{{\dot r}_i}{c}\Big)^2+(1+\gamma)\Big(\frac{{\dot r}_j}{c}\Big)^2-\frac{2(1+\gamma)}{c^2} (\dot{\bf r}_i \cdot \dot{\bf r}_j)-\nonumber\\
&&-~\frac{3}{2c^2}\Big[\frac{({\bf r}_i-{\bf r}_j)\cdot{\dot{\bf r}}_j}{r_{ij}}\Big]^2+\frac{1}{2c^2}\big(({\bf r}_j-{\bf r}_i)\cdot{\ddot{\bf r}}_j\big)\bigg\}+
\frac{1}{c^2}\sum_{j\not=i}\frac{\mu_j}{r_{ij}^3}
\Big\{\big({\bf r}_i-{\bf r}_j\big)\cdot\Big( (2+2\gamma){\dot {\bf r}}_i-(1+2\gamma){\dot {\bf r}}_j\Big)\Big\}({\dot{\bf r}}_i-{\dot{\bf r}}_j)+
\nonumber\\
&&+~\frac{3+4\gamma}{2c^2}\sum_{j\not=i}\frac{\mu_j{\ddot {\bf r}}_j}{r_{ij}}+\sum_{m=1}^3\frac{\mu_m({\bf r}_m-{\bf r}_i)}{r^3_{im}}+\sum_{c,s,m}{\bf F} +{\cal O}(c^{-4}).
\label{eq:4-26-mod}
\end{eqnarray}
The first term in (\ref{eq:4-26-mod}) represents the Newtonian gravitational acceleration, which must be computed using barycentric {\tt TDB}-compatible quantities. The subsequent terms are the post-Newtonian contributions. The penultimate term on the right-hand side of (\ref{eq:4-26-mod}) refers to the `Big3' asteroids—Ceres, Pallas, and Vesta—using the index $m$. The final term accounts for forces on the Earth, Moon, and Mars from 297 other asteroids, which are grouped into three taxonomic classes (C, S, M), see \cite{Standish_Williams_2008} for details. For a planetary orbiter, it is usual to integrate the difference between the orbiter's acceleration and the planet's acceleration, both derived from the EIH equations. Typically, only the relativistic corrective terms with the index $j$ matching either the central planet or the Sun have significant effects.

\subsubsection{Light time equation}
\label{sec:lighttime}

To determine the orbits of planets and spacecraft, one must also account for the propagation of EM signals between any two points in space. The corresponding light-time equation is derived from the metric tensor in (\ref{eq:metric}) as follows:
{}
\begin{equation}
\label{eq:light-time}
t_2-t_1 = \frac{r_{12}}{c} + (1+\gamma)\sum_{\tt B}\frac{GM_{\tt B}}{c^3}\ln\left[\frac{r_1^{\tt B}+r_2^{\tt B}+r_{12}}{r_1^{\tt B}+r_2^{\tt B}-r_{12}}\right]+ \mathcal{O}(c^{-5}),
\end{equation}
where $t_1$ and $t_2$ refer to the signal's emission and reception coordinate times, respectively, $\vec r_1\equiv \vec r(t_1)$ and $\vec r_2\equiv \vec r(t_2)$ are the barycentric position vectors of emitter and receiver, and $r_{12}\equiv |\vec r_2-\vec r_1|$. For each gravitating body ${\tt B}$ in the sum, $r_1^{\tt B}\equiv |\vec r_1-\vec x_{\tt B}(t_1)|$ and $r_2^{\tt B}\equiv |\vec r_2-\vec x_{\tt B}(t_2)|$ are the emitter/receiver distances to ${\tt B}$. The Shapiro term is proportional to $(1+\gamma)$, with $\gamma=1$ in general relativity.

The Shapiro time delay is proportional to $(1+\gamma)$. The average solar potential at a distance of 1 AU is $\left<U\right>/c^2 = 0.99 \times 10^{-8}$. The mean delay caused by the Sun for GPS signals is  $\sim2.66$ ns, equivalent to  $\sim79.8$ cm. The annual variation is $\pm7$ mm. The delay due to Earth’s gravity can reach up to 42.3 ps (or $\sim1.3$ cm), and it varies with the elevation angle. Therefore, only the Sun and Earth must be included in the model  (\ref{eq:light-time}). At L-band, solar plasma delay is negligible except near conjunction; apply standard transformations for troposphere/ionosphere in the observable model. Most of the solar term behaves as an overall scale in range and cancels in differenced observables, but should be retained for absolute light-time consistency at the centimeter level. 

For increased fidelity near conjunctions, evaluate the Shapiro term in \eqref{eq:light-time} with a first-order
(retarded) approximation for the gravitating-body position:
\[
\mathbf{x}_{\tt B}^{\mathrm{ret}}(t_i) \;\approx\;
\mathbf{x}_{\tt B}(t_i) \;-\; \frac{\bigl\|\mathbf{r}_i-\mathbf{x}_{\tt B}(t_i)\bigr\|}{c}\,\mathbf{v}_{\tt B}(t_i),
\qquad i\in\{1,2\}.
\]
This captures the leading $\mathcal{O}(v_{\tt B}/c)$ effect without introducing an iterative implicit definition. The light-time and its derivatives used here are consistent with the
time-transfer-function (TTF) formalism; see e.g., \cite{Hees-etal:2014} for the moving-axisymmetric-body generalization used in interplanetary radioscience.

When an endpoint fixed to the rotating Earth is supplied in a terrestrial rotating frame rather than as an inertial {\tt GCRS}/{\tt BCRS} position evaluated at its emission or reception epoch, the equivalent one-way rotating-frame correction is the Sagnac term
\begin{equation}
\Delta t_{\rm sag} =
\frac{1}{c^2}\Big(\vec\Omega_\oplus\cdot
\big(\vec r_R(t_r)\times \vec r_T(t_e)\big)\Big),
\label{eq:sagnac}
\end{equation}
so that, in this rotating-coordinate implementation only,
\(\Delta t=\Delta t_{\rm geom}+\Delta t_{\rm Shapiro}+\Delta t_{\rm sag}\).
Equation~(\ref{eq:sagnac}) must not be added separately if \(\vec r_R(t_r)\) and \(\vec r_T(t_e)\) have already been transformed to inertial {\tt GCRS}/{\tt BCRS} coordinates at their respective epochs; in that case the rotation of the ground endpoint is already contained in the endpoint kinematics. Numerically, for a typical {\tt GNSS} MEO geometry with \(|\vec r_R|\simeq6378\)~km, \(|\vec r_T|\simeq26560\)~km, and \(|\vec\Omega_\oplus|=7.292115\times10^{-5}\,{\rm s^{-1}}\), one finds \(|\Delta t_{\rm sag}|\lesssim1.4\times10^{-7}\)~s, or approximately \(41\)~m in range, so the effect is mandatory whenever rotating terrestrial endpoint coordinates are used \cite{Ashby-lrr-2003-1}. The term (\ref{eq:sagnac}) leads to a correction
\begin{equation}
\frac{\partial}{\partial t_{\rm r}}\Delta t_{\rm sag}
= \frac{1}{c^{2}}\Big(\boldsymbol{\Omega}_\oplus\cdot\Big(\big[\,\mathbf{v}_{\rm R}(t_{\rm r}) \times \mathbf{r}_{\rm T}(t_{\rm e})\big]
- \big[\mathbf{v}_{\rm T}(t_{\rm e}) \times \mathbf{r}_{\rm R}(t_{\rm r}) \big]\Big)\Big),
\label{eq:sagnac_rate}
\end{equation}
that is required wherever one forms two-way frequency/phase models in the same rotating-coordinate convention. For a rotating ground site this term is at the $10^{-12}-10^{-11}$  fractional level for L-band. 

\subsubsection{Frequency transfer observable at \texorpdfstring{$\mathcal{O}(c^{-3})$}{O(c^{-3})}}

For one-way transfer from emitter {\tt A} to receiver {\tt B},
\begin{equation}
\frac{\nu_{\tt A}}{\nu_{\tt B}} =
\frac{(u^{0})_{\tt A}}{(u^{0})_{\tt B}}\,\frac{q_{\tt A}}{q_{\tt B}} + \mathcal{O}(c^{-4}),
\label{eq:freq-transfer}
\end{equation}
where the ratio $q_A/q_B$ follows from differentiating the light-time equation \eqref{eq:light-time} with respect to emission time, retaining terms through $\mathcal{O}(c^{-3})$  \cite{Blanchet-etal:2001,Turyshev-etal:2013}. 

For GNSS applications the relevant observable is the one-way downlink. An explicit post-Newtonian expansion of the propagation factor $q_A/q_B$ through $\mathcal{O}(c^{-3})$, consistent with Eq.~(\ref{eq:light-time}), is given in Ref.~\cite{Blanchet-etal:2001} (see also Refs.~\cite{Moyer:2003,Hees-etal:2014}). In our implementation we compute $(u^0)_A$ and $(u^0)_B$ from the adopted metric and obtain $q_A/q_B$ from the derivative $\mathrm{d}(t_2-t_1)/\mathrm{d}t_1$ evaluated along the modeled light-time solution. These expressions are consistent with our light-time Eq.~(\ref{eq:light-time}) and close the propagation part of the range-rate and frequency-transfer model through \(\mathcal{O}(c^{-3})\). A strict \(10^{-16}\)-class absolute frequency comparison additionally requires the \(\mathcal{O}(c^{-4})\) proper-time terms bounded in Sec.~\ref{sec:properT-TDB}; the present closure test does not validate those omitted clock-rate contributions.

\subsection{Equations of motion in the scaled {\tt TT}-compatible {\tt GCRS}}

\subsubsection{ {\tt GCRS}: Practically-relevant formulation}
\label{sec:rec-GCRS}

In practical {\tt GCRS} implementations, one includes all post-Newtonian terms up to \(\mathcal O(c^{-4})\) in \(G_{00}\), \(\mathcal O(c^{-3})\) in \(G_{0\alpha}\), and \(\mathcal O(c^{-2})\) in \(G_{\alpha\beta}\), but discards any metric perturbations smaller than \(5\times10^{-18}\).  Accordingly, the metric tensor of {\tt GCRS} (see \cite{Turyshev-etal:2025,Turyshev-scales:2025})  retaining only \(\lvert\delta G_{mn}\rvert\gtrsim5\times10^{-18}\), sufficient for high-precision time-keeping applications,  becomes
{}
\begin{eqnarray}
G_{00}(T,{\bf X})  &=& 1 - \frac{2}{c^2} \Big\{W_{\tt E}(T,{\bf X}) + W_{\rm tid}(T,{\bf X})\Big\} + \frac{2}{c^4}W^2_{\rm E}(T,{\bf X}) + {\cal O}\Big(c^{-5}; 6.61\times 10^{-25}\Big), 
\label{eq:G00tr}\\
G_{0\alpha}(T,{\bf X})  &=& -{2 G\over c^3}  \frac{[ {\vec J}_{\tt E}\times{\vec X}]_\alpha}{R^3}+ {\cal O}\Big(c^{-5}; 2.79\times 10^{-19}\Big),   
\label{eq:G0atr}\\
G_{\alpha\beta}(T,{\bf X})  &=& \eta_{\alpha\beta}\Big(1 + \frac{2}{c^2} \Big\{W_{\tt E}(T,{\bf X}) + W_{\rm tid}(T,{\bf X})\Big\} \Big) + {\cal O}\Big(c^{-4}; 5.57\times 10^{-20}\Big),
\label{eq:Gabtr}
\end{eqnarray}
where \(\vec J_{\tt E}\) is the Earth's angular momentum, \(J_{\tt E}\simeq 5.86\times10^{33}\,{\rm kg\,m^2\,s^{-1}}\). We reserve \(\vec S_{\tt E}\equiv \vec J_{\tt E}/M_{\tt E}\) for the corresponding specific spin angular momentum, with units \({\rm m^2\,s^{-1}}\), used in the equations of motion below. Also, $W_{\tt E}(T,{\bf X}) $  is the Earth's gravitational potential 
{}
\begin{eqnarray}\label{BD-WE-sphe}
W_{\tt E}(T,\vX)
&=&
{G M_{\tt E} \over R}
\Big\{ 1 + \sum_{\ell= 2}^\infty \sum_{m=0}^{\ell}
\Big( {R_{\tt E}\over R} \Big)^\ell P_{\ell m}(\cos \theta)
\Big(C^{\tt E}_{\ell m}(T,R) \cos m \phi
+ S^{\tt E}_{\ell m}(T,R) \sin m \phi \Big) \Big\} + \cO(c^{-4}),
\end{eqnarray}
where $M_{\tt E}$ and $R_{\tt E}$ are the Earth's mass and equatorial radius, respectively, while $P_{\ell k}$ are the associated Legendre-polynomials \cite{Abramovitz-Stegun:1965}.  $W_{\rm tid}(T,{\bf X})$ is  the Newtonian the tidal potential by external bodies
{}
\begin{equation}
\label{W-tidal}
W_{\rm tid}(T,\vX) =
U_{\rm ext}(\ve{x}_{\tt E} + \vX) - U_{\rm ext}(\ve{x}_{\tt E}) - \big(\vX \cdot
\vec \nabla  U_{\rm ext}(\ve{x}_{\tt E})\big),
\end{equation}
where $\vec{r}_{\tt BE}=\vec x_{\tt E}-\vec x_{\tt B}$ is the vector connecting the center of mass of body $\tt B$ with that of the Earth, with ${r}_{\tt BE}=|\vec{r}_{\tt BE}|$ and $\vec{n}_{\tt BE}=\vec{r}_{\tt BE}/{r}_{\tt BE}$, also $\widehat{\vec { X}}={\vec { X}}/{ X}$ and $\cos\theta_{\tt BE} = ( { {\vec  n}}_{\tt BE}\cdot \widehat{\vec { X}})$, with  $P_\ell\bigl(\cos\theta\bigr)$ being the Legendre polynomials. Naturally, the quadratic term (i.e., $\sim {\cal O}(X^2)$) in the resulting expression for $W_{\rm tidal}$ is the dominant one.

The error bounds in  \eqref{eq:G00tr}–\eqref{eq:Gabtr}  are due to  the dominant omitted corrections evaluated at GPS altitude, specifically:  $\delta G_{00}^{\rm (mix)}=-4c^{-4}
W_{\tt E}\,W_{\rm tid}\simeq 4c^{-4}(GM_{\tt E}/r_{\tt GPS})(GM_{\tt S}/{\rm AU}^3+Gm_{\tt M}/r_{\tt EM}^3)r_{\tt GPS}^2\simeq 6.61\times10^{-25}$;
$\delta G_{0\alpha}^{\rm (tid)}=-4c^{-3}W_{\rm tid}^\alpha \simeq 4c^{-3} (GM_{\tt M}/r_{\tt EM}^3)v_{\tt M} r_{\tt GPS}^2\simeq2.79\times10^{-19}$; and $\delta G_{\alpha\beta}^{\rm (2PN)}=\eta_{\alpha\beta}\,\tfrac32 c^{-4} W_{\tt E}^2\simeq \tfrac32c^{-4} (GM_{\tt E}/r_{\tt GPS})^2\simeq  4.78\times10^{-20}$, see details in  \cite{Turyshev-scales:2025}. Although we evaluated the metric components in \eqref{eq:G00tr}–\eqref{eq:Gabtr} at GPS altitude (i.e., where tidal contributions exceed those at the surface), these expressions remain valid for all Earth‐orbit regimes from MEO through GEO.

\subsubsection{Equations of motion in the {\tt GCRS}}
\label{sec:eq-mot-GCRS}

The trajectory of an Earth satellite in the {\tt GCRS} is written using the geocentric state \((\vec X,\dot{\vec X},\ddot{\vec X})\), consistent with the notation used in Sec.~\ref{sec:pos-vel-acc}. It is governed by the following equation of motion \cite{Huang-etal:1990,Moyer:2003,Ashby-lrr-2003-1,Godard-etal:2012,Turyshev-Toth:2023-grav-phase}:
{}
\begin{eqnarray}
\ddot {\vec X}&=& \ddot {\vec X}_{\tt E}+\ddot {\vec X}_{{\tt E}J_2}+\ddot {\vec X}_{\tt tidal}+\ddot {\vec X}_{\tt GR}+\ddot {\vec X}_{\tt SRP}+\ddot {\vec X}_{\tt {\tt E} elast}+\ddot {\vec X}_{\tt {\tt E} rad}+\ddot {\vec X}_{\tt custom},
\label{eq:eqmot_tot}
\end{eqnarray}
where \(\ddot{\vec X}_{\tt E}\) is the Newtonian acceleration due to the Earth's gravity field; \(\ddot{\vec X}_{{\tt E},J_2}\) is the acceleration due to Earth's oblateness; \(\ddot{\vec X}_{\tt tidal}\) is the tidal acceleration induced by external bodies; \(\ddot{\vec X}_{\tt GR}\) is the relativistic acceleration; \(\ddot{\vec X}_{\tt SRP}\) is the direct/indirect solar-radiation-pressure acceleration; \(\ddot{\vec X}_{\tt E elast}\) is the acceleration due to elastic Earth response, pole, and ocean tides; \(\ddot{\vec X}_{\tt E rad}\) is the acceleration due to Earth albedo and infrared radiation; and \(\ddot{\vec X}_{\tt custom}\) denotes any custom force model. All quantities in Eq.~(\ref{eq:eqmot_tot}) are \({\tt TT}\)-compatible unless explicitly stated otherwise.

The acceleration of a spacecraft at $\vec{X}$, due to the Earth's gravity field in (\ref{eq:eqmot_tot}) is computed as usual
{}
\begin{eqnarray}
\ddot {\vec X}_{\tt E}= {\vec \nabla}U_{\tt E}(\vec X)+{\cal O}(c^{-4})&=&-\frac{GM_{\tt E}}{R^3}{\vec X}\Big(1+{\cal O}\Big(\frac{R^2_{\tt E}}{R^2}\Big)\Big)
\simeq-\frac{GM_{\tt E}}{a^2}{\vec n}_R \approx 0.57~{\rm m/s}^2,
\label{eq:newt}
\end{eqnarray}
where $\vec a=a \vec n_R$ is the semi-major axis of the spacecraft orbit, ${\vec n}_R=\vec{X}/R$ is the unit vector in the direction of the spacecraft from the {\tt GCRS} with $R=|\vec X|$. In this estimate, we truncated the Earth relativistic gravitational potential $U_{\tt E}(\vec X)\equiv W_{\tt E}(T,\vX)$ from (\ref{BD-WE-sphe}) to the monopole term. Clearly, this is the dominant term in the equations of motion of an Earth-orbiting satellite (\ref{eq:eqmot_tot}). 

Let $\mu_{\tt E} = G M_{\tt E}$, $R_{\tt E}$ the Earth's mean equatorial radius, $J_{2,\tt E}$ the quadrupole coefficient, ${\vec V}=\dot {\vec X}$, $V=|{\vec V}|$, and $\hat{\mathbf{S}}$ the unit spin axis of the Earth. The 1PN gravitoelectric acceleration due to the oblateness is
\begin{align}
\ddot {\vec X}_{{\tt E}, J_2}= & \frac{3 J_{2,\tt E}\,\mu_{\tt E} R_{\tt E}^{2}}{2\,c^{2}\,R^{4}}
\Big( \big[ 5\,{\mathbf{n}}_R(\hat{\mathbf{S}}\!\cdot\!{\mathbf{n}}_R)^{2} - 2\,\hat{\mathbf{S}}(\hat{\mathbf{S}}\!\cdot\!{\mathbf{n}}_R) - {\mathbf{n}}_R\big]\big[ V^{2} - \frac{4\mu_{\tt E}}{R} \big] -\nonumber\\
&\;\; -\, 4 \big[ 5(\mathbf{n}_R\!\cdot\!\mathbf{V})(\hat{\mathbf{S}}\!\cdot\!{\mathbf{n}}_R)^{2} - 2(\hat{\mathbf{S}}\!\cdot\!\mathbf{V})(\hat{\mathbf{S}}\!\cdot\!{\mathbf{n}}_R) - ({\mathbf{n}}_R\!\cdot\!\mathbf{V}) \big]\mathbf{V}
\;-\; \frac{4\mu_{\tt E}}{R}\big[ 3(\hat{\mathbf{S}}\!\cdot\!{\mathbf{n}}_R)^{2} - 1 \big]{\mathbf{n}}_R \Big).
\label{eq:aJ2rel}
\end{align}
In the worst-case orientation and using $J_{2,{\tt E}}=1.08263\times 10^{-3}$, $R_{\tt E}=6378~\mathrm{km}$, $\mu_{\tt E}=3.986\times 10^{14}~\mathrm{m^{3}s^{-2}}$, this acceleration was evaluated to contribute:
LEO:  $\sim 7\times 10^{-11}$~m\,s$^{-2}$, MEO: $\sim 10\times 10^{-13}$~m\,s$^{-2}$, cislunar: $\sim 5\times 10^{-16}$~m\,s$^{-2}$.  Thus, for LEO its omission biases precise orbit and gravity solutions, while for MEO it is small but systematic. A useful scaling is $\ddot{\vec X}_{\tt EJ2\,rel}\propto r^{-4}$, so the effect rapidly decreases for higher Earth orbits. As a rough displacement scale over a one-day arc, $\tfrac12 |\ddot{\vec X}|\,\Delta t^2$ with $\Delta t=86400$\,s gives $\sim 0.3$\,m (LEO, using $7\times10^{-11}\,{\rm m/s^2}$) and $\sim 4$\,mm (GPS-like MEO, using $1\times10^{-12}\,{\rm m/s^2}$); at lunar distance it falls below the micron level. Accordingly, the 1PN $J_2$ term is mandatory for LEO POD, can matter for mm-level MEO processing, and is negligible for lunar-distance dynamics.

The tidal acceleration due to the gravity fields of the Sun, Venus and Moon are also well known. Using $u^{\tt tidal}_{\tt E}(\vec X)\equiv W_{\tt tidal}(\vec X)$ from (\ref{W-tidal}), we compute it as usual:
{}
\begin{eqnarray}
\ddot {\vec X}_{\rm tidal}&=& {\vec \nabla}u^{\tt tidal}_{\tt E}(\vec X)+{\cal O}\Big(\frac{X^3}{r^4_{\tt BE}},c^{-2}\Big)\simeq\sum_{b\not={\rm E}}\frac{GM_b}{r^3_{\tt BE}}\Big(3(\vec{n}^{}_{\tt BE}\cdot\vec{X})\vec{n}^{}_{\tt BE}-\vec{X}\Big)\leq\nonumber\\
&\leq&
\frac{2GM_{\tt S} a}{a^3_{\tt E}}(\vec{n}_{\tt S}\cdot\vec{n}_X)\vec{n}_{\tt S}+
\frac{2GM_{\tt M} a}{a^3_{\tt M}}(\vec{n}_{\tt M}\cdot\vec{n}_X)\vec{n}_{\tt M} +
\frac{2GM_{\tt V} a}{a^3_{\tt V}}(\vec{n}_{\tt V}\cdot\vec{n}_X)\vec{n}_{\tt V}
\approx \nonumber\\
&\approx & 2.11\times 10^{-6}~{\rm m/s}^2+4.59\times 10^{-6}~{\rm m/s}^2+2.45\times 10^{-10}~{\rm m/s}^2,
\label{eq:tidal}
\end{eqnarray}
where $a_{\tt E}, e_{\tt E}$, and ${\cal E}_{\tt E}$ and $a_{\tt V}, e_{\tt V}$, and ${\cal E}_{\tt V}$ are the semi-major axis, eccentricity and eccentric anomaly characterizing the heliocentric orbits of the Earth and Venus  correspondingly, while $a_{\tt M}, e_{\tt M}$, and ${\cal E}_{\tt M}$, are the same quantities for the geocentric orbit of the Moon. We will use the value of $a_{\tt M}=384,400$~km to denote the Earth-Moon mean distance.

In addition, we estimate the maximal accelerations from the tides introduced by Jupiter and Saturn:
{}
\begin{eqnarray}
|\ddot {\vec X}^{\tt J}_{\rm tidal}|&\lesssim&\frac{2GM_{\tt J}a}{(a_{\tt J}-a_{\tt E})^3}\approx 2.70\times 10^{-11}~{\rm m/s}^2,
\qquad
|\ddot {\vec X}^{\tt Sa}_{\rm tidal}|\lesssim\frac{2GM_{\tt Sa}a}{(a_{\tt Sa}-a_{\tt E})^3}\approx 9.52\times 10^{-13}~{\rm m/s}^2,
\label{eq:tidal2a}
\end{eqnarray}
with contributions from Uranus, Neptune being below $10^{-14}$~m/s$^2$. Note that these tidal forces are not  accounted for in the equations of motion used to process GPS observables. Due to the long orbital periods of these planets, they contribute a nearly constant background that is removed along with other constant terms during data analysis.

The relativistic corrections to the accelerations of Earth satellites written in a local inertial {\tt GCRS} that is kinematically non-rotating read as \cite{Brumberg-Kopeikin:1989}:
{}
\begin{eqnarray}
\ddot {\vec X}_{\tt GR}&=&\frac{GM_{\tt E}}{c^2R^3}\bigg\{\Big[2(\beta+\gamma)\frac{GM_{\tt E}}{R}-\gamma\dot{\vec X}^2\Big]\vec X+2(1+\gamma)({\vec X}\cdot\dot{\vec X})\dot{\vec X}\bigg\}+\nonumber\\
&&+\, (1+\gamma)\frac{GM_{\tt E}}{c^2R^3}\Big[\frac{3}{R^2}[\vec X\times\dot{\vec X}](\vec X\cdot \vec S_{\tt E})+[\dot{\vec X}\times\vec S_{\tt E}]\Big]+
(1+2\gamma)\frac{GM_{\tt S}}{c^2r^3_{\tt E}}\big[[{\vec r}_{\tt E}\times \dot{\vec r}_{\tt E}]\times\dot{\vec X}\big],
\label{eq:rel_Eqmot}
\end{eqnarray}
where \(GM_{\tt E}\) and \(GM_{\tt S}\) are the standard Earth and Sun gravitational parameters; $\beta,\gamma$ are PPN parameters equal to 1 in GR; $c$ is the speed of light; $\vec X, \dot{\vec X}$ are the position and velocity of a satellite with respect to the geocenter; \(\vec r_{\tt E}, \dot{\vec r}_{\tt E}\) are the position and velocity of the Earth with respect to the Sun; \(\vec S_{\tt E}=\vec J_{\tt E}/M_{\tt E}\) is the Earth's specific spin angular momentum. Introducing \(\mathcal{R}\) as the rotation matrix from ECEF to the inertial axes, we write
\[
\vec S_{\tt E} \simeq 0.33068\,R_{\tt E}^{2}\,\omega_{\tt E}\,\mathcal{R}\,\hat{\mathbf z},
\]
where the numerical factor is the normalized terrestrial moment of inertia used for this estimate.

The first line describes the so-called Schwarzschild term \cite{Schwarzschild:1916,Schwarzschild:2003}, the next term is the frame-dragging gravitomagnetic Lense-Thirring effect \cite{Lense-Thirring:1918}, whereas the third corresponds to the de Sitter effect \cite{deSitter:1917}. Other relativistic effects are related to, e.g., other celestial bodies, which can be approximated as small corrections to the non-relativistic tidal forces, Thomas precession, or Earth's oblateness quadruple term; however, they are currently not considered in the precise {\tt GNSS} orbit solutions, due to their very small magnitudes \cite{Huang-etal:1990, Brumberg-Kopeikin:1989,Soffel-etal:2003,DSX-I,Godard-etal:2012}.

We evaluate the Schwarzschild term in (\ref{eq:rel_Eqmot}) as:
{}
\begin{eqnarray}
\ddot {\vec X}_{\rm Sch}&=&\frac{GM_{\tt E}}{c^2R^3}\bigg\{\Big[2(\beta+\gamma)\frac{GM_{\tt E}}{R}-\gamma\dot{\vec X}^2\Big]\vec X+2(1+\gamma)({\vec X}\cdot\dot{\vec X})\dot{\vec X}\bigg\}
\simeq
\frac{3(GM_{\tt E})^2}{c^2a^3}{\textstyle\frac{1}{3}}(2\beta+\gamma)\vec n_R
\leq  2.83\times 10^{-10}~{\rm m/s}^2.~~~~
\label{eq:rel_Eqmot_S}
\end{eqnarray}
For {\tt GNSS} MEO orbit determination, planetary tidal terms beyond Sun/Moon are typically negligible at the cm level over daily arcs and are largely absorbed by empirical parameters; Eq.~(\ref{eq:tidal2a}) quantifies that even worst-case Jupiter/Saturn tidal accelerations remain $\lesssim 10^{-11}$ m/s$^2$.

Next, we evaluate the frame-dragging gravitomagnetic Lense-Thirring effect  in (\ref{eq:rel_Eqmot}):
{}
\begin{eqnarray}
\ddot {\vec X}_{\tt LT}&=&(1+\gamma)\frac{GM_{\tt E}}{c^2R^3}\Big[\frac{3}{R^2}[\vec X\times\dot{\vec X}](\vec X\cdot \vec S_{\tt E})+[\dot{\vec X}\times\vec S_{\tt E}]\Big]
\simeq
1.98\Big(\frac{GM_{\tt E}}{a}\Big)^\frac{3}{2}\frac{R^2_{\tt E} \omega_{\tt E}}{c^2a^2}\, 
{\textstyle\frac{1}{2}}\sin2 i_0\sin i_{\tt ecl}\leq
0.98\times 10^{-12}~{\rm m/s}^2,~~~~~~
\label{eq:rel_Eqmot-LT}
\end{eqnarray}  
where we used the unit vectors ${\vec n}_{[\vec X\times\dot{\vec X}]}=[\vec X\times\dot{\vec X}]/|\vec X\times\dot{\vec X}|$, $\vec n_R=\vec{X}/X$ and $\vec n_V=\vec{V}/V$. In addition, we introduce the unit vector ${\vec n}'_Z= \hat{\vec T}{\vec n}_Z$ relating the Earth's angular momentum with the direction to the north pole. Also, $i_0$ is the spacecraft's orbital inclination, $i_0=55^\circ$ for GPS orbits, and where $i_{\tt ecl}=23.5^\circ$ is Earth's axial tilt.

Finally, we consider the de Sitter effect due to the Sun in (\ref{eq:rel_Eqmot}):
{}
\begin{eqnarray}
\ddot {\vec X}^{\tt S}_{\tt dS}&=&(1+2\gamma)\frac{GM_{\tt S}}{c^2r^3_{\tt E}}\big[[{\vec r}_{\tt E}\times \dot{\vec r}_{\tt E}]\times\dot{\vec X}\big]
\simeq
\frac{3}{c^2a_{\tt E}}\Big(\frac{GM_{\tt S}}{a_{\tt E}}\Big)^\frac{3}{2}\sqrt{\frac{GM_{\tt E}}{a}}\, {\textstyle\frac{1}{3}}(1+2\gamma)
\sin i_{\tt ecl}\leq 
8.92\times 10^{-12}~{\rm m/s}^2.
\label{eq:rel_Eqmot-DS}
\end{eqnarray}

The de~Sitter term written here is the dominant solar contribution for Earth satellites. Quantitatively, the dominant contribution to the geodetic (de~Sitter) term in
Eq.~(\ref{eq:rel_Eqmot}) is solar. For a near-Earth satellite, a simple magnitude estimate gives
\begin{equation*}
\frac{a^{\rm Moon}_{\rm dS}}{a^{\rm Sun}_{\rm dS}}
\sim
\frac{GM_{\tt M}\,v_{\tt EM}/r_{\tt EM}^{2}}{GM_{\tt S}\,v_{\tt ES}/r_{\tt ES}^{2}}
\approx 2\times 10^{-4},
\end{equation*}
where $(\vec r_{\tt ES},\vec v_{\tt ES})$ and $(\vec r_{\tt EM},\vec v_{\tt EM})$ denote the Earth--Sun and
Earth--Moon relative state vectors. Thus, if the solar de~Sitter acceleration is $\sim 10^{-11}$~m/s$^2$ (\ref{eq:rel_Eqmot-DS}), the corresponding Moon term is $\lesssim 10^{-15}$~m/s$^2$, and other planets are smaller still.

Thus, relativistic contributions presented above are important and must be kept in the model.

\subsubsection{Transforming equations of motion from {\tt GCRS} to {\tt BCRS}}
 
The metric tensor (\ref{eq:G00tr})--(\ref{eq:Gabtr}) along with the gravitational potentials (\ref{BD-WE-sphe})--(\ref{W-tidal}) describes spacetime in the {\tt GCRS}, which we adopt to formulate the relativistic model for timing and frequency observables. Detailed derivations of the {\tt GCRS} formulation can be found in \cite{Soffel-etal:2003, Turyshev-etal:2013, Turyshev-Toth:2013, Turyshev:2014dea}. As a result, within the local {\tt GCRS}, the following gravitational accelerations due to mass monopoles must be considered:
{}
\begin{itemize}
\item The Newtonian gravitational acceleration from the Earth, given by (\ref{eq:newt}).
\item The relativistic correction to the Earth's gravitational acceleration, provided by the leading term in  (\ref{eq:rel_Eqmot}).
\item The accelerations associated with De Sitter and Lense-Thirring precessions, given by the 2-nd and 3-rd terms in (\ref{eq:rel_Eqmot}).
\item The third-body Newtonian acceleration due to tidal effects from all other solar system bodies, described by (\ref{eq:tidal}).
\end{itemize}
 
Accurate evaluation of third-body accelerations in a {\tt TT}-compatible {\tt GCRS} propagation requires consistent use of barycentric ephemerides (typically delivered in {\tt TDB}-compatible form) and the corresponding {\tt TT}$\leftrightarrow${\tt TDB} time mapping. The procedure, as outlined in \cite{Moyer:2003}, Sec.~4.5.1, involves the following steps:
 {}
\begin{itemize}
\item Convert the integration time from {\tt TT} to {\tt TDB} along the spacecraft's worldline using
Eqs.~(\ref{eq:coord-tr-RR1a})--(\ref{eq:coord-TTa2b}). Although ${\tt TT}-{\tt TDB}$ is periodic with amplitude
$\lesssim 2$~ms, an unconverted ephemeris query can correspond to a barycentric epoch error that maps to
$v_{\tt E}\Delta t \lesssim \SI{60}{m}$ and can produce $\sim$cm-level differences over a 24~h arc in third-body accelerations; we therefore treat the time-scale conversion as part of the deterministic force model.
 \item Obtain the {\tt TDB}-compatible {\tt BCRS} state vectors for both the Earth and the third body from the planetary ephemeris at the corresponding {\tt TDB} epoch. 
\item Determine the position vectors from the spacecraft to the third body and from Earth to the third body using the {\tt TDB}-compatible {\tt BCRS} states of the Earth and the third body, combined with the {\tt TT}-compatible {\tt GCRS} state of the spacecraft.
\item Compute the third-body acceleration using these position vectors and the mass parameter of the third body from the {\tt BCRS}, ensuring compatibility with the {\tt TDB} time scale. 
\item Apply a scaling factor to the computed third-body acceleration by multiplying it by $(1-L_{\tt C})$, where $L_{\tt C}$ accounts for the relativistic corrections.
\noindent\emph{Implementation guard:} Apply the $(1-L_{\tt C})$ factor only if the third-body term was built from {\tt TT}-compatible constants. If you compute the term with {\tt TDB}-compatible states and GM values (e.g., DE440), do not apply an additional $(1-L_{\tt C})$; the scaling in Sec.~\ref{sec:GCRS-pvac} already accounts for it.
\end{itemize}

Note that relativistic corrections to third-body perturbations are typically small and often disregarded. The main effect is due to the Sun's gravitational influence, with smaller contributions from other planets, causing the local inertial frames near Earth to experience De Sitter precession relative to the {\tt BCRS} in the {\tt TDB} frame. Nevertheless, the {\tt TT}-compatible {\tt GCRS} remains kinematically non-rotating relative to the {\tt TDB}-compatible {\tt BCRS}. This non-rotation creates a non-inertial reference frame, requiring the inclusion of Coriolis acceleration.
 
The De Sitter acceleration (\ref{eq:rel_Eqmot-DS}), is a relativistic correction to 3rd-body acceleration and is implicitly included when using the EIH equations in the {\tt BCRS} (see \cite{Moyer:2003}, Sec.~4.4.2). It is important to note that Lense-Thirring precession (\ref{eq:rel_Eqmot-LT}) (see also \cite{Moyer:2003}, Secs.~4.4.3 and 4.5.4), is an independent effect caused by the angular momentum of the Earth (or a planet) and is not included in the EIH equations. The full set of relativistic correction terms is provided in  (\ref{eq:rel_Eqmot}).
  
 \subsubsection{Considering the gravitational harmonics of the Earth}
 
The Earth's motion around the Sun induces relativistic distortions in the gravitational spherical harmonics when observed in the {\tt BCRS}. These distortions complicate the accurate computation of accelerations arising from these harmonics within the barycentric system. To mitigate these effects, it is necessary to apply relativistic corrections, which involve several key steps detailed in \cite{Moyer:2003}, Sec.~4.4.5:
 {}
 \begin{itemize}
\item Transform the spacecraft's position vector from the {\tt TDB}-compatible {\tt BCRS} to the {\tt TT}-compatible {\tt GCRS} using a Lorentz transformation (refer to \cite{Moyer:2003}, Eq.~(4-11)).
\item Calculate the acceleration due to gravitational spherical harmonics in the {\tt TT}-compatible {\tt GCRS}.
\item Convert the computed acceleration back to the barycentric frame using the method outlined in \cite{Moyer:2003}, Eq.~(4-55).
\end{itemize}

This method can be applied to other planets as well. Given the relativistic distortions present in the barycentric reference frame, performing spherical harmonic expansions in a planetocentric system provides improved precision. Additionally, it is imperative to confirm that the reference radius and mass parameters used in the spherical harmonic expansion are properly scaled to account for relativistic effects. If these parameters do not reflect the appropriate relativistic scaling, they must be adjusted to ensure accuracy in the calculations.

This method accounts for the relativistic adjustments that arise from transformations between different space-time coordinate systems. Incorporating these corrections significantly improves the precision of the calculated acceleration due to the geopotential in the geocentric reference frame. Notably, the relativistic correction associated with the acceleration from Earth's quadrupole moment, $J_2$, is extensively discussed in \cite{Soffel:1989}, Sec. 4.2.3.

\subsection{Practical implementation and validation}  
\label{sec:implemnt}

In this Section, we describe an internal frame-closure validation of the implemented {\tt GCRS}$\leftrightarrow${\tt BCRS} dynamics and state transformations. The test compares a trajectory integrated in the {\tt TT}-compatible {\tt GCRS} with a trajectory obtained by transforming the same initial state to the {\tt TDB}-compatible {\tt BCRS}, integrating the barycentric equations of motion, and mapping the result back to the {\tt GCRS}. This closure tests numerical and modeling consistency of the implementation under matched force-model assumptions; it is not an independent validation against tracking observations or against a separate software implementation. Table~\ref{tab:closure-setup} gives the configuration used for the closure test.

\begin{figure}
\begin{minipage}[b]{.329\linewidth}
\includegraphics[width=1.0\linewidth]{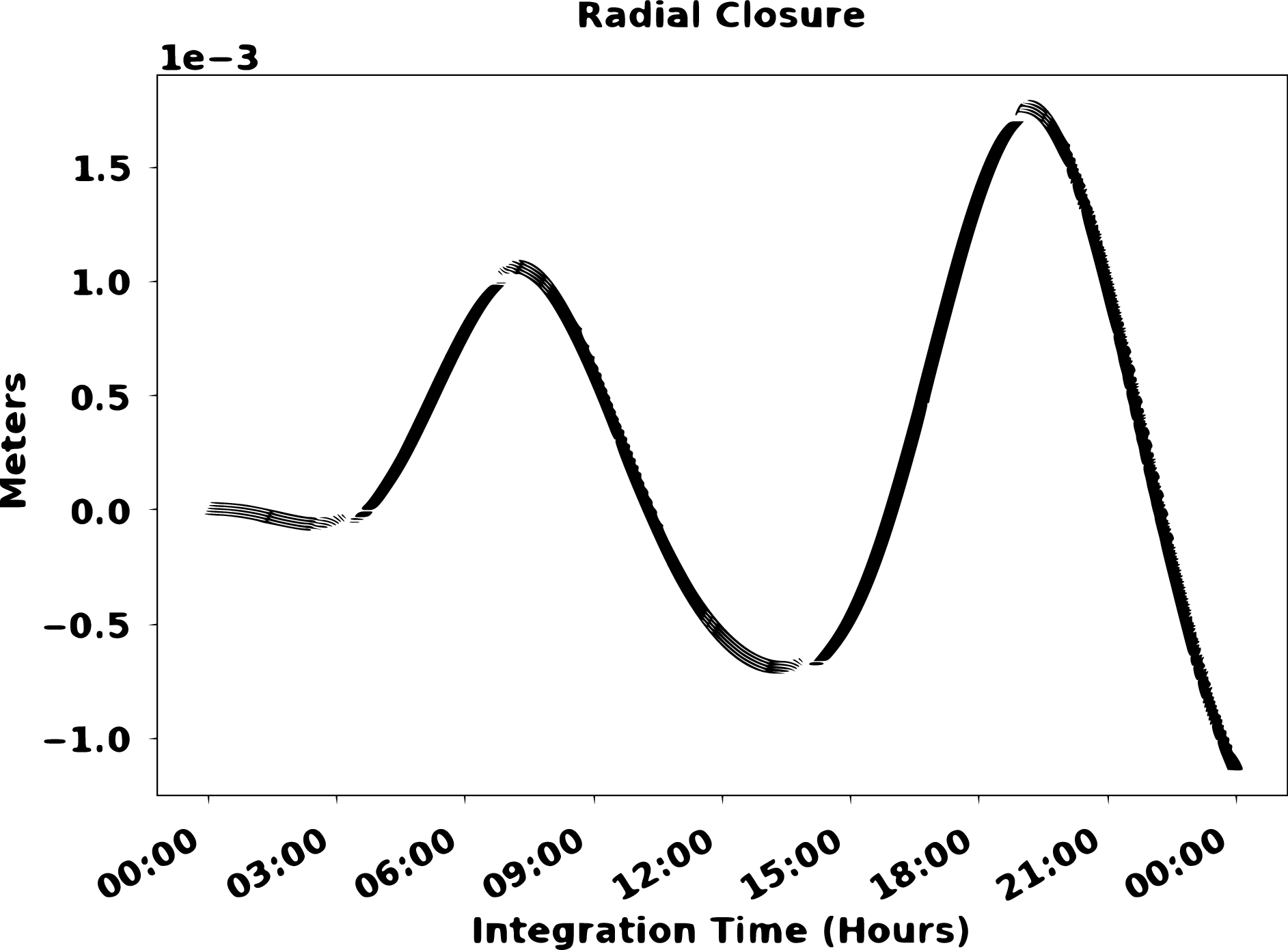}
\end{minipage}
\begin{minipage}[b]{.329\linewidth}
\includegraphics[width=1.00\linewidth]{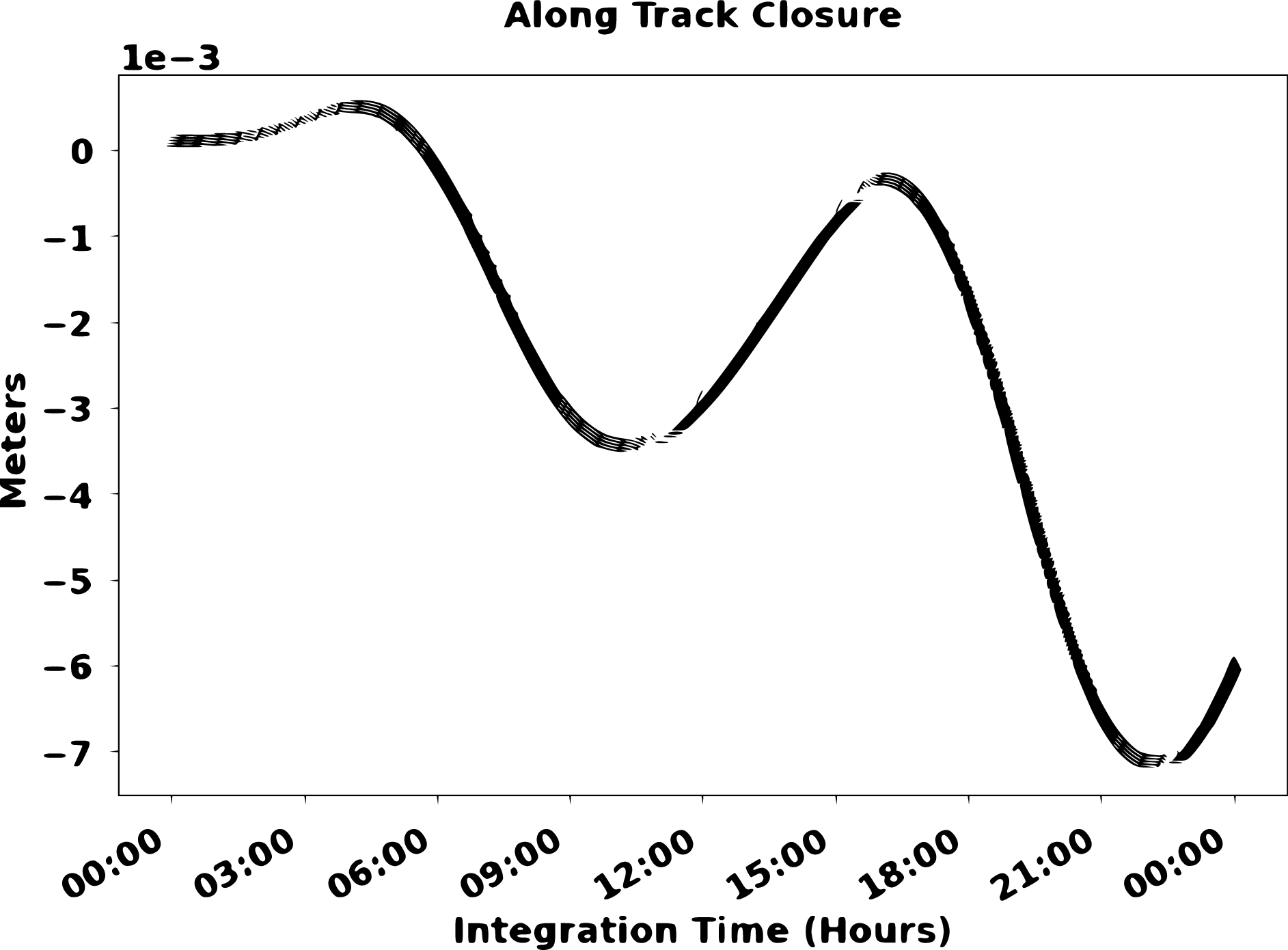}
\end{minipage}
\begin{minipage}[b]{.329\linewidth}
\includegraphics[width=1.00\linewidth]{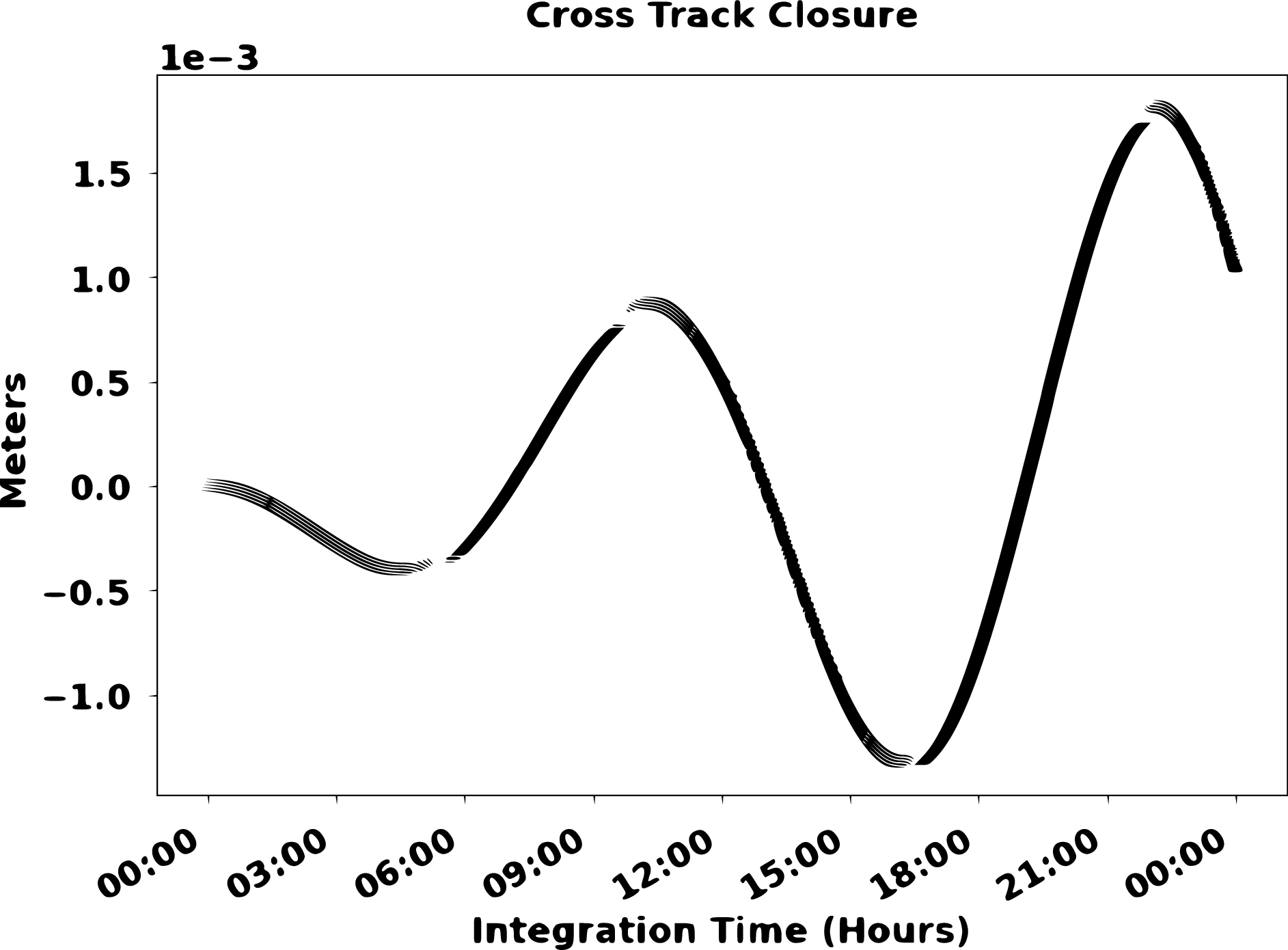}
\end{minipage}
  \vspace{-10pt}
  \caption{Millimeter‑level {\tt GCRS}$\rightarrow${\tt BCRS}$\rightarrow${\tt GCRS} frame-closure test of the {\tt BCRS} implementation in JPL GipsyX. Panels show radial/along‑track/cross‑track residuals $\Delta\mathbf{r}(T)$ (m; note the $10^{-3}$ scale) over 24~h. The test uses DE440 ({\tt TDB} argument) and identical force models in both runs (Table~\ref{tab:closure-setup}); states are mapped using the screened operational transformations 
Eqs.~(\ref{eq:pos-approx-LEO}), (\ref{eq:vel-tran2}), (\ref{eq:coord-acc_sum-L}).}
\label{fig:closure}
  \vspace{-5pt}
\end{figure}

We define the {\it frame-closure residuals} as
\begin{equation}
\label{eq:earth-closure}
\Delta \vec r(T)\equiv \vec X_{{\tt BCRS}\rightarrow{\tt GCRS}}(T)-\vec X_{\tt GCRS}(T),\qquad
\Delta \dot{\vec r}(T)\equiv \dot{\vec X}_{{\tt BCRS}\rightarrow{\tt GCRS}}(T)-\dot{\vec X}_{\tt GCRS}(T),
\end{equation}
where $\vec X_{\tt GCRS}(T)$ is obtained by integrating the {\tt GCRS} equations of motion, and $\vec X_{{\tt BCRS}\rightarrow{\tt GCRS}}(T)$ is obtained by transforming the initial state into the {\tt BCRS}, integrating the corresponding barycentric equations of motion, and transforming the state back into the {\tt GCRS} at each epoch. Figure~\ref{fig:closure} shows the components of $\Delta\vec r(t)$ in the local radial/along-track/cross-track frame.

\begin{table}[h]
\centering
\caption{Configuration used for the closure test shown in Fig.~\ref{fig:closure}. }
\label{tab:closure-setup}
\begin{tabular}{@{}ll@{}}
\hline
Item & Value \\
\hline\hline
Ephemeris & JPL {\tt DE440} (argument: {\tt TDB}) \cite{Park-etal:2021} \\
Earth gravity & Spherical harmonics to degree/order $N=M=20$ ({\tt TT}-compatible) \\
Third bodies & Sun and Moon tidal terms (Newtonian) \\
Relativity ({\tt GCRS}) & Schwarzschild, de~Sitter, Lense--Thirring per Eq.~(\ref{eq:rel_Eqmot}) \\
Relativity ({\tt BCRS})  & EIH point-mass terms per Eq.~(\ref{eq:4-26-mod}) plus transformed local Earth-spin contribution  \\
 & used for the geopotential/spin model \\
SRP/tides & Standard GipsyX models (identical in both frames) \\
Span & 24~h; start epoch UTC: 2025-01-01 00:00:00 \\

State transformations & Screened operational forms:
Eqs.~(\ref{eq:pos-approx-LEO}) [position],
(\ref{eq:vel-tran2}) [velocity],
(\ref{eq:coord-acc_sum-L}) [acceleration] \\
\hline
\end{tabular}
\end{table}

Table~\ref{tab:test-vectors} provides a minimal \emph{template} for regression tests of the \({\tt GCRS}\leftrightarrow{\tt BCRS}\) transformations. It is not itself a numerical validation data set. Independent implementations can use the template to generate check cases once a reference epoch, ephemeris, Earth-orientation series, arithmetic precision, and input state are specified. Barycentric states are referenced to DE440 with \({\tt TDB}\) as the independent argument. Terrestrial station states should use the adopted ITRF realization and the IERS Conventions for the ITRF-to-\({\tt GCRS}\) transformation \cite{Petit-Luzum:2010}. 

Earth spin terms are included in both branches: in the {\tt GCRS} through the Lense--Thirring term in Eq.~(\ref{eq:rel_Eqmot}), and in the {\tt BCRS} through the transformed local Earth-spin acceleration used for the geopotential/spin model. The EIH point-mass terms alone do not contain this contribution. The closure test in Table~\ref{tab:closure-setup} intentionally uses a reduced third-body set containing only the Sun and Moon in both branches; therefore the Venus/Jupiter/Saturn estimates in Eqs.~(\ref{eq:tidal})--(\ref{eq:tidal2a}) quantify omitted operational perturbations but are not part of the closure validation shown in Fig.~\ref{fig:closure}.

One can validate independent implementations by reproducing the transformation test vectors at a fixed epoch using the same ephemeris (DE440), Earth-orientation series, and arithmetic precision. We therefore recommend publishing (as ancillary material) the \emph{numerical} input/output vectors for
Table~\ref{tab:test-vectors} together with the stated tolerances.

\begin{table*}[h]
\centering
\caption{Template regression test vectors for {\tt GCRS}{\,$\leftrightarrow$\,}{\tt BCRS} transformations.}
\label{tab:test-vectors}
\begin{tabular}{@{}llll@{}}
\hline
Item & Input & Output & Notes \\
\hline\hline
Time map & ${\tt TT}=\text{{\tt TT}(UTC 2025-01-01 00:00:00)}$ & ${\tt TDB}$ computed from Eq.~(\ref{eq:coord-TTa1a}) & report \({\tt TDB}-{\tt TT}\) [s] \\
Station position & specified \(\vec X_{\tt TT}\) and epoch & \((\vec r_{\tt E})_{\tt TDB}\) from Eq.~(\ref{eq:pos-approx-LEO}) & report \(\Delta\vec X\) [m] \\
MEO state & specified \((\vec X,\dot{\vec X},\ddot{\vec X})_{\tt TT}\) and epoch & \((\vec r_{\tt E},\dot{\vec r}_{\tt E},\ddot{\vec r}_{\tt E})_{\tt TDB}\) & Eqs.~(\ref{eq:pos-approx-LEO}),(\ref{eq:vel-tran2}),(\ref{eq:coord-acc_sum-L}) \\
\hline
\end{tabular}
\end{table*}

To validate the new relativistic framework we  have developed a rigorous development and testing plan with GipsyX \cite{Bertiger-etal:2020}. GipsyX is precision software developed by JPL for {\tt GNSS}-based positioning, navigation, timing, and Earth science applications. It supports GPS, GLONASS, BeiDou, and Galileo systems and offers precise, centimeter-level positioning for various platforms, including low-Earth orbiters and terrestrial stations. GipsyX integrates real-time and post-processing capabilities and is continuously updated to meet the evolving demands of space geodesy.\footnote{\label{foot:gypxy}For further details on GipsyX, please visit \url{https://gipsyx.jpl.nasa.gov/}}

We have undertaken a comprehensive implementation approach to enhance the accuracy of {\tt GNSS} positioning by incorporating advanced computational models:
\begin{itemize}
\item Imported and used the JPL planetary ephemeris {\tt DE440} \cite{Park-etal:2021} for high-precision celestial mechanics.
    \item Implemented the {\tt BCRS}, integrating state-of-the-art relativistic models to ensure precise trajectory computations.
    \item Developed multiple centers of integration to facilitate robust data handling and computational efficiency.
    \item Integrated models for lunar orientation, lunar gravity, and lunar albedo to enhance lunar navigation capabilities.
\end{itemize}

To validate the implementation of the barycentric coordinate reference frame  and the associated transformations, we conducted a series of rigorous tests:
\begin{enumerate}
    \item Integrated a GPS orbit within the {\tt GCRS} to establish a baseline for comparison.
    \item Transformed the orbital state from the  {\tt GCRS} to the {\tt BCRS}, while maintaining high accuracy in the transformation process.
    \item Re-integrated the GPS orbit in the 
    {\tt BCRS}, starting from the initial condition transformed from the  {\tt GCRS}.
    \item Compared the integrated orbit in the {\tt BCRS} (from Step 3) with the transformed  orbit in the {\tt GCRS} (from Step 2) to assess the precision and accuracy of the transformations.
\end{enumerate}
This testing procedure demonstrates millimeter-level internal consistency of the implemented state transformations and matched force models for the tested GPS-like MEO trajectory. It does not by itself establish end-to-end navigation accuracy from tracking data.

The numerical analysis conducted between the {\tt GCRS} and the {\tt BCRS} demonstrates that orbit state vectors can be consistently mapped between the two reference systems when the appropriate space-time transformations, time-scale conversions, and relativistic acceleration corrections are applied. In the barycentric system, the relativistic corrections to acceleration are an explicit part of the propagated dynamics; in the geocentric formulation, the same physics is represented through the local relativistic terms and the {\tt GCRS}$\leftrightarrow${\tt BCRS} transformation chain. 

The frame-closure residuals defined in Eq.~(\ref{eq:earth-closure}) are evaluated in the {\tt GCRS} after mapping the {\tt BCRS}-integrated state back to the same epoch. For the configuration in Table~\ref{tab:closure-setup}, the residuals remain in the millimeter regime over the 24~h test interval (Fig.~\ref{fig:closure}), with the largest excursion occurring in the along-track component (peak-to-peak \(\sim7\)~mm). These residuals quantify internal consistency of the transformation and force-model implementation. They should not be interpreted as an end-to-end orbit, clock, or observable residual, which would require simulated or real tracking data and an estimation strategy.

\section{Conclusions and next steps}
\label{sec:conclude}

We conducted a comprehensive analysis of general relativistic effects on the dynamics of near-Earth satellites within the {\tt GCRS} and {\tt BCRS} frameworks. Our 24~h closure test demonstrates that, under matched force-model assumptions and consistent time-scale handling, the implemented {\tt GCRS} and {\tt BCRS} formulations are mutually consistent at the few-millimeter level for the tested GPS-like MEO trajectory. To keep relativistic modeling biases below the sub-cm level in range-equivalent units and below the $\mathcal{O}(10)$~ps level in one-way time transfer, it is necessary to apply the {\tt TT}$\leftrightarrow${\tt TDB} and {\tt GCRS}$\leftrightarrow${\tt BCRS} transformations consistently with the dynamical and signal-propagation models, which are especially critical for long-duration and deep-space missions. In this paper we verified internal software consistency via mm-level 24~h {\tt GCRS}$\leftrightarrow${\tt BCRS} frame closure (Fig.~\ref{fig:closure}); end-to-end orbit/clock estimation performance remains a separate observational validation problem depending on the measurement model, data quality, and estimation strategy.

Upcoming cislunar missions will require a formally defined Moon-centered reference system and associated time scales, e.g., \cite{Bar-Sever-etal:2024,UNOOSA:2018,Delepaut-etal:2020,Israel-etal:2020-LunaNet,Parker-etal:2022}. Appendix~\ref{sec:LCRS} defines an {\tt LCRS} coordinate time {\tt TCL} (Luni-centric coordinate time, in analogy to {\tt TCG}) together with a scaled lunar-surface time {\tt TL} (in analogy to {\tt TT}); both are related to {\tt TCB}/{\tt TDB} via the same $1$PN transformation structure, see e.g.,  \cite{Turyshev-etal:2025,Turyshev-scales:2025}. A future step toward operational realization is to specify the conventional origin and scaling needed for timekeeping across surface and orbital assets (i.e., a realized coordinated lunar time for clocks), and to validate the {\tt BCRS}$\leftrightarrow${\tt LCRS} chain via dedicated cislunar time-transfer and tracking experiments.

Several of the necessary transformations, particularly between the \({\tt GCRS}\) and \({\tt BCRS}\), have already been implemented in JPL software used for planetary and lunar ephemerides \cite{Standish_Williams_2008,Williams-Boggs:2015}. For precision lunar applications, however, the lunar gravity, orientation, and surface constants should be tied to post-GRAIL determinations and modern lunar ephemerides rather than to pre-GRAIL engineering compilations. In this paper, pre-GRAIL constants are cited only as historical operational context; the precision lunar terms in Appendix~\ref{sec:time-M} use post-GRAIL or modern ephemeris sources where available \cite{Konopliv-etal:2013,Williams-etal:2014,Smith-etal:2017,Park-etal:2021}. 

The methodology used to develop and validate the relativistic {\tt GNSS} navigation framework in {\tt GCRS} and {\tt BCRS} can be extended to the {\tt LCRS}; demonstrating millimeter-level orbit closure and picosecond-level timing consistency in cislunar geometries will require the {\tt NRHO}-specific screening and regression vectors described in Sec.~\ref{sec:cislunar-demo}. Looking ahead, we identify three potential pathways to mature the {\tt LCRS} into a flight-ready capability:
{}
\begin{enumerate}
\item\textit{Lunar {\tt GNSS} Pathfinder Missions:}  Integrate a compact {\tt GNSS} receiver and atomic clock payload (e.g.\ NASA’s Deep Space Atomic Clock DSAC–II or JPL’s Next-Gen On-Orbit Navigator) aboard a Commercial Lunar Payload Services (CLPS) lander or Lunar Gateway module to measure real-time {\tt LCRS} time offsets and validate centimeter-scale positioning \emph{in situ}.  
\item \textit{Lunar Relay Satellite Demonstrations:}  Deploy a smallsat in a halo orbit about the Earth–Moon L2 point, equipped with dual-frequency {\tt GNSS} antennas synchronized to a space‐qualified hydrogen maser.  This will exercise the {\tt LCRS} transformations under large Sun–Moon gravitational gradients and characterize residual frame-bias errors below 1 cm.  
\item \textit{Clock‐Experiment Campaigns:}  Perform precision time‐transfer tests between DSAC-class clocks on Earth‐orbiting {\tt GNSS} platforms and {\tt LCRS}-referenced clocks on lunar orbiters, quantifying relativistic scale factors and assessing long-term stability at the \(10^{-17}\) level using an extended relativistic framework (e.g., \cite{Turyshev-scales:2025}).
\end{enumerate}

By executing such demonstrations---first in MEO and then in representative Earth--Moon geometries---one can validate the end-to-end {\tt GCRS}/{\tt BCRS}/{\tt LCRS} reference-system infrastructure (state transformations, time scales, and $1$PN light-time) in a mission-relevant configuration. This infrastructure is a prerequisite for cm-class cislunar navigation and ps-level time-transfer; achieved user-level performance will additionally depend on signal availability, measurement quality, and the estimation strategy.

Ongoing work focuses on refining the relativistic corrections and coordinate transformations across {\tt GCRS}, {\tt BCRS}, and {\tt LCRS}, and on integrating them into operational systems such as JPL’s GipsyX. These efforts are essential for upcoming missions targeting the Earth-Moon system and beyond that demand stringent positional accuracy and timing precision. Results, when available, will be reported in subsequent publications. 

\section*{Data and code availability}
\label{sec:data-code}

The relativistic transformations and closure tests reported here were implemented in JPL's GipsyX. The DE440 ephemeris used in the tests is publicly distributed by JPL/NAIF. The GipsyX source code and some operational configuration details are not distributed with this paper. To support independent regression of the reference-frame transformations, an ancillary archive will include, subject to institutional release approval, the reference \({\tt UTC}\), \({\tt TT}\), and \({\tt TDB}\) epochs, the initial \({\tt GCRS}\) state used for the MEO transformation check, and the corresponding transformed \({\tt BCRS}\) state. These ancillary data are intended to test the state transformations independently of GipsyX. Reproducing the full 24~h closure residuals in Fig.~\ref{fig:closure} additionally requires the matched force-model configuration summarized in Table~\ref{tab:closure-setup}; consequently, Fig.~\ref{fig:closure} should be interpreted as an internal software-verification result rather than as a fully independent external reproduction package.

\section*{Acknowledgments}

The work described here was carried out at the Jet Propulsion Laboratory, California Institute of Technology, Pasadena, California, under a contract with the National Aeronautics and Space Administration.

\appendix

\section{Coordinate Transformation  for the Moon System}
\label{sec:time-M}

\subsection{Introducing the {\tt LCRS}}
\label{sec:LCRS}

The International Astronomical Union has adopted a standard Lunar Celestial Reference System ({\tt LCRS}) and Lunar Coordinate Time ({\tt TCL}) \cite{IAU2024ResII}. In this work, we follow that framework and the detailed development in \cite{Turyshev-scales:2025} to provide implementation-ready coordinate and time transformations between the {\tt BCRS} and the {\tt LCRS}. The {\tt LCRS} is realized as a kinematically non-rotating reference system centered at the Moon and aligned with the ICRS axes, analogous to the relation between the {\tt BCRS} and the {\tt GCRS}. The realization of lunar body-fixed orientation and cartographic elements follows the IAU Working Group recommendations; see, e.g., \cite{ArchinalEtAl2018}. These conventions enter the LCRS through the adopted lunar rotation model and the
definition of surface site coordinates. 

Following the structure of the {\tt GCRS} metric \cite{Turyshev-scales:2025}, the {\tt LCRS} with coordinates ($\cal{T}$ = \texttt{TCL}, $\vec{\cal{X}}$) is described by a metric of similar form. The components of this metric are tailored to account for the Moon's gravitational field, tidal effects from nearby massive bodies, and the relativistic corrections required for high-precision trajectory modeling can be expressed in the form (see details in \cite{Turyshev-etal:2025}):
{}
\begin{equation}
ds^2_{\tt LCRS}= \Big\{1 - 2c^{-2}\Big(U_{\tt M}({\cal T}, \vec  {\cal X}) + U^\star_{\rm tid}({\cal T}, \vec  {\cal X})\Big)   +{\cal O}(c^{-4})\Big\}c^2d{\cal T}^2 - \Big\{1 + 2c^{-2}\Big(U_{\tt M}({\cal T}, \vec {\cal X}) + U^\star_{\rm tid}({\cal T}, \vec {\cal X})\Big)+{\cal O}(c^{-4})\Big\} d\vec {\cal X}^2,
\label{eq:metric-CM}
\end{equation}
where $U_{\tt M}({\cal T}, \vec{\cal{X}})$ represents the Newtonian gravitational potential of the isolated Moon, and $U^\star_{\rm tid}({\cal T}, \vec{\cal{X}})$ denotes the tidal potential generated by all other solar system bodies (excluding the Moon), both evaluated at the origin of the  {\tt LCRS}. The omitted terms, arising from $U_{\tt M}^2$, may contribute up to  $\simeq8.43 \times 10^{-19}$, which is negligible for our purposes.

Analogous to the coordinate transformations between  {{\tt GCRS}} and  {{\tt BCRS}} described earlier, the time and position transformations between the  {\tt LCRS} ($\cal{T}$ = \texttt{TCL}, $\vec{\cal{X}}$) and the {{\tt BCRS}} ($t$ = \texttt{TCB}, $\vec{x}$) can be expressed as follows:
{}
\begin{eqnarray}
{\cal T} &=& t-c^{-2}\Big\{\int^{t}_{t_{0}}\Big( {\textstyle\frac{1}{2}}v_{\tt M}^2+\sum_{{\tt B}\not= {\tt M}} {G M_{\tt B} \over r_{\tt BM}}   \Big)dt + (\vec v_{\tt M} \cdot \vec r_{\tt M}) \Big\} +
 {\cal O}\Big(1.30\times 10^{-16}\Big)(t-t_0),
 \label{eq:coord-tr-T1-recM}\\[4pt]
\vec {\cal X} &=& \Big(1+  c^{-2}\sum_{{\tt B}\not= {\tt M}} {G M_{\tt B} \over r_{\tt BM}} \Big)\vec r_{\tt M}+ c^{-2}\Big\{{\textstyle\frac{1}{2}}( \vec v_{\tt M} \cdot\vec r_{\tt M})\vec v_{\tt M}  +(\vec a_{\tt M}\cdot \vec r_{\tt M})\vec r_{\tt M}- {\textstyle\frac{1}{2}}r^2_{\tt M}\vec a_{\tt M}\Big\}+ {\cal O}\Big(1.28\times 10^{-12}~{\rm m}\Big),
\label{eq:coord-tr-XrecM}
\end{eqnarray}
where \(\vec r_{\tt M}\equiv\vec x-\vec x_{\tt M}(t)\), while \(\vec x_{\tt M}\) and \(\vec v_{\tt M}=d\vec x_{\tt M}/dt\) are the Moon's position and velocity vectors in the {\tt BCRS}. For the solar part of the omitted \({\cal O}(c^{-4})\) rate terms, let \(r_{\tt SM}=|\vec x_{\tt S}-\vec x_{\tt M}|\) denote the Sun--Moon distance. The dominant omitted contribution in (\ref{eq:coord-tr-T1-recM}) then scales as $ c^{-4}\{-\frac{1}{8}v_{\tt M}^4-\frac{3}{2}v^2_{\tt M}G M_{\tt S}/r_{\tt SM}+{\textstyle\frac{1}{2}}(G M_{\tt S}/r_{\tt SM})^2\}\lesssim -1.22\times 10^{-16}$, corresponding to an accumulated time magnitude of  \(\approx\)\(10.50\)~ps/d. The uncertainty in (\ref{eq:coord-tr-XrecM}) is set by the omitted contribution from the solar quadrupole moment. The inverse transformations to Eqs.~(\ref{eq:coord-tr-T1-recM})--(\ref{eq:coord-tr-XrecM}) are obtained, to the retained post-Newtonian order, by moving the displayed \(c^{-2}\)-terms to the other side of the equations with the corresponding sign changes.

Using (\ref{eq:metric-CM}) and (\ref{eq:coord-tr-T1-recM})--(\ref{eq:coord-tr-XrecM}), we will derive the new relativistic time scales applicable to the Moon.

\subsection{Transformation between {\tt TL} and  {\tt TDB} }
\label{sec:tm-tdb}

Analogous to (\ref{eq:LG}), the time transformation from Lunar Time ({\tt TL}) at the Moon’s surface   to Luni-centric Coordinate Time  ({\tt TCL}) involves a change in the rate of time  (see nomenclature discussion in \cite{Turyshev-etal:2025})
{}
\begin{align}
\frac{d{\tt TCL}}{d{\tt TL}} = \frac{1}{1  -  L_{\tt L}} = 1 +  \frac{L_{\tt L}}{1  -  L_{\tt L}}\qquad {\rm or} \qquad
\frac{	d{\tt TL}}{d{\tt TCL}} = 1  -  L_{\tt L}.
\label{eq:(26)}
\end{align}

For the Moon, constant $L_{\tt L}$ is derived from its gravity and rotational dynamics as below (see discussion in \cite{Turyshev-scales:2025})
{}
\begin{align}
L_{\tt L} = \frac{1}{c^2}\Big\{ \frac{GM_{\tt M}}{R_{\tt MQ}} \big( 1 + {\textstyle\frac{1}{2}}J_{2\tt M}\big) + {\textstyle\frac{1}{2}} R_{\tt MQ}^2  \omega_{\tt M}^2 \Big\}+{\cal O}\big(3.27 \times 10^{-15}\big).
\label{eq:(29)}
\end{align}

The adopted lunar reference radius for gravity is $R_{\tt MQ} = 1738.0$~km, which is larger than the mean radius of $R_{\tt M} = 1737.1513$~km \cite{Smith-etal:2017}. The lunar gravitational constant $GM_{\tt M} = 4902.800118~{\rm km}^3/{\rm s}^2$ (DE440, \cite{Park-etal:2021}), and the gravity harmonic $J_{2 \tt M}$ is $2.033\times 10^{-4}$ \cite{Williams-etal:2014}, with a rotational angular velocity  $\omega_{\tt M} = 2\pi/(27.32166~{\rm d} \times 86400~{\rm s/d})=2.6616996\times 10^{-6}~{\rm s}^{-1}$. The corresponding value of $L_{\tt L}$, with the larger radius $R_{\tt MQ} $, is estimated to be $L_{\tt L}=3.139\,054\,1 \times 10^{-11}\simeq 2.7121 \,\mu$s/d. The error bound in (\ref{eq:(29)}) is determined by the omitted term involving the tesseral harmonics  $C_{22}=3.4673798\times 10^{-5}\pm1.7\times 10^{-9}$ of the Moon's gravitational field  \cite{Konopliv-etal:2013}.  

The constant  $L_{\tt L}$ allows for the scaling of coordinates and mass parameters to maintain the invariance of the speed of light and the equations of motion within the {\tt LCRS} during the transformation from {\tt TCL} to {\tt TL}. Thus, rather than using the coordinate time ${\cal T} = {\tt TCL}$, spatial coordinates $\vec{\cal X}$, and mass factors $(G M)_{\tt TCL}$ associated with the {\tt LCRS}, we will apply the following scaling for the relevant quantities on the lunar surface:
{}
\begin{equation}
{\tt TL} = {\tt TCL}- L_{\tt L}({\tt TCL}-{\tt T}_{\tt L0}), \qquad
\vec {\cal X}_{\tt TL} = (1-L_{\tt L})\vec {\cal X}_{\tt TCL}, \qquad
(GM)_{\tt TL} = (1-L_{\tt L})(GM)_{\tt TCL},
\label{eq:LSCRS}
\end{equation}
where $ {\tt T_{L0}} $  is the initial lunar time, which we will leave unspecified for the time being.

The transformation from {\tt TCL} to {\tt TCB} 
is determined using (\ref{eq:coord-tr-T1-recM}):
{}
\begin{align}
{\tt TCB}  -  {\tt TCL} = \Big\{\frac{1}{c^2}\int \Big( {\textstyle\frac{1}{2}}v_{\tt M}^2 +  \sum_{{\tt B}\not={\tt M}}{G M_{\tt B} \over r_{\tt BM}}  \Big)dt + \frac{1}{c^2}\big(\vec v_{\tt M} \cdot \vec r_{\tt 0M}\big)\Big\}_{\tt TCB}  +{\cal O}(c^{-4}),
\label{eq:(30-n)}
\end{align}
where \(\vec v_{\tt M}\) denotes the solar-system barycentric velocity vector of the Moon, \(v_{\tt M}\) is its scalar magnitude, \(U_{\tt M,\rm ext}=\sum_{{\tt B}\ne{\tt M}}GM_{\tt B}/r_{\tt BM}\) is the external potential evaluated at the Moon's center, and \(\vec r_{\tt 0M}\) is the Moon-centered position vector of the surface clock or site, expressed in the adopted {\tt LCRS} axes. The integral is taken from the chosen reference epoch \(t_0\) to the current {\tt TCB} epoch. The kinetic term uses the Moon's barycentric velocity in the adopted {\tt BCRS} time coordinate, while the external potential is evaluated at the Moon's center; the site-dependent term is expressed with the {\tt LCRS} surface vector. The dot product reaches an annual maximum of \(\pm0.58~\mu\mathrm{s}\), with smaller variations of \(\pm21\)~ps occurring at the Moon's sidereal period of \(27.32166~{\rm d}\).

Eq.~(\ref{eq:(30-n)}) provides the mean rate between ${\tt TCL}$ and ${\tt TCB}$ as follows:
{}
\begin{equation}
\Big<\frac{d{\tt TCL}}{d{\tt TCB}}\Big> =1-L_{\tt H},
\label{eq:constTCGTCBH}
\end{equation}
with $L_{\tt H}$  estimated to be  $L_{\tt H} =1.4825362\times 10^{-8}$, where $\left<...\right>$ denote the long-time averaging, see \cite{Turyshev-etal:2025}.
Numerical realizations of the ({\tt TCL}--{\tt TCB})/{\tt TDB} relations consistent with the IAU definition are now available as lunar time ephemerides (e.g., {\tt LTE440} \cite{LuYangXie2025LTE440}), which can be used as an independent validation of the constants and periodic terms entering (\ref{eq:(30-n)}).

To express  ${\tt TCB}$ via ${\tt TL}$, we introduce a constant $L_{\tt M}$, which defines the rate between ${\tt TL}$  and ${\tt TCB}$.  In analogy with the introduction of  $L_{\tt B}$, constant $L_{\tt M}$ can be formally introduced as follows
{}
\begin{equation}
\Big<\frac{d{\tt TL}}{d{\tt TCB}}\Big> =1-L_{\tt M}.
\label{eq:LM}
\end{equation}
Then,  by using the chain of time derivatives, we can establish the relationships between the constants $L_{\tt M}, L_{\tt L}$ and $L_{\tt H}$, similar to (\ref{eq:constLBLCLG}). Following this approach, and with the use of (\ref{eq:(26)}), (\ref{eq:constTCGTCBH}) and (\ref{eq:LM}), we obtain the following: 
{}
\begin{equation}
\Big<\frac{d{\tt TL}}{d{\tt TCB}}\Big>=\Big(\frac{d{\tt TL}}{d{\tt TCL}}\Big)\Big<\frac{d{\tt TCL}}{d{\tt TCB}}\Big> \qquad \Rightarrow \qquad (1-L_{\tt M})=(1-L_{\tt L})(1-L_{\tt H}),
\label{eq:consTM}
\end{equation}
from this, the constant $L_{\tt M}$ is determined as $L_{\tt M} \simeq L_{\tt L}+L_{\tt H}-L_{\tt L}L_{\tt H}=
1.485\,675\,294\times 10^{-8}\approx 1.283\,62$~ms/d.

To develop the  $({\tt TCB} - {\tt TL})$ transformation, we apply (\ref{eq:(26)}), (\ref{eq:consTM}), and  (\ref{eq:(30-n)}), leading to the following result:
{}
\begin{equation}
{\tt TCB} - {\tt TL} = \frac{L_{\tt L}}{1-L_{\tt L}}({\tt TL}-{\tt T}_{\tt L0})
+\frac{1}{1-L_{\tt M}}\Big\{\frac{1}{c^2}\int_{\tt T_{\tt L0}}^{\tt TL} \Big( {\textstyle\frac{1}{2}}v_{\tt M}^2 +  \sum_{{\tt B}\not={\tt M}}{G M_{\tt B} \over r_{\tt BM}}  \Big) dt + \frac{1}{c^2}(\vec v_{\tt M}\cdot \vec r_{\tt 0M})\Big\}_{\tt TL},
\label{eq:(12)TMTM-int}
\end{equation}
where $\vec r_{\tt 0M}$ is the position vector of a specific site on the lunar surface in the {\tt LCRS}. 

Expression  (\ref{eq:(12)TMTM-int}) allows us to formulate $({\tt TDB}  -  {\tt TL})$ as a function of {\tt TL}.  To achieve this, we use (\ref{eq:TDBC}), (\ref{eq:consTM}), and (\ref{eq:(12)TMTM-int}), to express  ${\tt TDB}  -  {\tt TL}$ as a function of {\tt TL} as shown below 
{}
\begin{eqnarray}
{\tt TDB}  -  {\tt TL} &=&   {\tt TDB}_0- \frac{L_{\tt B}-L_{\tt L}}{1-L_{\tt L}} \,\big({\tt TL}-{\tt T}_0\big)  +L_{\tt B}({\tt T}_0-{\tt T}_{\tt L0})+ \nonumber\\
&+&\frac{1-L_{\tt B}}{1-L_{\tt M}}\Big\{\frac{1}{c^2}\int_{\tt T_{\tt L0}}^{\tt TL} \Big( {\textstyle\frac{1}{2}}v_{\tt M}^2 +  \sum_{{\tt B}\not={\tt M}}{G M_{\tt B} \over r_{\tt BM}}  \Big) d{\tt TL} + \frac{1}{c^2}(\vec v_{\tt M}\cdot \vec {r_{\tt 0M}}_{\tt TL})\Big\}+{\cal O}(c^{-4}).
\label{eq:(nonN2-int)}
\end{eqnarray}

It is useful to express the difference $({\tt TDB} - {\tt {\tt TL}})$ as a function of {\tt TDB}. This can be achieved by applying (\ref{eq:TDBC}), (\ref{eq:LSCRS}), (\ref{eq:(30-n)}), and (\ref{eq:consTM}), resulting in the following expression (see details in \cite{Turyshev-etal:2025,Turyshev-scales:2025}):
{}
\begin{eqnarray}
{\tt TDB}  -  {\tt TL} &= & \frac{1-L_{\tt L}}{1-L_{\tt B}}\,{\tt TDB}_0
 -\frac{L_{\tt B}-L_{\tt L}}{1-L_{\tt B}} \big( {\tt TDB}-{\tt T}_0\big)  +L_{\tt L}({\tt T}_0-{\tt T}_{\tt L0})+ \nonumber\\
 &+&\frac{1-L_{\tt L}}{1-L_{\tt B}}\Big\{\frac{1}{c^2}\int_{{\tt T_0+TDB_0}}^{\tt TDB}\Big( {\textstyle\frac{1}{2}}v_{\tt M}^2 +  \sum_{{\tt B}\not={\tt M}}{G M_{\tt B} \over r_{\tt BM}}  \Big) d{\tt TDB} +\frac{1}{c^2}(\vec v_{\tt M}\cdot \vec {r_{\tt 0M}}_{\tt TDB})\Big\}+{\cal O}(c^{-4}).
\label{eq:(nonN2-in+)}
\end{eqnarray}

\subsection{Relationships between {\tt TL} and {\tt TT}}

Using expression (\ref{eq:coord-tr-RR1a}) together with (\ref{eq:(nonN2-in+)}) and considering the Moon-related terms,  it is instructive to express the {\tt BCRS}  position vector between a body {\tt B} and the Moon as $\vec r_{\tt BM}=\vec r_{\tt BE}+\vec r_{\tt EM}$, where $\vec r_{\tt BE}=\vec x_{\tt E}-\vec x_{\tt B}$ is the position vector from the body {\tt B} to the Earth, and  $\vec r_{\tt EM}=\vec x_{\tt M}-\vec x_{\tt E}$ is the Earth-Moon relative position vector, also $r_{\tt BM}\equiv|\vec x_{\tt BM}|$, $r_{\tt EM}\equiv|\vec x_{\tt EM}|$.  Also, by representing $\vec v_{\tt M}=\vec v_{\tt E}+\vec v_{\tt EM}$, where $\vec v_{\tt E}$ is the {\tt BCRS} velocity of the Earth and $\vec v_{\tt EM}$ is the Earth-Moon relative velocity, we present expression $(\tt TL-TT)$ as a function of {\tt TT} in the following form \cite{Turyshev-etal:2025,Turyshev-scales:2025}
{}
\begin{eqnarray}
{\tt TL} -   {\tt TT} &=&\frac{L_{\tt G}-L_{\tt L}}{1-L_{\tt B}}\big( {\tt TT}-{\tt T_0}\big) -L_{\tt L}({\tt T}_0-{\tt T}_{\tt L0})-\nonumber\\
  &-&   \frac{1}{c^2}\int_{\tt T_0}^{\tt TT} \Big\{
{\textstyle\frac{1}{2}}v_{\tt EM}^2 +  \frac{GM_{\tt E}-2GM_{\tt M}}{r_{\tt EM}} +W^{\tt S}_{\tt EM}\Big\} d{\tt TT} -\frac{1}{c^2}(\vec v_{\tt EM}\cdot \vec {\cal X}_{\tt TT})+ 
 {\cal O}\Big(c^{-5}; 4.76 \times 10^{-19} ({\tt TT}-{\tt TT}_0)\Big),~~~~~
\label{eq:(nonN2-in+2)-TT-int4-fin}
\end{eqnarray} 
where the solar tidal potential at the Moon $W^{\tt S}_{\tt EM}$ is given by {}
\begin{eqnarray}
\label{eq:(nonN2-in+2)-tid}
W^{\tt S}_{\tt EM}
 &=&\sum_{\ell=2}^{3}
\frac{GM_{\tt S}}{r_{\tt SE}} \Big(\frac{r_{\tt EM}}{r_{\tt SE}}\Big)^\ell P_\ell(\vec n_{\tt SE}\cdot \vec n_{\tt EM}) +{\cal O}\Big(4.30\times 10^{-19}\Big),
\end{eqnarray}
where we kept the solar octupole tidal term  $\ell =3$. The magnitude of this term was estimated to be $\simeq 1.68\times 10^{-16}$, which is small, but large enough to be part of the model.  The error bound here is set by the solar $\ell=4$ tidal contribution, evaluated to be $c^{-2}GM_{\tt S}/r_{\tt SE}(r_{\tt EM}/r_{\tt SE})^4\simeq 4.30\times 10^{-19}$. 

We can present result (\ref{eq:(nonN2-in+2)-TT-int4-fin}) in a different form by introducing the constant $L_{\tt EM}$ and periodic terms $P_{\tt EM}$ as below
{}
\begin{align}
 \frac{1}{c^2} \Big\{
{\textstyle\frac{1}{2}}v_{\tt EM}^2 +  \frac{GM_{\tt E}-2GM_{\tt M}}{r_{\tt EM}} +W^{\tt S}_{\tt EM}\Big\}
=
L_{\tt EM}+\dot P_{\tt EM}(t)+{\cal O}\big(c^{-5}; 4.76\times 10^{-19}\big),
\label{eq:expra=10+}
\end{align}
where the constant rate $L_{\tt EM}\simeq L_{\tt H}-L_{\tt C}$ and the  periodic terms $\dot P_{\tt EM}(t) \simeq \dot P_{\tt H}(t)-\dot P(t)$ are given as below:
{}
\begin{eqnarray}
L_{\tt EM}&=&
\frac{1}{c^2} \Big\{
\Big<{\textstyle\frac{1}{2}}v_{\tt EM}^2 +  \frac{GM_{\tt E}-2GM_{\tt M}}{r_{\tt EM}}\Big> +
\big<W^{\tt S}_{\tt EM}\big>\Big\},
\label{eq:expRR1+}\\
\dot P_{\tt EM}(t)&=&
\frac{1}{c^2} \Big\{{\textstyle\frac{1}{2}}v_{\tt EM}^2 +  \frac{GM_{\tt E}-2GM_{\tt M}}{r_{\tt EM}}-\Big<{\textstyle\frac{1}{2}}v_{\tt EM}^2 +  \frac{GM_{\tt E}-2GM_{\tt M}}{r_{\tt EM}}\Big>+
W^{\tt S}_{\tt EM}-\big<W^{\tt S}_{\tt EM}\big> \Big\}.
\label{eq:expRR2+}
\end{eqnarray}

Result (\ref{eq:expra=10+}), together with (\ref{eq:expRR1+}) and (\ref{eq:expRR2+}), provides valuable insight into the structure of the constant term \( L_{\tt EM} \) and the periodic terms \( P_{\tt EM}(t) \). These expressions can be used to explicitly establish the structure of the series \( P_{\tt EM}(t) \). 

Finally, using   (\ref{eq:expra=10+}) in (\ref{eq:(nonN2-in+2)-TT-int4-fin}), we express $({\tt TL} -   {\tt TT})$ as a function of {\tt TT}:
{}
\begin{align}
{\tt TL} -   {\tt TT} =\frac{L_{\tt G}-L_{\tt L}-L_{\tt EM}}{1-L_{\tt B}}\big( {\tt TT}-{\tt T_0}\big) &- L_{\tt L}({\tt T}_0-{\tt T}_{\tt L0}) - \Big(P_{\tt EM}({\tt TT})-P_{\tt EM}({\tt T_0} )\Big)-\frac{1}{c^2}(\vec v_{\tt EM}\cdot \vec{\cal X}_{\tt TT})\,+\nonumber\\
&+ 
 {\cal O}\Big(4.76 \times 10^{-19} ({\tt TT}-{\tt T}_0)\Big),
\label{eq:(nonN2-in+2)-TT-RR}
\end{align}
 where $\vec{\cal X}_{\tt TT}$ is the {\tt TT}-compatible lunicentric position of the lunar clock.  
 
Considering the  \(O(c^{-2})\) term in \(L_{\tt EM}\) (\ref{eq:expRR1+}), we use Moon–Earth relative speed of 
$v_{\tt EM}\approx1022$ m/s, so the kinematic dilation contributes
$c^{-2}\,\big<\tfrac12\,v_{\tt EM}^2\big>
\simeq5.81\times10^{-12},$
well above our \(5\times10^{-18}\) cutoff.
Taking $r_{\tt EM}$ to be the instantaneous Earth–Moon separation, the Newtonian monopole term at the Earth–Moon distance was estimated to contribute up to  
$c^{-2}(GM_{\tt E}-2GM_{\tt M})/r_{\tt EM}\simeq1.13\times10^{-11}$. The solar quadrupole tide \(W^{\tt S}_{\tt EM}\) yields
$c^{-2}\big<W^{\tt S}_{\tt EM}\big>
\simeq c^{-2}\tfrac14\,{GM_{\tt S}\,r_{\tt EM}^2}/{r_{\tt SE}^3}\simeq1.63\times10^{-14}.$
As a result,  collecting all the contributions, the secular‐drift coefficient for the Earth–Moon system is
\begin{equation}
\label{eq:L_EM}
L_{\tt EM}
=1.709\,390\,6 \times10^{-11}
=1.4769\ \mu\mathrm{s}/\mathrm{d}.
\end{equation}

With $L_{\tt EM}\simeq L_{\tt H}-L_{\tt C}=1.4769~\mu$s/d,  the total constant rate between the clock on or near the lunar surface and its terrestrial analogue to the accepted level of accuracy is estimated to be  
{}
\begin{eqnarray}
L_{\tt B}-L_{\tt M}\simeq L_{\tt G}-L_{\tt L}-L_{\tt EM}=\big(60.2146-2.7121-1.4769\big)~\mu{\rm s/d}=56.0256~\mu{\rm s/d}.
\label{eq:(RR)}
\end{eqnarray}

Note that, if the smaller value for the lunar radius $R_{\tt M}$ is used in (\ref{eq:(29)}) instead of $R_{\tt MQ}$, the value of $L_{\tt L}$ is estimated to be $L_{\tt L}=3.140\,587\,7\times 10^{-11}\simeq 2.7135 \,\mu$s/d. With this value, the total rate in (\ref{eq:(RR)}) is $L_{\tt B}-L_{\tt M}=56.0242~\mu{\rm s/d}$. Also, if the selenoid value of $W_{\tt gM}=2821713.3~\mathrm{m}^2\mathrm{s}^{-2}$ from \cite{Martinec-Pec:1988} is used to determine  $L_{\tt L} = 3.139\,579\,5 \times 10^{-11} \simeq 2.7126~\mu\mathrm{s}/\mathrm{d}$, the value of $L_{\tt B}-L_{\tt M}=56.0251~\mu{\rm s/d}$. This dispersion highlights the need for further studies of the lunar constants. 

Now we consider the periodic term $P_{\tt EM}$, see (\ref{eq:expRR2+}). From the vis–viva relation for the Moon’s motion about the Earth–Moon barycenter given as $v^2_{\tt EM}(r)=(GM_{\tt E}+GM_{\tt M})\big({2}/{r_{\tt EM}}-{1}/{a_{\tt EM}}\big)$, with $r_{\tt EM}=a_{\tt EM}\bigl(1 - e_{\tt M}\cos E\bigr)$, with $e_{\tt M}=0.0549$ being the Moon’s orbital eccentricity With these quantities, the orbital part of  integrand in (\ref{eq:expra=10+}) reads
\[
\frac{1}{c^2}\Big\{\tfrac12\,v_{\tt EM}^2
  +\frac{GM_{\tt E}-2GM_{\tt M}}{r_{\tt EM}}\Big\}
=\frac{1}{c^2}\Big\{\frac{2GM_{\tt E} - GM_{\tt M}}{r_{\tt EM}}
  -\frac{GM_{\tt E}+GM_{\tt M}}{2a_{\tt EM}}\Big\}.
\]
Using this result in (\ref{eq:expRR2+}) and 
expanding  $r_{\tt EM}$ to first order in $e_{\tt M}$, we can write  $r_{\tt EM}=a_{\tt EM}\big(1-e_{\tt M}\cos[\omega_{\tt M}(t-t_0)] +{\cal O}(e^2_{\tt M})\big)$, where $\omega_{\tt M}$ is the Moon’s mean orbital angular rate, we have  \cite{Turyshev-etal:2025,Turyshev-scales:2025}
\begin{align}
\label{eq:orb-P-dot}
\delta \dot P_{\tt EM}(t)&=
\frac{1}{c^2} \Big\{{\textstyle\frac{1}{2}}v_{\tt EM}^2 +   \frac{GM_{\tt E}-2GM_{\tt M}}{r_{\tt EM}}-\Big<{\textstyle\frac{1}{2}}v_{\tt EM}^2 +   \frac{GM_{\tt E}-2GM_{\tt M}}{r_{\tt EM}}\Big>\Big\}=\nonumber\\
&=
\frac{1}{c^2}\big(2GM_{\tt E} - GM_{\tt M}\big)\Big(\frac{1}{r_{\tt EM}}-\frac{1}{a_{\tt EM}}\Big)\simeq 
\frac{1}{c^2}\big(2GM_{\tt E} - GM_{\tt M}\big)\frac{e_{\tt M}}{a_{\tt EM}}\cos[\omega_{\tt M}(t-t_0)]=\nonumber\\
&=1.259\,047 \times 10^{-12}\cos[\omega_{\tt M}(t-t_0)]=0.109\,\cos[\omega_{\tt M}(t-t_0)]~\mu{\rm  s/d}.
\end{align}

Integrating this result in time gives
\begin{align}
\label{eq:orb-P}
\delta  P_{\tt EM}(t)&\simeq -
\frac{1}{c^2}\big(2GM_{\tt E} - GM_{\tt M}\big)\frac{e_{\tt M}}{a_{\tt EM}\omega_{\tt M}}\sin[\omega_{\tt M}(t-t_0)]=-0.473\, \sin[\omega_{\tt M}(t-t_0)]~\mu{\rm  s}.
\end{align}

Thus, as shown in (\ref{eq:(RR)}), there is a secular trend between a clock on the surface of the Moon and one on the Earth's surface. This trend occurs at a rate of $ (L_{\tt B}-L_{\tt M})/(1-L_{\tt B})=6.484\,440\,4  \times 10^{-10}=56.0256~\mu$s/d, with the lunar clock running faster than its terrestrial counterpart. Additionally, there are periodic terms represented by the series $P_{\tt EM}(t)$ with the largest contribution of $0.473\, \sin[\omega_{\tt M}(t-t_0)]~\mu{\rm  s}$, as given by (\ref{eq:orb-P}) with many smaller terms with various periods \cite{Turyshev-scales:2025}. Lastly, small site-dependent terms also contribute at specific periods \cite{Turyshev-etal:2025}. 

\subsection{Position transformations}
\label{sec:pos-M}

To develop the position transformation for the Moon, we define $\vec {\cal X}_{\tt TCL}$ as the position vector of a site on the lunar surface in the {\tt LCRS} and  ${\vec r_{\tt M}}$  as the corresponding vector observed from the {\tt BCRS} (denoted ${\vec r_{\tt M}}_{\tt TCB}$).  The goal is to express the relationship between $\vec {\cal X}_{\tt TCL}$ and ${\vec r_{\tt M}}_{\tt TCB}$  in terms of the {\tt TL} and {\tt TDB} timescales. To achieve this, we use (\ref{eq:LSCRS}) to express the $\vec {\cal X}_{\tt TCL}$  via ${\tt TL} $ and (\ref{eq:TDBC}) to relate ${\vec r_{\tt M}}_{\tt TCB}$ to ${\vec r_{\tt M}}_{\tt TDB}$. This process leads to the following result:
{}
\begin{eqnarray}
\vec {\cal X}_{\tt TL} &=& \Big(1+ L_{\tt B} -L_{\tt L} +c^{-2} \sum_{{\tt B}\not= {\tt M}} {G M_{\tt B} \over r_{\tt BM}} \Big){\vec r_{\tt M}}_{\tt TDB}+ c^{-2} \Big\{{\textstyle\frac{1}{2}}( \vec v_{\tt M} \cdot\vec r_{\tt M})\vec v_{\tt M} \Big\}_{\tt TDB}  +
 {\cal O}\Big(1.00\times 10^{-7}~{\rm m}\Big),~~~~
\label{eq:coord-tr-Xrec=M}
\end{eqnarray}
where $\vec r_{\tt M} \equiv \vec x - \vec x_{\tt M}(t)$  and similar to (\ref{eq:coord-tr-XrecM}), the error is determined by the omitted acceleration-dependent terms.  The inverse transformation to that in (\ref{eq:coord-tr-Xrec=M}) is straightforward; it is performed by shifting the $c^{-2}$-terms to the other side of the equation with the opposite signs. Note that the two leading corrections—the $c^{-2}\!\sum_{\tt B\ne M} GM_{\tt B}/r_{\tt BM}$ scale term and the
$\tfrac12 c^{-2}( \mathbf v_{\tt M}\!\cdot\!\mathbf r_{\tt M})\,\mathbf v_{\tt M}$ Lorentz term—mirror their Earth analogues  in (\ref{eq:coord-tr-XrecRR}).

The first group of terms in (\ref{eq:coord-tr-Xrec=M}) represents a scale change. With $L_{\tt B}-L_{\tt L}=1.547\,380\,65\times 10^{-8}\approx 1.336\,94$~ms/d and $c^{-2}\big<\sum_{\tt B\not=M} G M_{\tt B} / r_{\tt BM}\big> = 0.988438\times 10^{-8}$, the total is $2.53582 \times 10^{-8}$, corresponding to a scale change of $4.41$~cm for the Moon's radius. The annual variation is $\pm1.65\times 10^{-10}$, or $\pm0.29$~mm. Additionally, the Earth's and Moon's annual variations will affect the surface-to-surface distance along the center-to-center line, resulting in an annual variation of $1.34~{\rm mm}|\cos l'|$.

The last term in (\ref{eq:coord-tr-Xrec=M}) accounts for the Lorentz contraction. For a site on the Moon's surface, this contraction amounts to $8.7$ mm. In the direction of motion, the combined effect of the scale change and Lorentz contraction totals $53$ mm, while perpendicular to the motion, the scale change is $44$ mm. These adjustments are essential for accurately determining the positions of retroreflectors used in lunar laser ranging (LLR), see \cite{Williams-Boggs:2015}.

 \subsection{Expressions to transform position, velocity and acceleration}
 
Based on the analysis above and considering realistic measurement uncertainties, we use the magnitudes of the terms for a low-lunar-orbit (LLO) orbiter as the basis of our recommendation. Using the same approach as  in Sec.~\ref{sec:pos-vel-acc}, we present the expressions recommended for position, velocity, and acceleration transformations for the {\tt LCRS}\footnote{Note that these expressions are analogous to those developed for the {\tt GCRS} in (\ref{eq:pos-approx-LEO}), (\ref{eq:vel-tran2}), and (\ref{eq:coord-acc_sum-L}), respectively.}:     
 {}
\begin{eqnarray}
(\vec x-\vec x_{\tt M})_{\tt TDB}  &=&\vec {\cal X}_{\tt TL}- (L_{\tt B} -L_{\tt L})\vec {\cal X}_{\tt TL}-c^{-2}\Big\{ {\textstyle\frac{1}{2}} ( \vec v_{\tt M} \cdot\vec {\cal X})\vec v_{\tt M}+
\gamma U_{\tt M,\rm ext}\, \vec {\cal X}\Big\}_{\tt TL}   + {\cal O}\Big(3.34\times 10^{-7}~{\rm m}\Big),
\label{eq:pos-approx-LEO-rec}\\[4pt]
(\dot{\vec x}-\dot{\vec x}_{\tt M})_{\tt TDB}&=&\dot{\vec {\cal X}}_{\tt TL}-c^{-2}\Big\{\Big(
{\textstyle\frac{1}{2}} v_{\tt M}^{2} +(1+\gamma)U_{\tt M,\rm ext}  
+\big(\vec v_{\tt M} \cdot \dot {\vec {\cal X}}\big)\Big)\dot{\vec {\cal X}}+ 
{\textstyle\frac{1}{2}}( \vec v_{\tt M} \cdot \dot{\vec {\cal X}})\vec v_{\tt M}
\Big\}_{\tt TL} +{\cal O}\Big(7.82\times 10^{-9}~{\rm m/s}\Big),~~~~~~
  \label{eq:vel-tran2-rec}\\[4pt]
    (\ddot{\vec x}-\ddot{\vec x}_{\tt M})_{\tt TDB}&=&\ddot{\vec {\cal X}}_{\tt TL} +(L_{\tt B} -L_{\tt L})\ddot{\vec {\cal X}}_{\tt TL} -c^{-2}\Big\{\Big(
{v_{\tt M}^{2}} +(2+\gamma)U_{\tt M,\rm ext}  + 3\big( \vec a_{\tt M} \cdot \vec {\cal X}\big)+2\big(\vec v_{\tt M} \cdot \dot {\vec {\cal X}}\big)\Big)\ddot{\vec {\cal X}}+
\nonumber\\
 &+&
\Big(
(\vec v_{\tt M}\cdot\vec a_{\tt M}) +(1+2\gamma)\dot U_{\tt M,\rm ext}  
+4\big( \vec a_{\tt M} \cdot \dot{\vec {\cal X}}\big)+\big(\vec v_{\tt M} \cdot \ddot {\vec {\cal X}}\big)\Big)\dot{\vec {\cal X}}+
\Big(( \vec a_{\tt M} \cdot \dot{ \vec {\cal X}})+{\textstyle\frac{1}{2}} ( \vec v_{\tt M} \cdot \ddot{\vec {\cal X}})\Big)\vec v_{\tt M}+
\nonumber\\
 &+& 
\Big(( \vec a_{\tt M} \cdot\vec {\cal X})+ ( \vec v_{\tt M} \cdot \dot{\vec {\cal X}})-(\dot{\vec {\cal X}}\cdot\dot{\vec {\cal X}})-({\vec {\cal X}}\cdot\ddot{\vec {\cal X}})\Big)\vec a_{\tt M}+
(\vec a_{\tt M}\cdot \ddot{\vec {\cal X}})
\vec {\cal X}\Big\}_{\tt TL}   +
 {\cal O}\Big(3.99\times 10^{-15}~{\rm m/s}^2\Big),
   \label{eq:vel-tran2-rec-ddot}
 \end{eqnarray}
 where  ${\vec  v}_{\tt M}=d{\vec  x}_{\tt M}/dt$, ${\vec  a}_{\tt M}=d^2{\vec  x}_{\tt M}/dt^2$, and $ U_{\tt M,\rm ext} $ and $\dot U_{\tt M,\rm ext} $ are given by (\ref{eq:Uext0}) and (\ref{eq:Uext}), correspondingly by replacing {\tt E} with {\tt M}, see \cite{Turyshev-scales:2025}. Also, the error terms are determined by the magnitude of similar terms as in  (\ref{eq:pos-approx-LEO}), (\ref{eq:vel-tran2}), and (\ref{eq:coord-acc_sum-L}).

As a result, for cislunar {\tt GNSS} use, retain Eqs.~(\ref{eq:pos-approx-LEO-rec})--(\ref{eq:vel-tran2-rec-ddot}) and re-screen the omitted terms for the relevant lunar-orbit or transfer envelope. The numerical remainders displayed in Eqs.~(\ref{eq:pos-approx-LEO-rec})--(\ref{eq:vel-tran2-rec-ddot}) are tied to the low-lunar-orbit screening adopted here and should not be reused unchanged for {\tt NRHO} or lunar-transfer states. In the {\tt BCRS}$\leftrightarrow${\tt LCRS} map, \(\vec v_{\tt M}\) is the Moon's barycentric velocity; the scale term \((L_{\tt B}-L_{\tt L})\vec{\cal X}_{\tt TL}\), the external-potential scale term \(c^{-2}U_{\tt M,\rm ext}\vec{\cal X}\), and the Lorentz term \({\textstyle\frac{1}{2}}c^{-2}(\vec v_{\tt M}\!\cdot\!\vec{\cal X})\vec v_{\tt M}\) dominate the position transformation. The largest periodic term in \(({\tt TL}-{\tt TT})\) is \(\approx 0.473\,\mu\)s at the lunar sidereal period, Eq.~(\ref{eq:orb-P}).

%

\end{document}